\documentclass[a4paper, 12pt]{article}
\usepackage[utf8]{inputenc}
\usepackage[T1]{fontenc}
\usepackage{amsmath,bbm,comment}
\usepackage[numbers,sort&compress]{natbib}
\setcitestyle{citesep={,}}
\usepackage{multirow}
\usepackage{amsfonts}
\usepackage{amssymb}
\usepackage{graphicx}
\usepackage{transparent}
\usepackage{notoccite}
\usepackage[normalem]{ulem}
\usepackage[left=2cm, right=2cm]{geometry}
\usepackage{todonotes}
\usepackage{xcolor}
\usepackage{physics}
\usepackage{soul}
\usepackage{ntheorem}
\usepackage{cancel}
\usepackage{setspace} 
\usepackage[overlay,absolute]{textpos}
\usepackage{hyperref}
\usepackage{array}

\newcommand{\be}{\begin{equation}}
	\newcommand{\ee}{\end{equation}}
\newcommand{\bea}{\begin{eqnarray}}
	\newcommand{\eea}{\end{eqnarray}}

\newcommand{\nocitebrackets}[1]{\@cite{\@citex[][#1]}}
\newcommand{\plaincite}[1]{\begingroup
  \let\@cite\@firstofone
  \cite{#1}%
  \endgroup}
\definecolor{DS}{HTML}{629202}
\definecolor{AH}{HTML}{d91f05}
\definecolor{BH}{HTML}{f903d7}

\renewcommand{\dd}{\text{d}}

\newcommand{\Mp}{M_{\rm p}} 
\newcommand{\V}{\mathcal{V}} 
\newcommand{\sgn}{\text{sgn}} 

\DeclareMathOperator\arctanh{arctanh}

\usepackage{framed}

\counterwithin*{equation}{section}
\renewcommand{\baselinestretch}{1.2}

\begin{document}

\thispagestyle{empty}

\rightline{}

\vspace{-1cm}

\begin{center}
{\huge {\bf 
The Lorentzian Geometry of Tunneling in Global de Sitter at Late Time}\\[12pt]}
\bigskip
\bigskip
{\bf Thibaut Coudarchet}\textsuperscript{a,}\footnote{coudarchet@thphys.uni-heidelberg.de},
{\bf Ben Freivogel}\textsuperscript{b,}\footnote{benfreivogel@gmail.com},
{\bf Björn Hassfeld}\textsuperscript{c,}\footnote{hassfeld@wisc.edu}\\
{\bf and Arthur Hebecker}\textsuperscript{a,}\footnote{a.hebecker@thphys.uni-heidelberg.de}

\bigskip
\vspace{0.2cm}
\textsuperscript{a}{\it Institute for Theoretical Physics, Heidelberg University,\\ Philosophenweg 16\,}\&\,{\it 19, 69120 Heidelberg, Germany }\\[5pt] 

\textsuperscript{b}{\it ITFA and GRAPPA, University of Amsterdam, Science Park 904, 1098 XH\\ Amsterdam, the Netherlands}\\[5pt] 

\textsuperscript{c}{\it Department of Physics, University of Wisconsin-Madison, Madison, WI 53706, USA}\\[5pt] 

\bigskip
\bigskip
\end{center}

\begin{center}
\begin{minipage}[ht]{15.0cm}

It is widely believed that Coleman--De Luccia (CDL) instantons characterize tunneling transitions in a de Sitter multiverse.  Their most naive interpretation uses analytic continuation to Lorentzian de Sitter at the minimal size of the spatial three-sphere. However, what one really wants is a geometry where a small bubble of new vacuum forms within the huge spatial sphere of an old parent de Sitter. Even by applying de Sitter isometries to the original CDL solutions, this cannot in general be achieved. In particular, it fails in the gravity-dominated regime, i.e.~for up-tunneling and for transitions with heavy domain walls. These cases remain pathological in that the whole multiverse is in the causal future of every single up-tunneling event. To solve this problem, we develop a Hamiltonian description of how a small off-shell bubble grows and eventually goes on shell within the large spatial sphere of late-time de Sitter. We provide the corresponding WKB analysis, recovering the CDL rate. We explain that our tunneling process and that of CDL are described by two different analytic continuations of a unique on-shell trajectory in the Hamiltonian treatment. Our analysis has crucial implications for the Larfors--Johnson problem, which questions the standard mechanism for populating the string-theoretic flux landscape on the basis of an instability of the relevant domain walls.
\end{minipage}
\end{center}

\newpage

\pagestyle{plain}
\renewcommand{\thefootnote}{\arabic{footnote}}
\setcounter{footnote}{0}

\renewcommand{\baselinestretch}{1.5}


\tableofcontents

\section{Introduction}

At present, the multiverse is arguably the best option for understanding why the cosmological constant is small \cite{Weinberg:1987dv}. However, this comes at a price: One has to quantify how all these vacua are realized dynamically, thereby making predictions possible \cite{Linde:1993nz,Vilenkin:2006xv,Freivogel:2011eg}. A key theoretical motivation for this multiverse paradigm comes from the large number of 4d solutions of string theory -- the so-called string landscape \cite{Bousso:2000xa,Susskind:2003kw,Denef:2004ze} (see \cite{Ibanez:2012zz,Denef:2008wq, Schellekens:2015zua} for reviews and~\cite{Hebecker:2020aqr} for an introduction with a focus on the multiverse paradigm).

In populating the multiverse, the process of creation from nothing \cite{Vilenkin:1982de,Hartle:1983ai,Linde:1983mx,Vilenkin:1984wp} as well as
vacuum transitions presumably play a role. In the present paper we focus on the latter.
The theory of vacuum transitions was first developed for field-theoretic decays \cite{Coleman:1977py,Callan:1977pt} and then extended by Coleman and De Luccia (CDL) \cite{Coleman:1980aw} to include gravitational effects (see also \cite{Parke:1982pm,Brown:1988kg,Eckerle:2020opg,Lavrelashvili:1985vn,Blau:1986cw,Berezin:1987bc,Berezin:1988dz}). CDL propose to look for an appropriate `bounce solution' or `instanton': a solution to the Euclidean equations of motion whose action determines the exponential suppression of the decay rate.
The Lorentzian evolution following the bubble formation may be obtained by analytic continuation of this instanton.
Other approaches to vacuum transitions rely on the Hamiltonian formalism \cite{Fischler:1989se,Fischler:1990pk,DeAlwis:2019rxg,Cespedes:2020xpn,Cespedes:2023jdk}, thermal tunneling \cite{Hawking:1981fz}, the creation of universes behind a black hole horizon \cite{Farhi:1989yr} or the analysis of a `tunneling potential' \cite{Espinosa:2018hue}. 
Whatever the dominant vacuum transition rates turn out to be, they could then be used to make cosmological predictions, as first attempted in \cite{Linde:1993xx,DeSimone:2008bq,Bousso:2006ev,Bousso:2008hz} and hence studied by many authors. For a selection of recent work see e.g. \cite{Hertog:2011ky,Nomura:2012zb,Westphal:2012up,Pedro:2013nda,Hartle:2016tpo,Denef:2017cxt,Carifio:2017nyb,Hassfeld:2022vzk,Khoury:2022ish,Hassfeld:2024ake, Halverson:2026vru}.

In field theory, vacuum decay can be understood as the spontaneous formation of a small patch of new vacuum, separated from the parent vacuum by a domain wall. 
This configuration expands off-shell until the energy gain from the vacuum energy difference outweighs the cost of the growing domain wall. At this point, the bubble configuration goes on-shell and we may say that the transition has been successfully accomplished.

\begin{figure}[h!]
        \centering
        \includegraphics[scale=0.65]{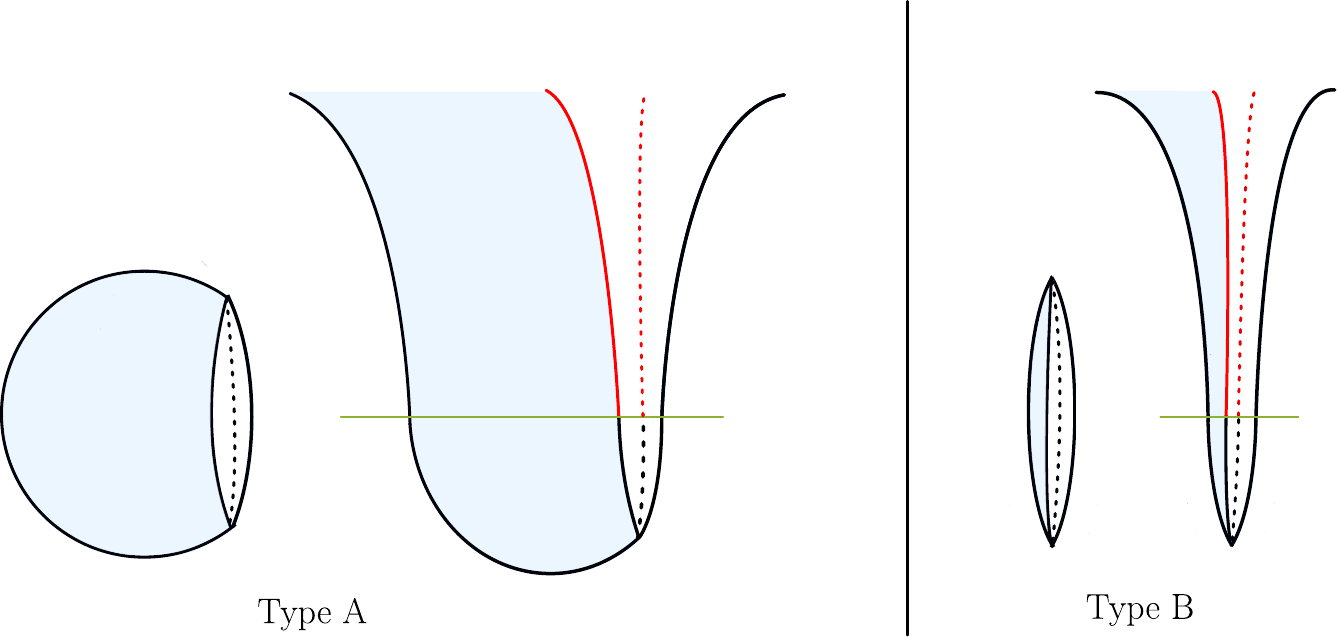}
        \caption{The Type--A and Type--B Euclidean instantons
        describing dS--to--dS transitions according to CDL. Their Lorentzian continuations `at the waist' are shown on the right.}
        \label{fig:cdl_A_B}
\end{figure}

For transitions between de Sitter spaces, this intuition does not apply since energy is not conserved. In particular, the expansion of the parent de Sitter can support the expansion of a new de Sitter even if the latter is not energetically favored. To discuss this in more detail, we will distinguish between Type--A and Type--B instantons, following the naming convention of \cite{Yang:2012cu}. The Type--A instanton is realized by gluing two patches of the different de Sitter four-spheres, with one patch being larger and the other smaller than the respective hemispheres. By contrast, in a Type--B instanton both patches of the four-spheres are smaller than a hemisphere, cf.~Fig.~\ref{fig:cdl_A_B}. In this figure, the parent de Sitter sides are colored blue. Hence, the figure illustrates Type--A down-tunneling on the left. The geometries on the right are supposed to illustrate either Type--B up- or down-tunneling (with the radii intentionally chosen to be similar). Clearly, the Type--A transition on the left can be understood within a dS static patch. It recovers the field theory intuition in the limit where the size of the white patch is much smaller than the blue-patch curvature radius. 

This is not true in the Type--B regime: Here, the field-theoretic logic of a small bubble forming and going on-shell within a larger parent space cannot be easily recovered. As we see on the r.h.~side of Fig.~\ref{fig:cdl_A_B}, the Lorentzian geometry arising from the Type--B instanton initially contains smaller-than-horizon-size spatial patches of both new and parent de Sitter. In other words, most of the parent de Sitter must disappear in the tunneling event, to be replaced by a small patch of the new de Sitter. This appears to be particularly problematic if we want to describe a late-time tunneling process, i.e.~if we want to start with an on-shell parent de Sitter sphere parametrically larger than its horizon radius. One may try to achieve this by boosting (or using a different time slicing for) the half-Lorentzian Type--B geometry shown in the figure. While one can thereby indeed arrange for a small spatial region of new de Sitter to appear in a large parent region, this does not resolve the problems. Most importantly, the whole future multiverse continues to be contained in the causal future of this single tunneling event. This strongly violates standard multiverse logic, as we will explain in more detail below. In summary, it is not clear how to understand a late-time Type--B tunneling event using the CDL instanton and its analytic continuation. Similar issues arise for a Type--A up-tunneling event, which the reader may visualize using the l.h.~side of the figure, but with parent and new de Sitter changing roles.

Motivated by this situation together with the importance of quantifying dS--to--dS transition rates in the asymptotic future, we proceed as follows: We analyze the nucleation of new de Sitter bubbles using a Hamiltonian approach explicitly adapted to the late-time limit of the parent de Sitter.  Our starting point is the standard General-Relativity action for the two de Sitter spaces separated by a thin domain wall. Assuming maximal symmetry, we derive from this a quantum mechanical model with the only dynamical variable being the domain wall radius. This simplifies further in the late-time limit, where global slicing becomes flat slicing.
Note that such a description is very different from the established Hamiltonian treatment of \cite{Fischler:1989se,Fischler:1990pk,Kolitch:1997bi,Bachlechner:2016mtp,DeAlwis:2019rxg,Cespedes:2020xpn,Cespedes:2023jdk}, which starts with a general parametrization but eventually, after imposing the relevant constraints, returns to the static-patch metric.

Our decay rates follow from the WKB treatment of the Wheeler--DeWitt (WDW) equation associated with the quantum-mechanical model just described. Specifically, the rate is governed by the exponentially suppressed value of the WDW wavefunction at the on-shell point. The relevant exponent follows from an explicit integral of the WKB wave number $\kappa=\Im(p)$. The integration range goes from zero bubble radius to the radius where our bubble goes on shell. In the Type--B and the Type--A up-tunneling cases, the latter is the horizon radius.

Throughout this paper, we use decays to nothing \cite{Witten:1981gj,Young:1984jv,Horowitz:2007pr,Yang:2009wz,Blanco-Pillado:2016xvf,GarciaEtxebarria:2020xsr,Dibitetto:2020csn,Draper:2021qtc,Blanco-Pillado:2023aom} as a toy model for vacuum-to-vacuum transitions. Indeed, such decays can be understood as the nucleation of a thin end-of-the-world (ETW) brane \cite{Hassfeld:2023kpu}, and the same conceptual problems as for Type--B vacuum decays arise. Crucially, the analysis is much simpler since the second geometry is absent. While we extensively use this toy model, we also derive all relevant formulae for the more general case with non-trivial geometries on both sides of the domain wall.

Let us now turn to our findings: Our analysis recovers the CDL rate in the Type--A down-tunneling regime. This is not surprising since the new bubble goes on-shell within the static patch of the parent de Sitter. It is then immaterial whether this happens at the minimal radius of the dS sphere, which we will refer to as the dS `waist', as in CDL, or at asymptotically late time.

As a key non-trivial result, we recover the CDL rate also in the Type--B and the Type--A up-tunneling regimes. This is surprising, since the relevant off-shell trajectory of the bubble is very different from what CDL suggests: The bubble is born at zero radius within a parametrically large sphere of the parent de Sitter. It then grows and goes on shell only after reaching the horizon size of the parent dS. The WKB suppression characterizing this tunneling process is found to agree precisely with the CDL result. There, by contrast, small bubbles of the parent and new de Sitter are born together and go on shell at a radius much below horizon size - cf.~the illustration on the very right in Fig.~\ref{fig:cdl_A_B}. 

The agreement of the two processes above can be understood as follows: In both cases, one is dealing with the same function Im$(p(R))$, with $R$ the bubble radius, in the large-$R$ on-shell regime. However, our tunneling process and that of CDL correspond to different analytic continuations. The two integrals calculating the tunneling exponents differ by a corresponding constant, which is offset by the famous CDL-dS-background subtraction.

Our results may have decisive implications for the string landscape. Indeed, in its best-understood corner, i.e.~type IIB flux compactifications on Calabi--Yau orientifolds, the relevant domain walls are D5/NS5-branes wrapped on three cycles of the internal space. Such domain walls have a large tension, such that the relevant CDL process is always of Type--B.  This is precisely the case where, as we argued, the applicability of CDL at late times appears questionable and where our new analysis applies. Even more critically, as shown in \cite{Johnson:2008kc,Johnson:2008vn,Aguirre:2009tp}, the above type IIB domain walls are unstable and their rapid decay threatens the existence of instantons altogether. We will call this phenomenon the `Larfors--Johnson problem'.  Even when assuming other (potentially more suppressed) decay channels, the peculiarity of Type--B transitions from the CDL picture point of view combined with the instability of the wall seems to endanger the very existence of a multiverse. 
We will see how this issue can be directly addressed and set the stage for quantifying how the decay rates are affected by the instability.

We note that the global interpretation of de Sitter tunneling events has been questioned before, both in the CDL and the Hawking--Moss context \cite{Weinberg:2006pc, Brown:2007sd, Miyachi:2023fss, Saito:2024acm}. In particular, \cite{Brown:2007sd} argues for a thermal interpretation of the CDL transition, purely within the static patch. This would fit our conceptual goal of understanding such processes at very late time. However, in addition to being technically quite distinct from our calculation, this appears to clash with our findings conceptually: As we will see, our late-time bubbles can {\it not} go on shell in the static patch - they must reach super-horizon size to start classical expansion.

The paper is organized as follows: In Sect.~\ref{sec:late_time} we briefly review the CDL approach to bubble nucleation and we highlight difficulties in connecting it to nucleation at late times. 
In Sect.~\ref{sec:trajectories}, we review why and how the only allowed Lorentzian wall trajectories are the CDL solutions and their boosts, assuming a pure-tension domain wall.
We study the behavior of the trajectories in detail.
In Sect.~\ref{sec:rates} and App.~\ref{app:dS_dS_solutions} we derive the action of our one-variable quantum mechanical system, starting from the Einstein--Hilbert action.
We can choose to take the parameter to be the radius of the domain wall when focusing on the late-time limit.
We perform a Hamiltonian analysis of the system and derive the tunneling rate using the WKB method.
Surprisingly, we find the same rate, even for Type--B transitions, as that derived by CDL. 
We discuss our results in Sect.~\ref{sec:discussion} and explain how our late-time Lorentzian method could be applied to address questions that are not covered by other approaches to bubble nucleation. We also discuss here the implications of our results to the Larfors--Johnson problem. Finally, we give our conclusions in Sect.~\ref{sec:conclusions}.

\section{Late-time Lorentzian bubble nucleation}
\label{sec:late_time}

We start this section by briefly recalling Coleman--De Luccia transitions in de Sitter space. 
We then review bubbles of nothing, which are vacuum decay processes that are, arguably, easier to describe.
We explain how, by boosting the Lorentzian domain wall trajectories that naively arise from such transitions, one may try to apply the CDL computation to describe late-time vacuum decay.
This turns out to be highly problematic and thus endangers our understanding of an eternally inflating multiverse.

\subsection{Coleman--De Luccia instantons}

We consider Coleman--De Luccia (CDL) \cite{Coleman:1977py,Coleman:1980aw} tunneling in the thin-wall regime. In what follows, we focus on ``dS--dS'' tunneling, but the problem we will raise and the solution we will propose straightforwardly generalize to the case where the created vacuum is Minkowski or AdS. 
The two dS spaces are labeled by $1$ and $2$, and are characterized by expansion rates $H_1$ and $H_2$. 
We will consider transitions from vacuum--1 to vacuum--2.
The CDL instanton is an $O(4)$-symmetric solution to the Euclidean equations of motion derived from the action 
\begin{align}
    S_{\rm E}=\sum_{i\,=\,1}^2\left[\,\Mp^2\int_{\mathcal{M}_i}\sqrt{g_i}\left(-\frac{\mathcal{R}_i}{2} +3H_i^2\right)-\Mp^2\int_{\partial\mathcal{M}_i}\sqrt{h}\mathcal{K}_i\right]+\int_{\partial \mathcal M_1}\sqrt{h}\sigma\,.\label{eq:S_euclidean}
\end{align}
Here $\mathcal{M}_i$, $i=1,2$, are two Euclidean spacetime regions, $\mathcal{R}_i$ are the corresponding Ricci scalars, and $\mathcal{K}_i$ are the extrinsic curvatures of the two boundaries: $\partial \mathcal{M}_1=-\partial \mathcal{M}_2$. The induced metric on the boundary is denoted by $h$. Finally, $\sigma$ is the tension of the domain wall connecting $\mathcal{M}_1$ and $\mathcal{M}_2$, and 
$\Mp=1/\sqrt{8\pi G_{\rm N}}$ is the reduced Planck mass.

The instanton consists of two patches of $4$-spheres with curvature radii $1/H_{1,2}$, glued together along a domain wall with the geometry of an $S^3$ (see Fig.~\ref{fig:cdl_A_B}). 
The metric for each sphere reads
\begin{align}
    \dd s_{1,2}^2=\frac{1}{H_{1,2}^2}\left(\dd\chi^2+\sin(\chi)^2\dd\Omega_3^2\right)\,,\label{eq:g_euclidean}
\end{align}
and the Israel conditions fix the location of the domain wall to a subspace defined by $\chi=\chi_{1,2}$ (different for each 4-sphere). The radius of the $S^3$ domain wall is given by
\begin{align}\label{eq:R_crit_dS_dS}
    R_{\rm crit}^2=\frac{4\xi^2}{(H_1^2-H_2^2)^2+2\xi^2(H_1^2+H_2^2)+\xi^4}\qquad\qquad\mbox{with}
    \qquad\qquad
    \xi\equiv \frac{\sigma}{2\Mp^2}\,.
    \end{align}

As explained in the introduction, there are two qualitatively different instanton geometries, which we will refer to as Type--A and Type--B:

Type--A instantons are characterized by one of the two sphere-patches being larger than a hemisphere and the other one being smaller. 
The l.h.~side of Figure~\ref{fig:cdl_A_B} shows the case in which the larger patch belongs to the `parent de Sitter' (in blue), i.e., $H_1>H_2$.
This is the geometry relevant for `down-tunneling processes'. 
For an `up-tunneling process', i.e., for $H_1<H_2$, the roles of the `small' and the `large' patches are reversed.

For Type--B instantons, both patches are smaller than a hemisphere, cf.~the r.h.~side of Fig.~\ref{fig:cdl_A_B}. 

The geometry where both patches are larger than a hemisphere requires a negative-tension domain wall. We will not consider such unstable configurations in the following.

In both Type--A and Type--B situations, the Lorentzian spacetime, after bubble nucleation, can be obtained by first reparameterizing the instanton such that each part of the four-sphere is given by a metric
\begin{align}
    \dd s^2=\frac{1}{H_{1,2}^2}\left(\dd \theta^2+\sin^2(\theta)(\dd \alpha^2+\sin^2(\alpha)\dd \Omega_2^2)\right)
\end{align}
and the location of the domain wall is determined by $\sin(\theta)\cos(\alpha)=\text{const}$.
Then, the Lorentzian spacetime is obtained by the analytic continuation
\begin{align}\label{eq:analyt_cont}
    \theta \to \frac{\pi}{2}-iH_{1,2}t\,.
\end{align}
The resulting geometry is shown in Fig.~\ref{fig:cdl_A_B}.
More precisely, the geometries shown are obtained by cutting both the Euclidean and the Lorentzian spacetimes in half and gluing them together.
In the Type--B version of this `half-Lorentzian' geometry, the initial on-shell spatial geometry is small compared to either dS radius. 
It can nevertheless grow because the large tension of the domain wall pulls the space to larger size.
In fact, if the tension is sufficiently large, the energy densities of the two dS spaces can be negligible during the initial stage of this on-shell expansion. 

\begin{figure}[h!]
        \centering
        \includegraphics[scale=0.25]{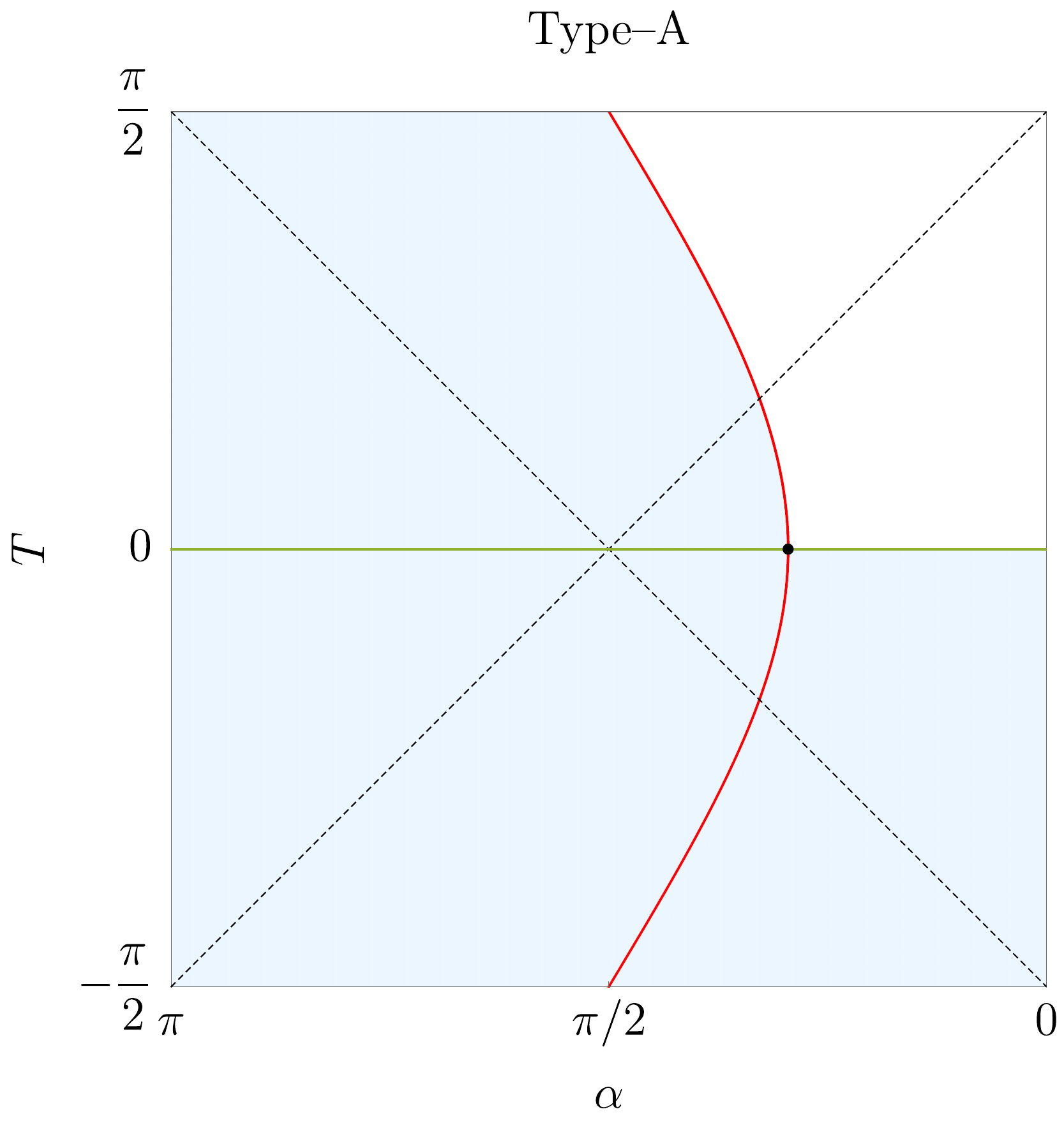}
        \quad
        \includegraphics[scale=0.25]{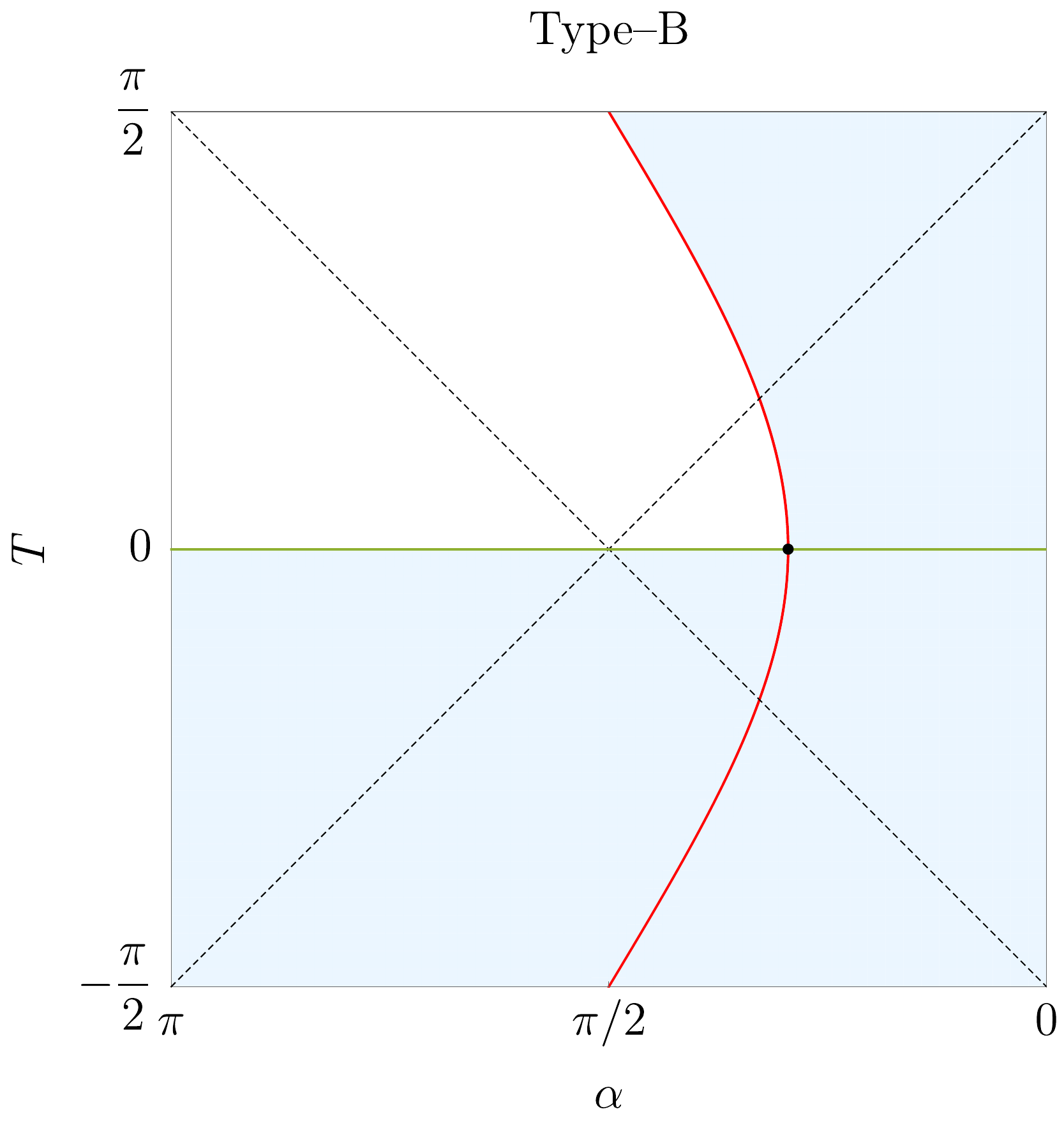}
        \caption{Penrose diagrams of dS tunneling events of Type--A (left) and Type--B (right). 
        The parent dS regions are shown in blue and the CDL domain wall in red. The waists, shown in green, have sections where the blue regions end and the new dS spaces must be attached.}
        \label{fig:typeAB}
\end{figure}

The spacetimes and the domain wall trajectories therein can be represented in Penrose diagrams, cf.~Fig.~\ref{fig:typeAB}.
Both diagrams show a dS space with the domain wall trajectory (red line) that corresponds to the analytic continuation of the CDL solution.
For a CDL transition, as described by the instanton and half-Lorentzian geometries from Fig.~\ref{fig:cdl_A_B}, it is natural to assume that the lower halves of the dS spaces remain the parent dS phase. The domain wall emerges only at the black dot -- it is simply absent below the `waist'. In both pictures, the blue shaded region belongs to the parent dS and the new dS (which is not shown) has to be glued into the white region.
We observe that Type--B transitions have the peculiar feature that most of the spatial 3-sphere of the parent dS disappears during the tunneling event.
Note that in both cases, the domain wall takes on its minimal radius at the point of nucleation.

For completeness, we state the global Lorentzian dS metric and its relation to the Penrose diagrams:
\begin{align}
    \dd s^2&=-\dd t^2+\frac{\cosh^2\left(H_1t\right)}{H_1^2}\left(\dd\alpha^2+\sin^2(\alpha)\dd\Omega_2^2\right)\,,\quad -\infty<t<\infty\,,\quad 0\leq\alpha\leq \pi\,,
    \label{eq:global_dS_coordinates}\\
    \dd s^2&=\frac{1}{H_1^2\cos^2(T)}(-\dd T^2+\dd\alpha^2+\sin^2(\alpha)\dd\Omega_2^2)\,,\qquad -\frac{\pi}{2}<T<\frac{\pi}{2}\,,\quad 0\leq \alpha\leq \pi\,.
    \label{eq:global_dS_coordinates_compact}
\end{align}
The coordinates $T$ and $\alpha$ define the axes of the Penrose diagrams.

The decay rate $\Gamma$ of a dS decay can be computed from the Euclidean actions of the instanton and the vacuum configurations:
\begin{align}\label{eq:Decay_rate_def}
    \Gamma \sim \exp(-B^{\rm CDL})\,,\qquad B^{\rm CDL}=S_{\rm E,instanton}-S_{\rm E,vacuum}\,.
\end{align}
For down-tunneling, i.e.,~$H_1>H_2$, the decay exponent $B^{\rm CDL}$ takes the form \cite{Coleman:1980aw,Parke:1982pm}
\begin{align}
\label{eq:B_rxy}
    B^{\rm CDL}=\frac{\pi^2\sigma^4}{2\Mp^6 (H_1^2-H_2^2)^3}r(x,y)=\frac{8\pi^2\Mp^2 x^3}{\xi^2}r(x,y) \,,
\end{align}
with the dimensionless variables $x,y$ and the function $r$ defined as
\begin{align}\label{eq:dim_less_variables}
    x=\frac{\xi^2}{H_1^2-H_2^2}\,,\quad y=\frac{H_1^2+H_2^2}{H_1^2-H_2^2}\,,\quad r(x,y)=2\frac{1+xy-\sqrt{1+2xy+x^2}}{x^2(y^2-1)\sqrt{1+2xy+x^2}}\,.
\end{align}
The instanton is of Type--A if $x<1$ and of Type--B if $x>1$.
For given Hubble constants $H_1$ and $H_2$ the value of $x$ is determined by the tension of the domain wall.
Small tensions lead to Type--A transitions while large tensions correspond to Type--B.
For $x\ll 1$, the decay occurs in the field-theoretic regime and $r(x,y)\to 1$.

The up-tunneling decay exponent $B_{\rm up}^{\rm CDL}$, for $H_1<H_2$, is determined by the `detailed-balance' relation
\begin{align}
    B_{\rm up}^{\rm CDL}=B_{\rm down}^{\rm CDL}+8\pi^2 \Mp^2 \left(\frac{1}{H_1^2}-\frac{1}{H_2^2}\right)\,.\label{eq:detailed_balance}
\end{align}
In terms of the variables $x,y$ defined in \eqref{eq:dim_less_variables}, $B_{\rm up}^{\rm CDL}$ takes the form \eqref{eq:B_rxy}, but with the function $r(x,y)$ now being replaced by
\begin{equation}
r(x,y)\to r_{\rm up}(x,y)=-2\frac{1+xy+\sqrt{1+2xy+x^2}}{x^2(y^2-1)\sqrt{1+2xy+x^2}}\,.
\end{equation}
Note that $x,y$ as well as $r_{\rm up}(x,y)$ are now negative. The prefactor in \eqref{eq:B_rxy} is negative as well such that the decay exponent remains positive and it is indeed larger than the corresponding down-tunneling exponent.
The absolute value of $x$ now determines if the instanton (which is the same as for the down-tunneling process) is of Type--A or Type--B.
The regime $|x|<1$ corresponds to Type--A whereas $|x|>1$ corresponds to Type--B.

\subsection{Bubbles of nothing (BoNs)}
Bubbles of nothing are special vacuum decay processes.
They were first discovered in \cite{Witten:1981gj} and subsequently studied in e.g.~\cite{Young:1984jv,Horowitz:2007pr,Yang:2009wz,Blanco-Pillado:2016xvf,GarciaEtxebarria:2020xsr,Dibitetto:2020csn,Draper:2021qtc,Blanco-Pillado:2023aom}.
Bubbles of nothing are technically simpler to analyze than dS--dS transitions, as there is only one dS space participating, and we will use them to develop our analysis in what follows.
De Sitter BoNs can exist whenever the fundamental theory contains an end-of-the-world (ETW) brane.
The decay process is still governed by the action \eqref{eq:S_euclidean}, but without the second ($i=2$) term in the sum and with $\sigma = 2\Mp^2\xi$ now denoting the ETW brane tension.
The bubble of nothing configuration is then given by a dS space with expansion rate $H$ with a spacetime boundary, the location of the ETW brane with tension $\sigma$. 
We note that the original Witten bubble \cite{Witten:1981gj} can be understood from an effective field theory point of view in precisely this way: It is a decay of Minkowski space triggered by the existence of a negative tension ETW brane \cite{Hassfeld:2023kpu}.
Unlike domain walls interpolating between two different vacua, negative tension ETW branes need not cause an instability. UV-complete ETW branes include O8--planes of string theory and the IR region of a Klebanov--Strassler throat \cite{Klebanov:2000hb}. The equivalence of the latter to the IR-brane of the Randall--Sundrum model 
\cite{Randall:1999ee,Giddings:2001yu} was made explicit in e.g.~\cite{Brummer:2005sh}.

\begin{figure}[h!]
        \centering
        \includegraphics[scale=0.65]{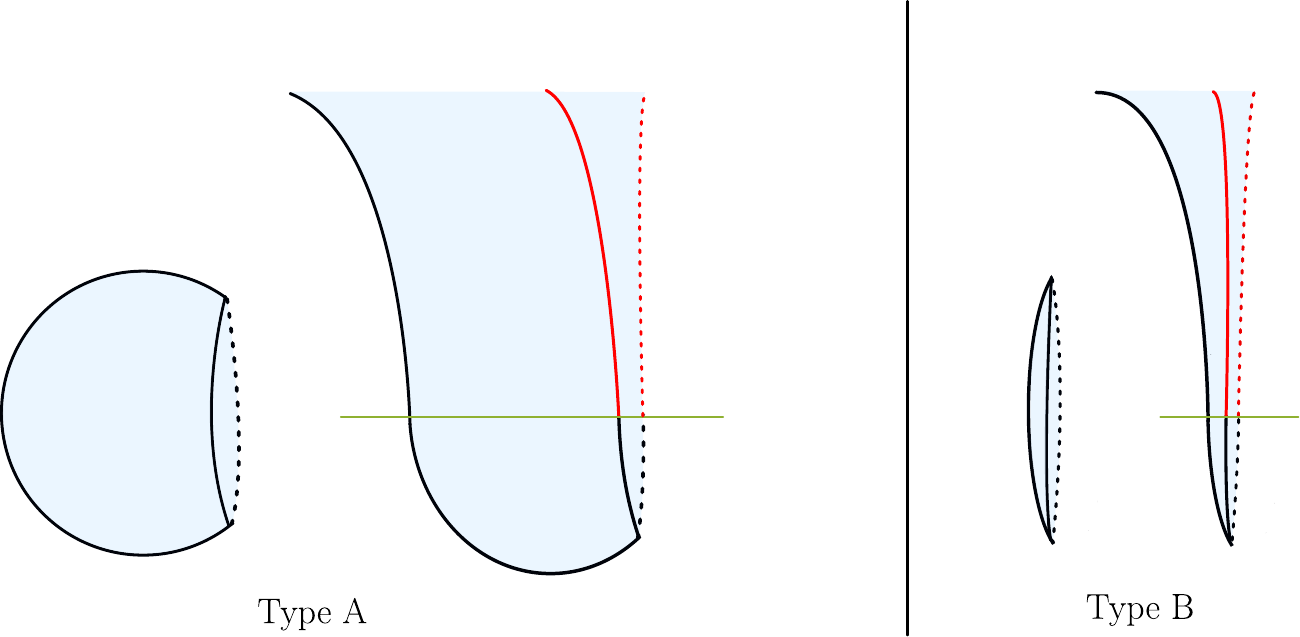}
        \caption{The Type--A and Type--B Euclidean instantons and their Lorentzian continuations at the waist for a bubble of nothing decay.}
        \label{fig:cdl_BoN}
\end{figure}

As before, to compute the decay rate, we are looking for an $O(4)$-symmetric instanton.
Its geometry is a patch of a four-sphere with an $S^3$ boundary. The latter is the location of the ETW brane.
The critical radius is determined by
\begin{align}\label{eq:R_crit_BoN}
    R_{\rm crit}^2=\frac{1}{\sigma^2/(4\Mp^4)+H^2}=\frac{1}{\xi^2+H^2}\qquad \mbox{or}\qquad \xi^2=\frac{1}{R_{\rm crit}^2}-H^2\,.
\end{align}
The transition is of Type--A, i.e., the instanton is larger than a hemisphere, if $\sigma<0$ and it is of Type--B if $\sigma>0$.
As depicted in Fig.~\ref{fig:cdl_BoN}, the analytic continuation of the instanton leads to a Lorentzian spacetime geometry with a timelike boundary. 
The decay rate to `nothing' can be computed by applying Eq.~\eqref{eq:Decay_rate_def}. The result is \cite{Hassfeld:2023kpu}
\begin{align}\label{eq:B_BoN}
    \Gamma\sim\exp(-B^{\rm CDL})\,,\qquad B^{\rm CDL}=\frac{4\pi^2\Mp^2}{H^2}\left(1+\frac{\xi}{\sqrt{H^2+\xi^2}}\right)\,.
\end{align}

\subsection{Challenges for the late-time tunneling interpretation of CDL}
A geometric connection between the CDL (or BoN) instantons and the corresponding Lorentzian spacetimes after tunneling can only be made through the gluing processes depicted in Figs.~\ref{fig:cdl_A_B} and~\ref{fig:cdl_BoN}.
The pictures naively suggest that dS vacuum decay always happens when the spatial dS sphere is at its minimal radius, i.e., at the dS waist.
However, it is to be expected that, at least when the decay occurs in the field-theoretic regime, dS decays can occur at any time. 
We are especially interested in decays at late times when the spatial dS sphere has become large, as we expect most of the decay processes to occur in this regime.

To investigate whether the decay rate computation is still appropriate for decays that occur at late times, one may first make use of the dS symmetries.
The CDL instanton is an $O(4)$ symmetric configuration which, after analytic continuation to Lorentzian signature, becomes an $O(1,3)$-symmetric spacetime.
The $O(1,3)$ group is generated by $SO(3)$ rotations of the spatial configuration and Lorentz boosts. 

\begin{figure}[h!]
        \centering
        \includegraphics[scale=0.25]{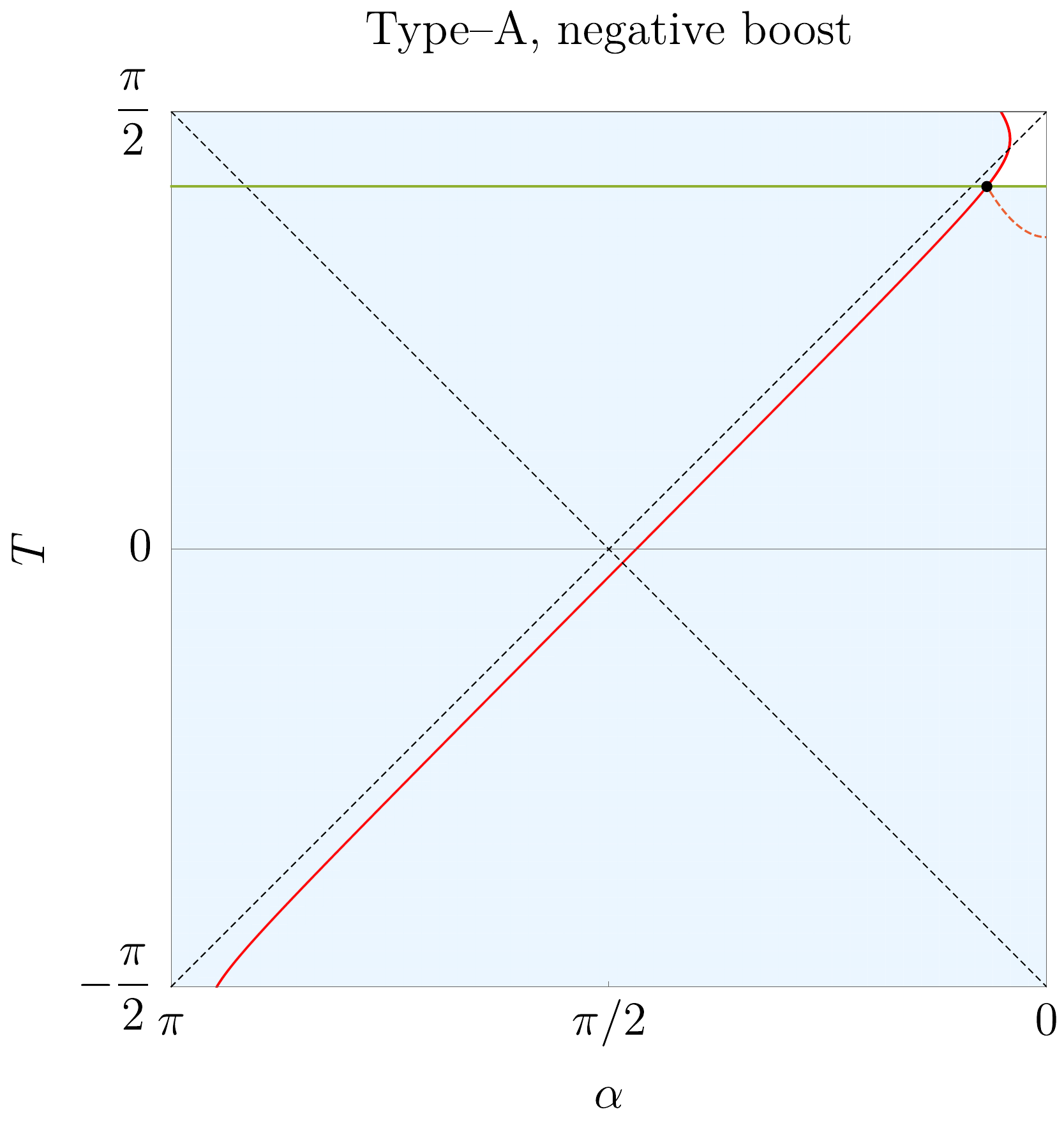}
        \quad
        \includegraphics[scale=0.25]{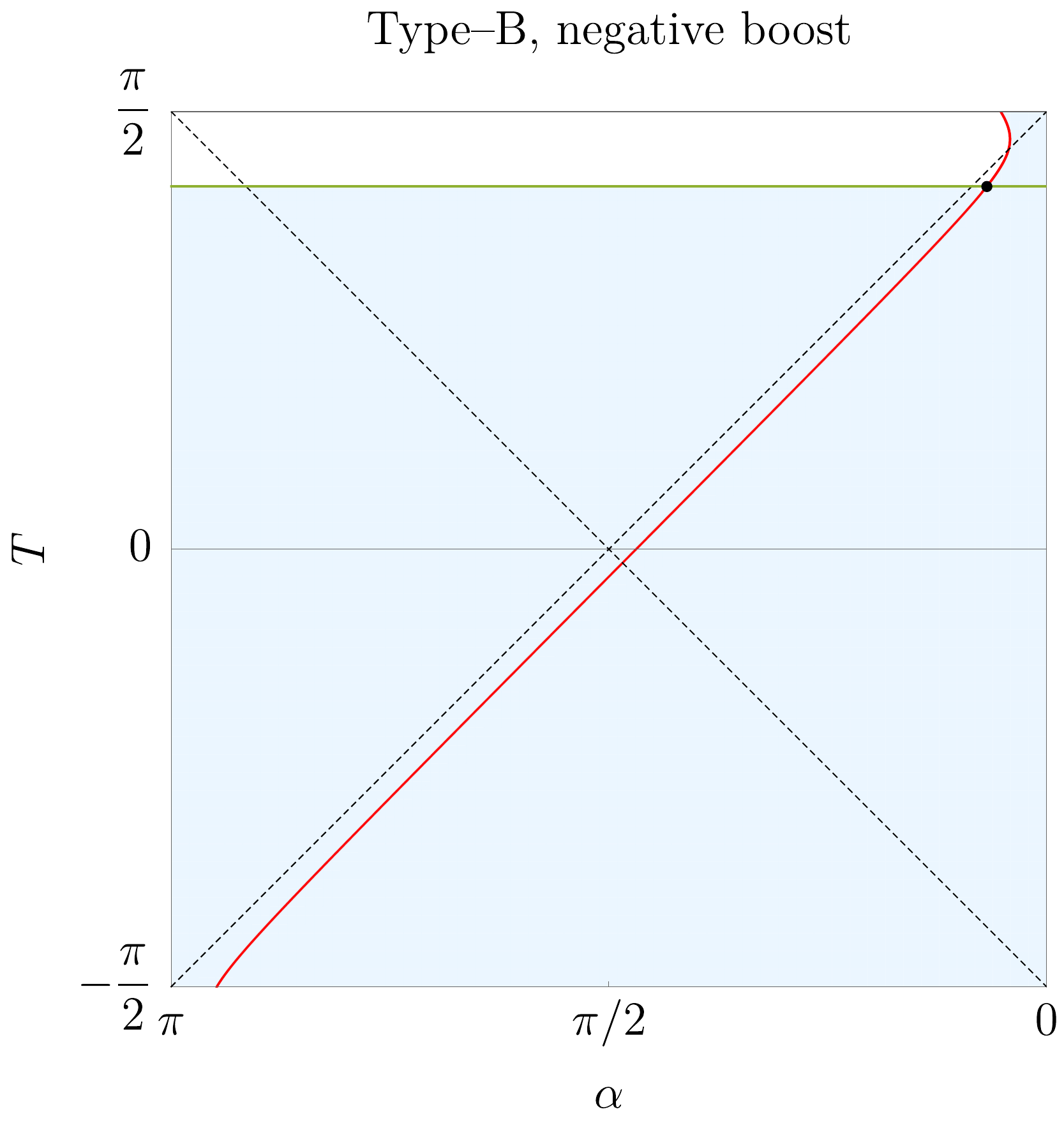}
        \caption{On the left: Negatively boosted CDL solution for Type--A, such that nucleation of a small bubble occurs on the late-time green line. The off-shell trajectory that we consider in our late-time approach is illustrated by the red dashed line. The point of nucleation has been boosted correspondingly so the picture remains conceptually the same as the unboosted case and we can trust the CDL decay rate computation. On the right: Negatively boosted solution for Type--B. In this case, however, this does not describe the creation of a small bubble at late times since on the green line a large part of the parent de Sitter spacetime disappears.}
        \label{fig:typeAB_boosted}
\end{figure}

In Fig.~\ref{fig:typeAB_boosted}, we depict the spacetime configurations arising from boosting the original CDL Type--A (left) and Type--B (right) Lorentzian geometries depicted in Fig.~\ref{fig:typeAB}.
We see that due to the boost, the point of minimal domain wall radius, depicted by a black dot as before, is pushed to later times. 
For the Type--A situation (left), we can imagine a vacuum transition to occur by first having a tiny region of new vacuum separated from the parent space by a domain wall being created as a quantum fluctuation, which then grows as an off-shell process until it is large enough to go on-shell. 
The presumed off-shell trajectory of the domain wall is illustrated by the red dashed line.
This is the natural instanton interpretation for field-theoretic decays.
It is hence totally plausible that this is also the correct interpretation for the Type--A CDL process and that the decay rate for such a transition is still given by \eqref{eq:B_rxy}.

However, consider now the Type--B geometry (right).
The same tunneling interpretation as before would lead us to the conclusion that the domain wall must undergo a very large off-shell transition.
Indeed, the domain wall would need to nucleate on the left side of the diagram and then transition all the way past the equator of the spatial dS sphere, which is very large at late times, to go on-shell at its minimal possible radius (black dot). 
It seems totally unreasonable for this transition to be the dominant decay process, as a huge portion of the parent spacetime disappears during the tunneling event.

\begin{figure}[ht!]
        \centering
        \includegraphics[scale=0.25]{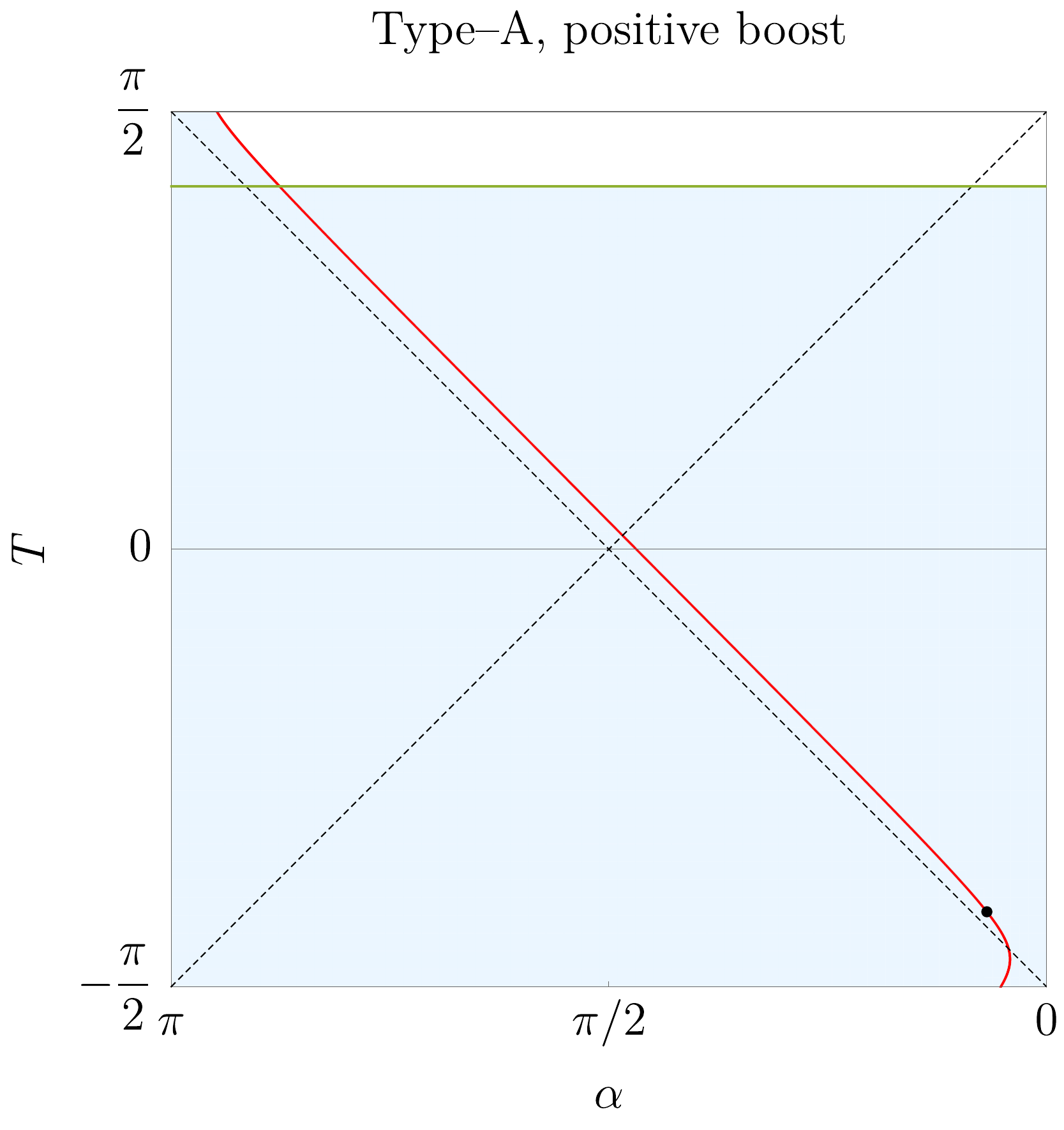}
        \quad
        \includegraphics[scale=0.25]{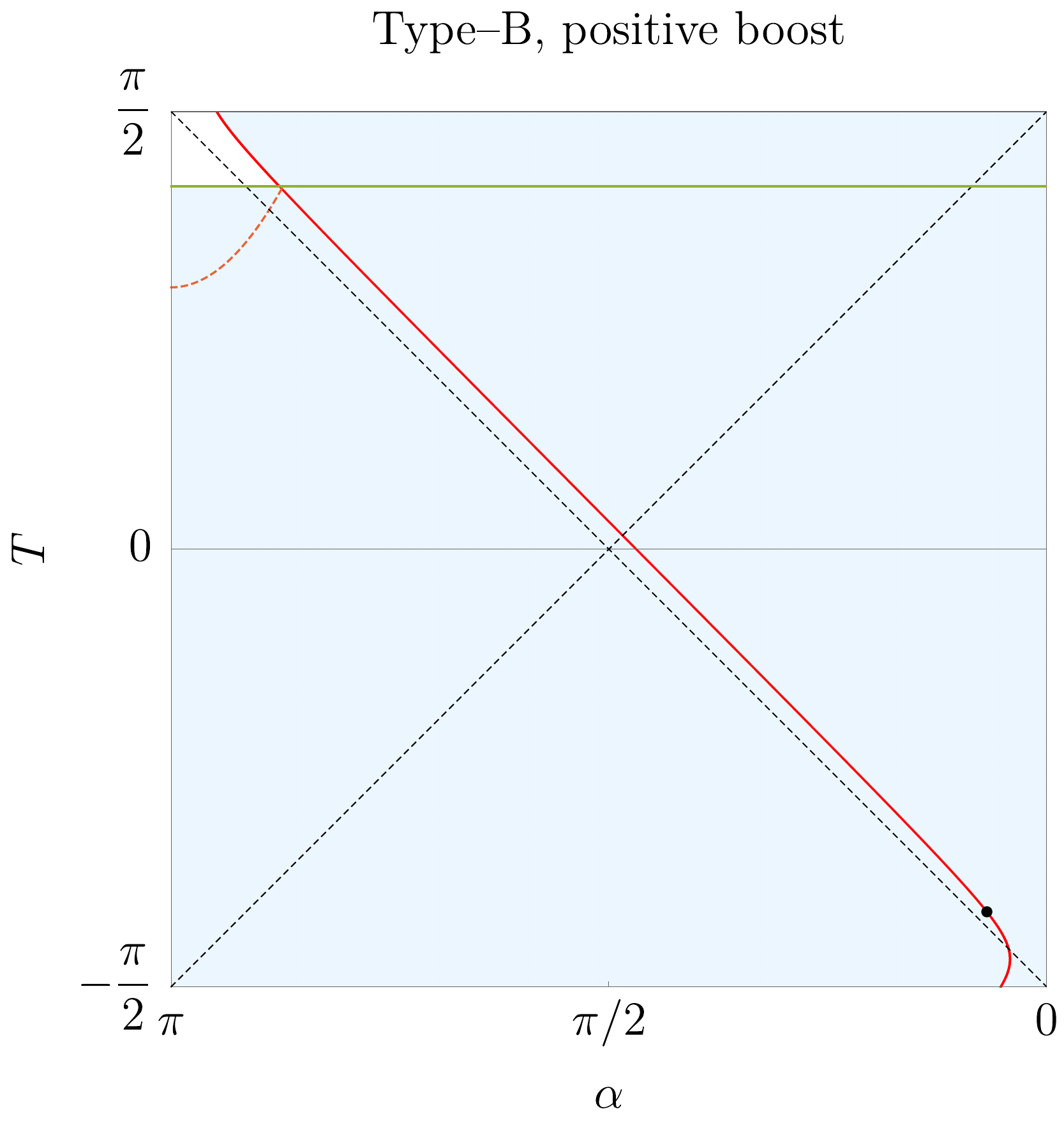}
        \quad
        \includegraphics[scale=0.25]{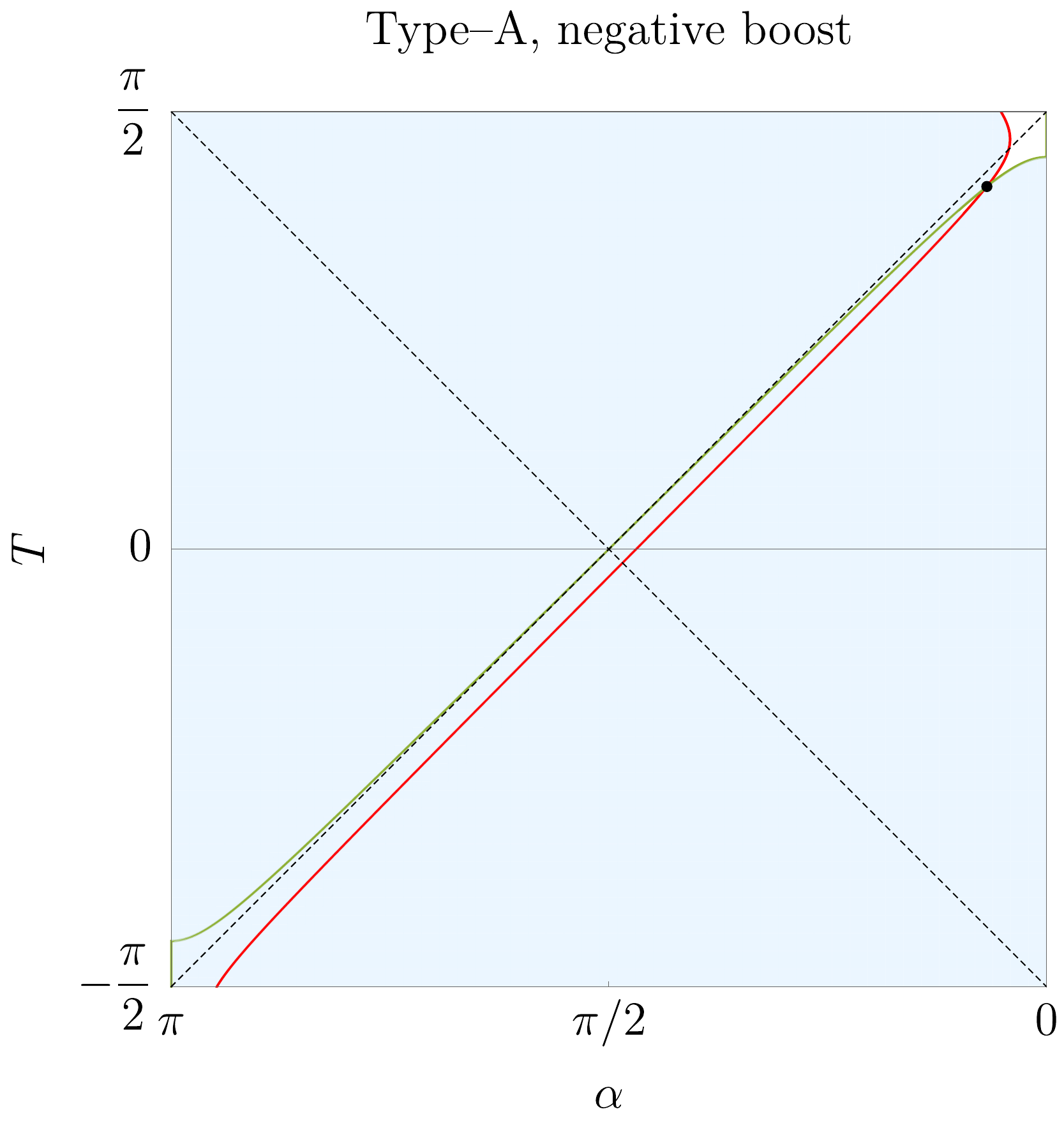}
        \quad
        \includegraphics[scale=0.25]{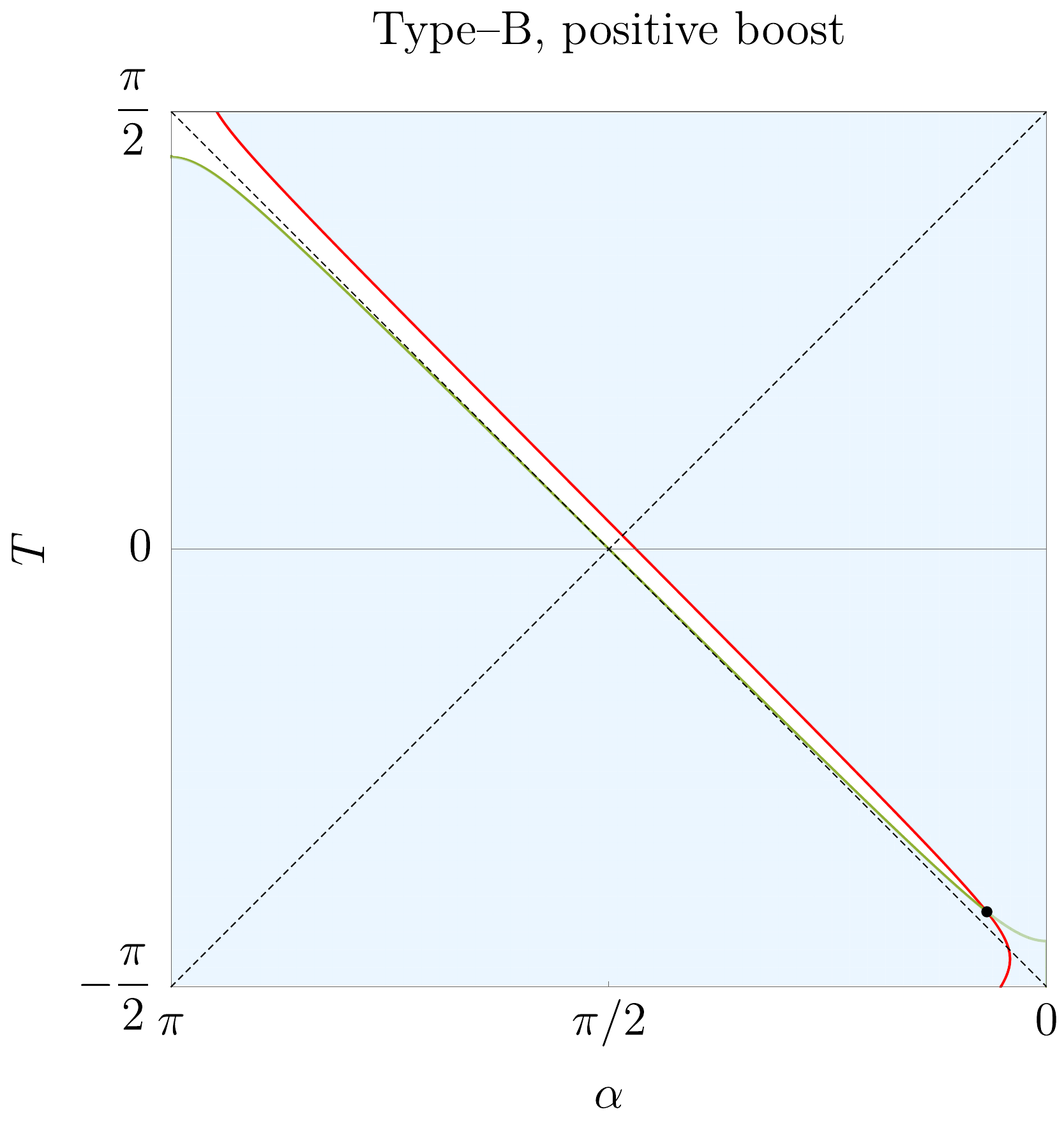}
        \caption{On the top left: Positively boosted CDL solution for Type--A. Like for Type--B with negative boost, this does not describe the creation of a small bubble at late times since on the green line a huge part of the parent de Sitter spacetime disappears. On the top right: Positively boosted CDL solution for Type--B. Here, as desired, a small bubble nucleates on the late-time green line. However, in this case the boosted nucleation point of CDL lies far in the past of the green line. Since we want nucleation to occur {\it on} the green line, we suspect that CDL may not capture the relevant physics. Bottom left: Negatively boosted Type--A CDL solution (red), but this time with the boosted waist highlighted in green. Bottom right: Positively boosted Type--B CDL solution (red), as on the right above, but this time with the boosted waist highlighted in green.}
        \label{fig:typeAB_boosted_bis}
\end{figure}

Rather, one may consider a boost in the opposite direction, as is depicted at the top of Fig.~\ref{fig:typeAB_boosted_bis}.
For the Type--A situation (left), this now leads to a presumably highly suppressed transition process.
For the Type--B case (right), on the other hand, the domain wall would still need to cross the parent dS horizon before going on-shell, but the total required off-shell region is reduced significantly.
However, the radius where the domain wall goes on shell is now very different from the CDL critical radius. 
Indeed, the minimal domain wall radius of the total trajectory is attained at very early times now.
If a bubble emerges in the way illustrated by the red dashed line, it appears to be very different from the original CDL process depicted in Fig.~\ref{fig:typeAB}.
It is, at this point, unclear how the corresponding decay rate for late-time Type--B transitions has to be modified from the CDL result \eqref{eq:B_rxy}.

If, on the other hand, CDL instructs us that the bubble has to go on-shell at its minimal possible radius, we face another puzzle. 
At the bottom of Fig.~\ref{fig:typeAB_boosted_bis}, a boost of the Type--B diagram in Fig.~\ref{fig:typeAB} is shown where the `gluing surface' (green line) now has been boosted as well. 
If the process where a bubble forms on the green surface were the correct interpretation of the CDL process, we would have to conclude that the future domain of influence of Type--B transitions, i.e., the forward lightcone of the black dot, is a large part of the parent dS space. 
One may now be worried that if many such transitions occur, they interfere with each other and hence threaten the classical existence of a global dS space altogether. 

In order to clarify the nature of bubble formation in dS space, we now want to study the process depicted on the left of Fig.~\ref{fig:typeAB_boosted} and on the top-right of Fig.~\ref{fig:typeAB_boosted_bis} and compute the resulting decay rates for late-time transitions. 
To do so, we first analyze in Sect.~\ref{sec:trajectories} all allowed Lorentzian trajectories of domain walls that interpolate between two dS spaces.
This analysis gives us an understanding of all the ways in which an off-shell domain wall can go on-shell.
To compute the decay rate for late-time dS transitions, we will study, in Sect.~\ref{sec:rates}, the quantum mechanics of the domain wall responsible for vacuum decay.

\section{Domain wall trajectories}
\label{sec:trajectories}

In this section, we study all possible Lorentzian trajectories of domain walls and ETW branes in de Sitter spacetime. 
It is clear that taking the CDL instanton and analytically continuing it to Lorentzian signature provides \emph{one} allowed domain wall trajectory. 
It is furthermore clear that using the $SO(1,4)$ symmetry of de Sitter space leads to further admissible solutions. 
In the following we argue that every domain wall trajectory allowed by the Hamiltonian constraint is of this form and we study its embedding into the parent dS space in detail. 
This result is important for understanding how off-shell domain wall geometries can go on-shell.

\subsection{dS--dS domain wall dynamics}

The Lorentzian action for a system of two dS spaces joined by a domain wall reads
\begin{align}
    S=\sum_{i\,=\,1}^2\left[\,\Mp^2\int_{\mathcal{M}_i}\sqrt{-g_i}\left(\frac{\mathcal{R}_i}{2}-3H_i^2\right)+\Mp^2\int_{\partial\mathcal{M}_i}\sqrt{-h}\mathcal{K}_i\right]-\int_{\partial M_1}\sqrt{-h}\sigma\,.\label{eq:S_Lorentzian}
\end{align}
As for the Euclidean action of \eqref{eq:S_euclidean}, the spacetime consists of two regions $\mathcal{M}_{1,2}$ glued at their common boundary $\partial\mathcal{M}$,  with induced metric $h$.
The Ricci scalars are denoted by $\mathcal{R}_i$ and the extrinsic curvature scalars by $\mathcal{K}_i$.
The equations of motion are given by the Einstein equations and the Israel junction conditions.
The latter take the form
\begin{align}
    \Mp^2 \Delta \mathcal{K}_{ab}=S_{ab}-\frac{1}{2}h_{ab}S\,,\label{eq:Israel}
\end{align}
with $\Delta \mathcal{K}_{ab}=\mathcal{K}_{1,ab}-\mathcal{K}_{2,ab}$ being the difference of the extrinsic curvature tensors and $S_{ab}$ the localized stress-energy tensor of the boundary.
For a pure-tension domain wall,
\begin{align}
    S_{ab}=-\sigma h_{ab}\qquad \mbox{implies} \qquad S=-3\sigma\qquad\mbox{and}\qquad S_{ab}-\frac{1}{2}h_{ab}S=\frac{1}{2}\sigma h_{ab}\,.
\end{align}
The Israel conditions hence take the form
\begin{align}
    \Delta\mathcal{K}_{ab}=\xi h_{ab}\,,\qquad \text{with} \qquad \xi\equiv\frac{\sigma}{2\Mp^2}\,.\label{eq:Israel_xi}
\end{align}

We now focus on geometries where the space away from the domain wall is pure de Sitter.
We can then use the Gauss--Codazzi equations to express the Ricci tensor $\mathcal{R}^{(3)}_{ab}$ of the three-dimensional domain-wall hypersurface as
\begin{align}
    \mathcal{R}^{(3)}_{ab}=2H_1^2 h_{ab}+\mathcal{K}_1 \mathcal{K}_{1,ab} -  \mathcal{K}_{1,ac}\mathcal{K}_{1,b}^{~~c}=2H_2^2 h_{ab}+\mathcal{K}_2 \mathcal{K}_{2,ab} -  \mathcal{K}_{2,ac}\mathcal{K}_{2,b}^{~~c}\,.
\end{align}
Using the condition \eqref{eq:Israel_xi}, we find
\begin{align}
    0=2(H_1^2-H_2^2)h_{ab}+\xi\mathcal{K}_{2,ab}+(2\xi^2+\xi\mathcal{K}_2)h_{ab}\,,
\end{align}
from which we conclude that
\begin{align}
    \mathcal{K}_{2,ab}=\frac{2(H_2^2-H_1^2-\xi^2)h_{ab}}{\xi}-\mathcal{K}_2h_{ab}\qquad \mbox{and hence}\qquad \mathcal{K}_2 = \frac{3(H_2^2-H_1^2-\xi^2)}{2\xi}\,.
\end{align}
Inserting this back into the previous equations, we find
\begin{align}
    \mathcal{K}_{1,ab} = \frac{H_2^2-H_1^2+\xi^2}{2\xi}h_{ab}\,\,\,,\qquad \mathcal{K}_{2,ab}= \frac{H_2^2-H_1^2-\xi^2}{2\xi}h_{ab}\,.\label{eq:Israel_separate}
\end{align}

We hence see that one may choose either of the dS spaces and study the motion of the domain wall entirely from the perspective of this side. The effect of the other side only comes in through $H_2^2$ on the l.h.~side of 
\eqref{eq:Israel_separate} or through $H_1^2$ on the r.h.~side.
The equation of motion for the ETW brane of a bubble of nothing has the same structure as \eqref{eq:Israel_separate}: it is simply $\mathcal{K}_{ab}=\xi h_{ab}$. We will now pick one of the two sides of our domain wall, call the corresponding Hubble scale $H$, and study the motion from the perspective of that side.

In \cite{Kodama:2002kj}, the geometry of spacetime boundaries whose extrinsic curvature is proportional to the induced metric, also known as `umbilic surfaces', was studied in detail. We see from \eqref{eq:Israel_separate} that this is precisely our case of interest.
The result is that these surfaces are 3d dS spaces and their embedding into the parent space can be characterized as follows:
Consider the parent dS as a submanifold of $\mathbb{R}^{1,4}$ with coordinates $(X^0,\dots,X^4)$:
\begin{align}
    X^0=\frac{1}{H}\sinh(Ht)\,,\qquad X^i=\frac{1}{H}\cosh(Ht)z^i(\alpha,\Omega_2)\,,\label{eq:dS_embed_t}
\end{align}
or, equivalently,
\begin{align}
    X^0=\frac{1}{H}\tan T\qquad\text{ and }\qquad X^i=\frac{z^i(\alpha,\Omega_2)}{H\cos T}\,.\label{eq:dS_embed_T}
\end{align}
Here, the $z^i(\alpha,\Omega_2)$ 
characterize an $S^3\subset \mathbb{R}^4$ and the resulting metric on the dS space takes the form \eqref{eq:global_dS_coordinates}, or \eqref{eq:global_dS_coordinates_compact}.
Then, \cite{Kodama:2002kj}\footnote{
See 
specifically Sec.~5.2.2.
} showed that the surfaces swept out by the domain wall take the form $X\cdot n =\text{const}$ with $n$ a spacelike unit vector.
Up to isometry transformations of the parent dS space, this can be written as $X^4=\text{const}$.
This is exactly the structure of the analytic continuation \eqref{eq:analyt_cont} of the CDL solution, which explicitly reads\footnote{We suppress the alternative solution $X^4\to -X^4$, which does not add new physics.}
\begin{align}\label{eq:traj_unboosted}
    X^4=\frac{1}{H}\sqrt{1-H^2R_{\rm crit}^2}\equiv\frac{w}{H}\qquad\mbox{with}\qquad 0<w<1\,.
\end{align}
Hence, all possible domain wall trajectories are given by the canonical CDL trajectory, up to parent dS isometries.

This can also be argued at a more elementary level as follows: First, let us pick one of the two dS spaces and describe the bubble from the perspective of that side. We assume an $SO(3)$-symmetric bubble wall and consider its trajectory $R(t)$. Crucially, there exists a time $t=t_0$ at which $R(t_0)<1/H$. This is unavoidable since, unless the bubble has emerged from a singularity with $R=0$, it must have passed through the `waist' of the dS. Thus, at least for some portion of the trajectory, we may use static patch coordinates
\begin{align}
    \dd s^2=-(1-H^2r^2)\dd\tilde t^2+\frac{\dd r^2}{1-H^2r^2}+r^2\dd\Omega_2^2\,.
\end{align}
The bubble trajectory $(r,\tilde t)=(R(\tilde t),\tilde t)$ obeys a first-order differential equation. For example, this may be given as 
\begin{align}
    \left(\frac{\dd R(\tau)}{\dd \tau}\right)^2=\frac{R(\tau)^2}{R_{\rm crit}^2}-1\,,
    \label{foei}
\end{align}
with $\tau$ the eigentime of the bubble.\footnote{
An 
equation analogous to \eqref{foei} appears e.g.~in \cite{Cespedes:2020xpn} (eq.~(5.21), with a presumably missing $G$ on the r.h.~side). Note also that, depending on the relation between dS radii and tension, the `preferred' dS with Hubble constant $H$ may either fill the space inside the bubble or between bubble and horizon.
}
It may be straightforwardly rewritten in terms of the static patch time $\tilde t$, with the resulting equation still first order.
A single initial condition then determines the trajectory. We see from \eqref{foei} that only initial values $R$ with $R_{\rm crit}<R<1/H$ are allowed. The resulting trajectories are then precisely those of CDL, up to translations in $\tilde t$. No other trajectories are possible inside the static patch and hence, by analyticity, in general. Moreover, each bubble wall can be described from either dS side and, in CDL, these geometries can be glued together consistently. Again, by the uniqueness argument above, this extends to all solutions in which two dS spaces are glued across an SO(3)-symmetric brane. Such geometries must then all be CDL.

Let us next study the functional form of all trajectories $R(t)$ or $R(T)$ in terms of the dS global time coordinate $t$ or $T$. For this, let us first return to the specific CDL domain wall geometry defined by \eqref{eq:traj_unboosted}. All further valid trajectories follow by applying SO(1,4) isometries of the parent dS. Of those, an SO(1,3) subgroup leaves the domain wall, which also has dS geometry, invariant. Furthermore, elements of the rotation subgroup SO(4)$\,\subset \,$SO(1,4) of the parent dS which are {\it not} in the rotation subgroup SO(3)$\,\subset\,$SO(1,3) of the domain wall change the solution, but in a trivial way: They simply rotate it without changing its position relative to the hyperplanes of constant time $t$ (or $T$). Thus, we may ignore them. This leaves us with a subgroup of dimension $10-6-(6-3)=1$. These are the boosts along $X^4$, which are indeed important since they change the position of our domain wall relative to constant time hypersurfaces. We may hence focus on the one-parameter set of solutions generated by such boosts.

\begin{figure}[ht]
        \centering
        \includegraphics[scale=0.35]{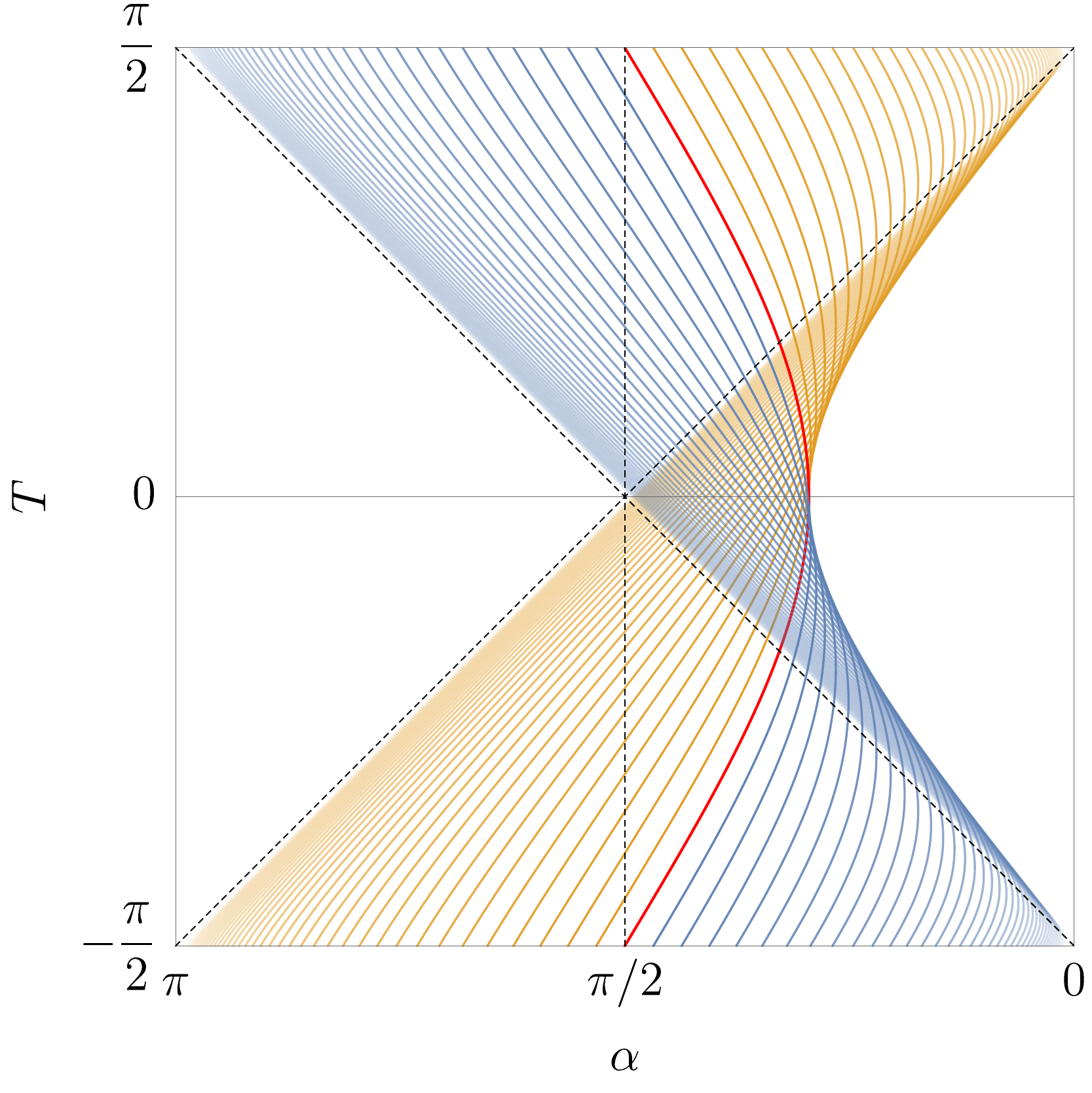}
        \caption{Boosted trajectories \eqref{eq:boost_rho_T} for different values of $\eta$. The red trajectory corresponds to $\eta=0$. Negative rapidities bend the late-time part of this curve towards the right (in orange) while positive rapidities bend it towards the left (in blue).}
        \label{fig:boost_rho_T}
\end{figure}

A boost along $X^4$ with rapidity $\eta$ takes the form
\begin{align}
    \begin{pmatrix}X'^{\,0}\\ X'^{\,4}\end{pmatrix}=\begin{pmatrix}\cosh\eta & -\sinh\eta\\ -\sinh\eta & \cosh\eta\end{pmatrix}\begin{pmatrix}X^{0}\\ X^{4}\end{pmatrix},
\end{align}
and the trajectory \eqref{eq:traj_unboosted} turns into
\begin{align}
\label{eq:X4}
    X'^{\,4}=\frac{w}{H\cosh\eta}-\tanh\eta\, X'^{\,0}\,.
\end{align}
It will now be convenient to choose the parameterization of the spatial sphere $S^3$ (cf.~\eqref{eq:dS_embed_T}) in terms of $\alpha$ and $\Omega_2$ such that the polar axis is aligned with the $X^4$-direction. This implies
\begin{align}
z'^{\,4}(\alpha,\Omega_2)=\cos(\alpha)\,,
\qquad
X'^{\,4}=\frac{1}{H}\cosh(Ht)\,\cos(\alpha)\,,
\qquad
X'^{\,0}=\frac{1}{H}\sinh(Ht)\,.
\label{z4def}
\end{align}
The embedding of the domain wall can then be characterized by functions $\alpha=\alpha(t)$ or $\alpha=\alpha(T)$ which are defined by
\begin{align}
    \cos(\alpha)\,=
    \,\frac{w}{\cosh(\eta)\cosh(Ht)}-\tanh(\eta)\tanh(Ht)
    \,=\,\frac{w\cos(T)}{\cosh(\eta)}-\tanh(\eta)\sin(T)\,.\label{eq:boost_rho_T}
\end{align}
A plot for different values of $\eta$ is shown in Fig.~\ref{fig:boost_rho_T}. We observe that
\begin{equation}
    \frac{\dd\alpha(T)}{\dd T}\left(T=\frac{\pi}{2}\right)=w\,,\qquad\text{ with }\qquad 0<w<1\,.
\end{equation}
Thus, all boosted trajectories reach the future boundary with the same slope.

\begin{figure}[ht]
        \centering
        \includegraphics[scale=0.27]{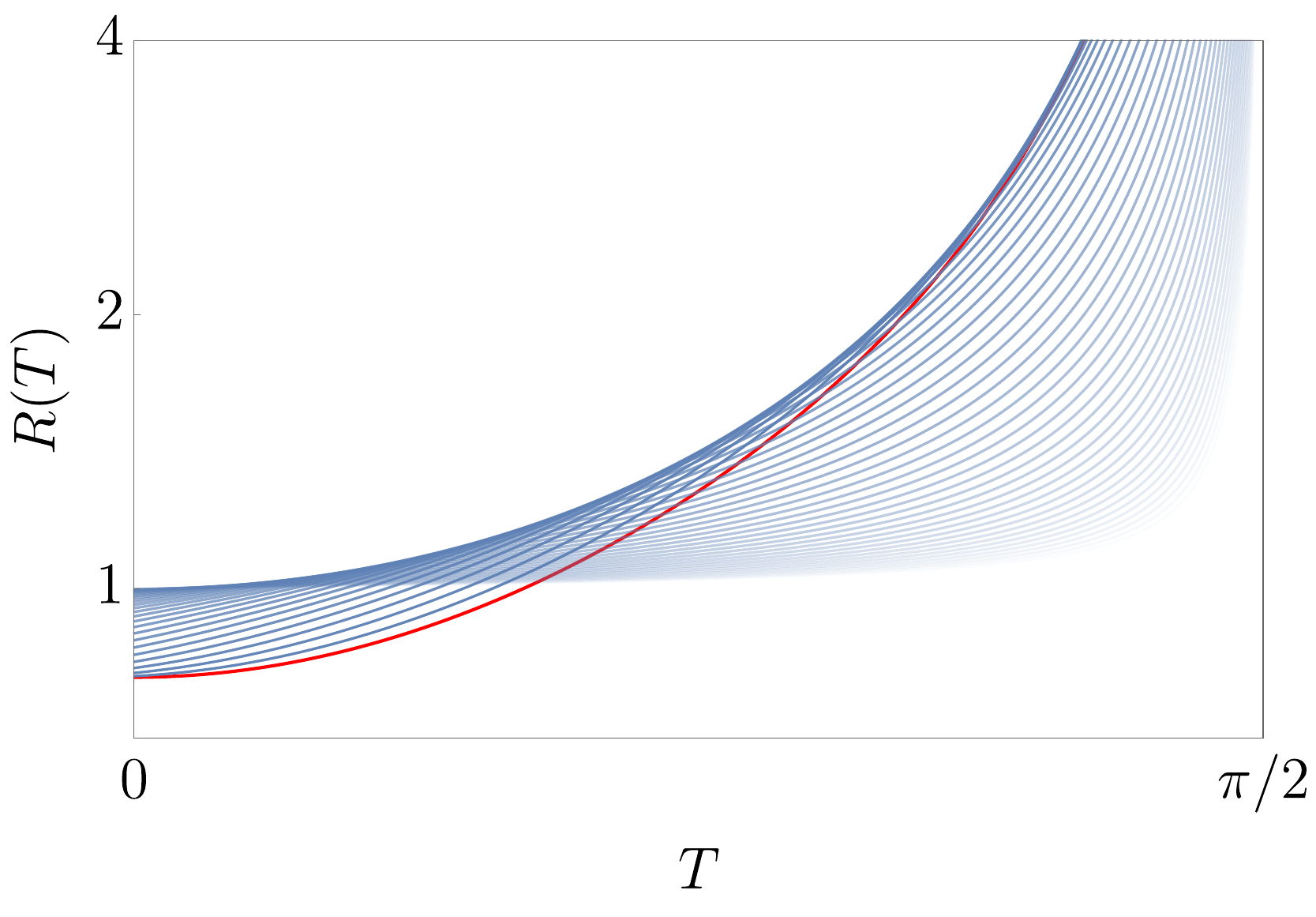}
        \quad
        \includegraphics[scale=0.27]{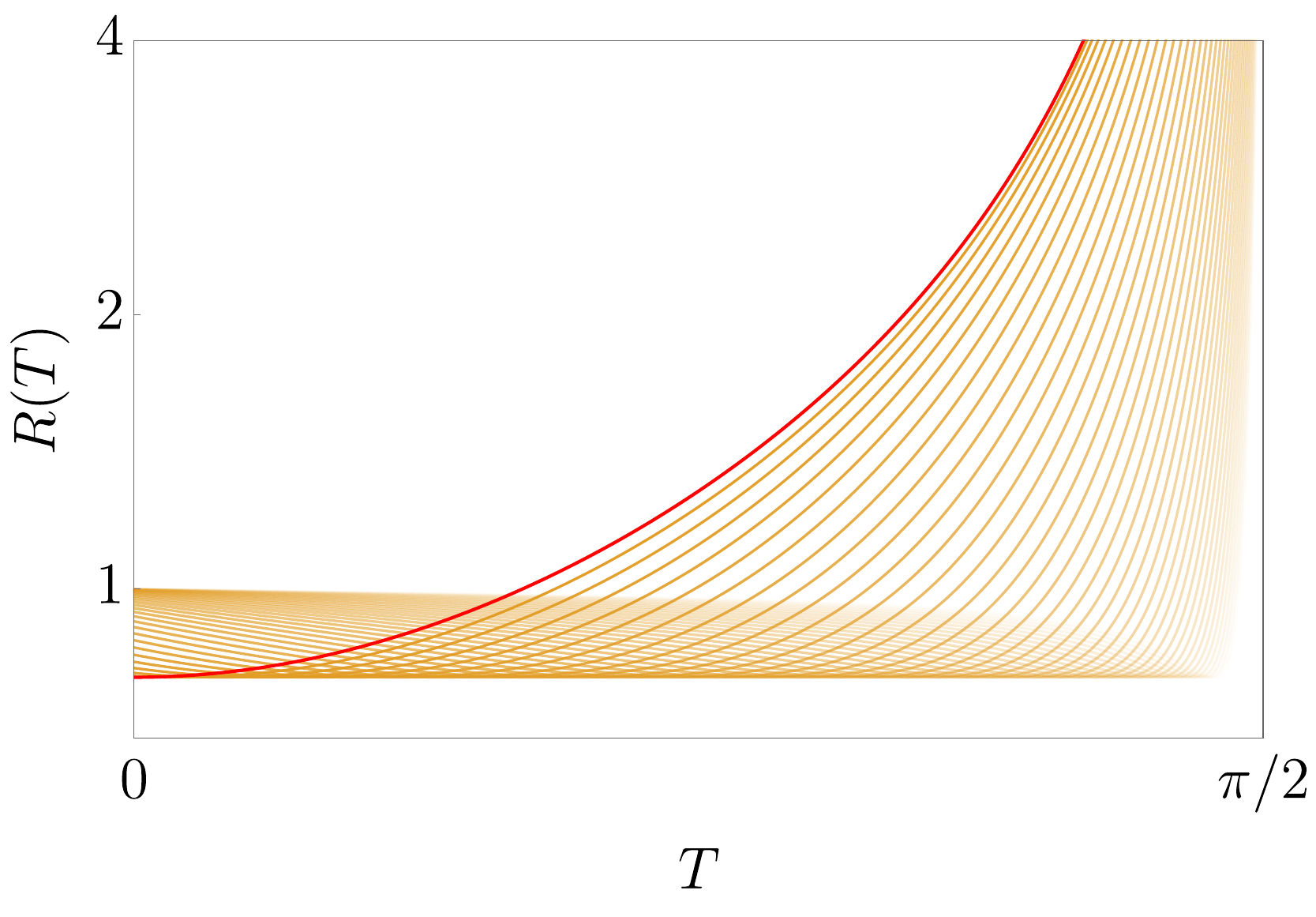}
        \caption{The wall radius $R$ as a function of the conformal time $T$ for positive rapidities (left panel) and negative rapidities (right panel). In both plots, the red curve corresponds to the unboosted solution and $H$ has been set to unity.}
        \label{fig:R_of_T}
\end{figure}

One can now compute the wall radius $R(T)=\sin\alpha/(H\cos T)$ or alternatively ${R(t) = \cosh(Ht) \sin(\alpha)/H}$ as a function of $T$ or $t$ by using \eqref{eq:boost_rho_T} only:
\begin{align}
    R^2(T)&=\frac{1-\left(\frac{ w\cos T}{\cosh\eta}-\tanh\eta\sin T\right)^2}{H^2\cos^2 T}\,,\\
    R^2(t)&=\frac{1}{H^2}\left[\cosh^2(Ht)-\frac{1}{\cosh^2(\eta)}\left(w-\sinh(\eta)\sinh(Ht)\right)^2\right]\,.\label{eq:R_t}
\end{align}
In Fig.~\ref{fig:R_of_T}, the evolution of the wall radius is displayed for different values of $\eta$. We will comment on these plots in the next subsection.

We continue by rewriting \eqref{eq:R_t} as
\begin{align}
    R^2(t)&=\frac{1}{H^2}\left[\cosh^2(Ht)(1-\tanh^2(\eta))+\tanh^2(\eta)-\frac{w^2}{\cosh^2(\eta)}+ \frac{2w\sinh(\eta)\sinh(Ht)}{\cosh^2(\eta)}\right]\nonumber\\
    &=\frac{1}{H^2\cosh^2(\eta)}\left[\cosh^2(Ht)+ 2w\sinh(Ht)\sinh(\eta)-w^2\right]+\frac{\tanh^2(\eta)}{H^2}\,.\label{eq:R_of_t_simplified}
\end{align}
One can immediately read off that $R(t)$ takes on its minimal value $R=R_{\rm crit}=\sqrt{1-w^2}/H$ at time $t_0$, determined by
\begin{align}
    \sinh(Ht_0)=- w\sinh(\eta)\,.\label{eq:min_radius}
\end{align}
For bubble-of-nothing solutions, the same expressions apply. While $w$ is still defined as in \eqref{eq:traj_unboosted}, the minimal radius $R_{\rm crit}$ is now given by \eqref{eq:R_crit_BoN}.

\subsection{Vacuum transitions at late times}
We now focus on the bubble trajectories at late times.

Let us start with Type--A down-tunneling transitions. 
As shown in Fig.~\ref{fig:typeAB}, we take the parent dS region to be defined by $\alpha(T)<\alpha< \pi$. This means that the parent dS lies to the left of the trajectories, also in Fig.~\ref{fig:boost_rho_T}.
For a bubble transition to occur as depicted in Fig.~\ref{fig:typeAB_boosted}, we require a boost parameter of $-\eta\gg 1$.
We expect that such bubbles are most likely to nucleate.
The corresponding bubble trajectories are depicted in orange in Fig.~\ref{fig:boost_rho_T}.
At late times, i.e., for $Ht\gg 1$, and for $-\eta\gg 1$, Eq.~\eqref{eq:R_of_t_simplified} simplifies correspondingly:
\begin{align}
    R^2\simeq\frac{e^{2(Ht+\eta)}- 2we^{Ht+\eta}+1}{H^2}\,.
    \label{eq:simp_A}
\end{align}
We see from \eqref{eq:min_radius} that for every time $t_0\gg 1/H$, there exists a value of $\eta$ such that $R=R_{\rm crit}$ at time $t=t_0$.
We strongly expect this to be the dominant bubble configuration to nucleate at late times.

We next consider a Type--B situation.
To be consistent with Fig.~\ref{fig:typeAB}, we consider the parent dS to be given by $0<\alpha<\alpha(T)$, i.e.~to fill the right side of Fig.~\ref{fig:boost_rho_T}.
As before, we want to consider transitions that erase as little as possible from the parent dS.
The relevant bubbles are characterized by $\eta\gg 1$ and thus their trajectories are depicted in blue in Figs.~\ref{fig:boost_rho_T} and~\ref{fig:R_of_T}.
Eq.~\eqref{eq:R_of_t_simplified} now simplifies as
\begin{align}
    R^2\simeq\frac{e^{2(Ht-\eta)}+ 2we^{Ht-\eta}+1}{H^2}\,.
    \label{simp}
\end{align}
We see that the bubble radius $R$ exceeds the dS radius at all times $t\gg 1/H$. Hence, even the smallest bubbles that can possibly nucleate at late times are larger than horizon size.\footnote{In fact, one can check using the full equation \eqref{eq:R_of_t_simplified} that for $\eta\gg 1$ the bubble radius exceeds $1/H$ already at very early times, as $\tanh^2(\eta\to\infty)\to 1$.}
More precisely, we see from \eqref{simp} that for $\eta\gg 1$ and late times there are two regimes: If $Ht\gg \eta\gg 1$, the bubble radius grows exponentially: $R\sim \exp(Ht)$.
This is the growth rate of the parent dS. Hence, one may say that the late-time expansion is solely due to the expansion of the ambient space.
By contrast, if $\eta\gg Ht\gg 1$, the bubble trajectory is simply $R\sim 1/H$.
The transition between these two regimes occurs when $Ht\sim \eta$. The middle term in \eqref{simp} is only relevant during the brief period of time when this relation holds. This overall behavior of the evolution of the bubble radius is nicely reflected in Fig.~\ref{fig:R_of_T}.

We note that for up-tunneling transitions, the situation is the same as in the Type--B down-tunneling case described above.
The minimal bubble that can nucleate at late times is the size of one Hubble patch of the parent dS.

\section{Decay rates}
\label{sec:rates}

Now that we have understood the structure of all possible Lorentzian domain wall trajectories, we turn to late-time decay rates.
As explained in Sect.~\ref{sec:late_time}, we expect to recover the CDL rate for Type--A down-tunneling events but not necessarily for all other transitions, as the domain wall has to grow beyond the parent dS horizon to go on shell.
We will set up a quantum mechanical model with one degree of freedom, the domain wall radius $R$, and derive the late-time decay rate.

\subsection{Quantum mechanical model}
\label{sec:QM_model}

Let us consider dS space in global coordinates,
\begin{align}
    \dd s^2=-\dd t^2+a(t)^2(\dd \alpha^2+\sin^2(\alpha)\dd\Omega_2^2)\,,\qquad a(t)=\frac{\cosh(Ht)}{H}\,,\label{eq:dS_metric}
\end{align}
and a domain wall embedding given by $\alpha=\alpha(t)$.
Starting from the action \eqref{eq:S_Lorentzian}, we want to derive a quantum mechanical model where $\alpha(t)$ is the only dynamical variable. The scale factor $a$ is thus not treated as a quantum mechanical variable and we simply consider the nucleation of a domain wall in the background of a fixed dS space. 
This approximation is justified at late times, when the spatial dS sphere is large.
We furthermore have opted not to introduce explicitly a lapse function in \eqref{eq:dS_metric} for better readability of the following equations. 
As is well known \cite{DeWitt:1967yk,Banks:1984cw}, the variation of the action with respect to the lapse yields the Hamiltonian constraint $\mathcal{H}=0$.
We will simply declare this to be the defining equation of our system.

To evaluate the gravitational part of the action \eqref{eq:S_Lorentzian} with metric \eqref{eq:dS_metric}, we first focus on one side of the domain wall and simply denote the corresponding Hubble parameter by $H$.
For a bubble of nothing scenario, this is already the full result, whereas for dS--dS transitions, the other side will have to be added later.
It is well known that
\begin{align}
    \sqrt{-g}\,\dd^4x=a^3(t)\sin^2(\alpha)\,\dd t\,\dd \alpha\,\dd\Omega_2\,,\qquad \mathcal{R}=6\left(\frac{\ddot{a}}{a}+\frac{\dot{a}^2+1}{a^2}\right)\,.
\end{align}
Hence, the bulk action can be written as
\begin{align}
    S_{\rm bulk}&=\frac{\Mp^2}{2}\int\dd\Omega_2\int\dd t \int_0^{\alpha(t)}\dd \tilde\alpha\,\sin^2(\tilde\alpha)\, \left(6[a^2\ddot a+a\dot{a}^2+a]-6a^3H^2\right)\,,\\
    &=12\pi \Mp^2 \int\dd t I(\alpha(t))\left(a^2\ddot a+a\dot{a}^2+a-a^3H^2\right)\,,
\end{align}
with $\tilde\alpha$ the integration variable belonging to $\alpha(t)$. We also defined
\begin{align}
I(\alpha(t))\equiv\int_0^{\alpha(t)}\sin(\tilde\alpha)^2\dd \tilde\alpha = \frac{\alpha(t)}{2}-\frac{\sin(\alpha(t))\cos(\alpha(t))}{2}\,.    
\end{align}
The term involving $\ddot{a}$ can be integrated using
\begin{align}
    \int\dd t Ia^2\ddot{a}=\int\dd t \frac{\dd}{\dd t}(Ia^2\dot a)-2\dot a^2 a I - \dot a a^2 \sin(\alpha)^2\dot \alpha\,. 
\end{align}
After dropping the total derivative term, this yields
\begin{align}
    S_{\rm bulk}= 12\pi \Mp^2\int\dd t \left(I(\alpha)[a-a\dot{a}^2-a^3H^2]-a^2\dot a \sin^2(\alpha)\dot \alpha\right)\,.
\end{align}

The extrinsic curvature scalar $\mathcal{K}$ can be calculated by defining the unit normal vector
\begin{align}
    n=n^\mu\partial_\mu = \frac{a(t)\dot\alpha (t)}{\sqrt{1-\dot\alpha^2a^2}}\partial_t+\frac{1}{a\sqrt{1-\dot\alpha^2a^2}}\partial_{\tilde\alpha}\,,
\end{align}
and evaluating the extrinsic curvature $\mathcal{K}=\nabla_\mu n^\mu$ on the subspace defined by $\tilde\alpha=\alpha(t)$:
\begin{align}
    \mathcal{K}=\frac{\ddot{\alpha}a+4\dot{a}\dot\alpha-3\dot a\dot\alpha^3a^2}{\sqrt{1-\dot\alpha^2a^2}^3}+\frac{2\cot(\alpha)}{a\cdot \sqrt{1-\dot\alpha^2a^2}}\,.
\end{align}
This leads, together with the expression
\begin{align}
    \sqrt{-h}\,\dd^3x=a(t)^2\sin^2(\alpha(t))\sqrt{1-a^2\dot\alpha^2}\dd\Omega_2\dd t\,,
\end{align}
to the boundary action
\begin{align}
    S_{\rm boundary}=4\pi\Mp^2\int\dd t \left(\sin^2(\alpha)\frac{\ddot{\alpha}a^3+4\dot{a}a^2\dot\alpha-3\dot a\dot\alpha^3a^4}{1-\dot\alpha^2a^2}+2a\cos(\alpha)\sin(\alpha)\right)\,.
\end{align}
Realizing that
\begin{align}
    \frac{\dd}{\dd t}\text{arctanh}(a\dot\alpha)=\frac{\dot a \dot\alpha + a \ddot \alpha}{1-a^2\dot\alpha^2}\,,
\end{align}
we can eliminate the $\ddot \alpha$ term in the action and arrive at
\begin{align}
\begin{split}
    S_{\rm boundary}= 4\pi\Mp^2 \int\dd t &\big[3\dot a a^2\dot\alpha \sin^2(\alpha)+a\sin(2\alpha)\\
    &- (2a\dot a \sin^2(\alpha)+a^2\sin(2\alpha)\dot\alpha)\arctanh(a\dot \alpha)\big]\,.
\end{split}
\end{align}

Altogether, combining bulk and boundary contributions from one side of the domain wall and the tension piece, we have
\begin{align}
    \begin{split}
        S=-24\pi\Mp^2&\int\dd t \left[a\dot a^2I(\alpha)\right]\\
    +4\pi\Mp^2 &\int\dd t \left[a\sin(2\alpha)- (2a\dot a \sin^2(\alpha)+a^2\sin(2\alpha)\dot\alpha)\arctanh(a\dot \alpha)\right]\\
    -4\pi\sigma &\int\dd t\, a^2\sin^2(\alpha)\sqrt{1-a^2\dot\alpha^2}\,.
    \end{split}\label{eq:S_before_small_angle}
\end{align}

\paragraph{Late-time\,/\,small-angle approximation:}

Since we are interested in describing the nucleation of small bubbles at late times, we can simplify our action by focusing on the small-angle limit. Depending on the case we want to describe, the relevant limit is either $\alpha\to 0$ or $\alpha\to\pi$. For the appearance of a small hole in a large sphere, corresponding to a bubble of nothing geometry, we need to consider $\alpha\to\pi$. We will treat this case in Sect.~\ref{sec:BON}. This is also the relevant limit to consider for the contribution of the parent dS in a dS--to--dS transition, as treated in Sect.~\ref{sec:dS-dS}.

We thus define $\beta\equiv\pi-\alpha$ and work under the assumption that $\beta\ll 1$ and $\dot a\simeq Ha\simeq e^{Ht}/2$. We get
\begin{equation}
\sin\alpha=\sin(\pi-\beta)\simeq\beta\,,\quad\sin 2\alpha\simeq -2\beta\,,\quad\cos\alpha\simeq -1+\frac{\beta^2}{2}\,,\quad I(\alpha)\simeq\frac{\pi}{2}-\frac{\beta^3}{3}\,.
\end{equation}
This leads to
\begin{align}
\begin{split}
S_{\alpha\to\pi}&=\cancel{-12\pi^2\Mp^2\int \dd tH^2a^3}+8\pi \Mp^2\int\dd t H^2a^3\beta^3\\
&+4\pi \Mp^2\int \dd t\left[-2a\beta +\left(2a^2H\beta^2+2a^2\beta\dot\beta\right)\arctanh(a\dot\beta)\right]\\
&-4\pi\sigma \int \dd ta^2\beta^2\sqrt{1-a^2\dot\beta^2}\,,
\end{split}\label{eq:S_before_subtraction}
\end{align}
with the first term crossed out since it does not involve $\beta$. We further write $R=a\sin\alpha\simeq a\beta$, from which we deduce
\begin{equation}
    a\dot\beta\simeq\dot R-HR\,,\qquad 2a^2\beta\dot\beta\simeq -2R(HR-\dot R)\,.
\end{equation}
Inserting these relations back into the action, we obtain
\begin{align}
\begin{split}
\label{eq:S}
\frac{S_{\alpha\to\pi}}{4\pi}&=2\Mp^2\int \dd t\left[H^2R^3-R+R\dot R\arctanh(\dot R-HR)\right]\\
&-\sigma \int\dd t R^2\sqrt{1-(\dot R-HR)^2}\,.
\end{split}
\end{align}

The $\alpha\to 0$ limit is relevant for describing the other side of a dS--to--dS transition. The necessary approximations in this situation are
\begin{align}
    \sin(\alpha)\simeq \alpha\,,\qquad \sin(2\alpha)\simeq 2\alpha\,,\qquad \cos(\alpha)\simeq 1-\frac{\alpha^2}{2}\,,\qquad I(\alpha)\simeq \frac{\alpha^3}{3}\,.
\end{align}
The action then takes the form
\begin{align}
    \begin{split}
        \frac{S_{\alpha\to 0}}{4\pi}&=2\Mp^2\int\dd t \left[-H^2 R^3 + R -R\dot R\arctanh(\dot R-HR)\right]\\
    &-\sigma \int\dd t R^2\sqrt{1-(\dot R-HR)^2}\,.
    \end{split}\label{eq:S_small_cap}
\end{align}
The late-time/small-angle approximation is equivalent to working in the flat slicing of dS
\begin{align}
    \dd s^2=-\dd t^2+e^{2Ht}\left(\dd r^2+r^2\dd\Omega_2^2\right)\,,\label{eq:metric_flat_expanding}
\end{align}
with an embedding $r=\tilde{r}(t)$ and $R(t)=e^{Ht}\tilde r(t)$.
 
We now remember that, due to time reparameterization invariance of the action \eqref{eq:S_Lorentzian}, the system is defined by the Hamiltonian constraint $\mathcal{H}=0$.
After quantization, we are hence dealing with a Wheeler--DeWitt (WDW) equation \cite{DeWitt:1967yk}, whose solutions are wavefunctions of the form $\Psi=\Psi(R)$.

The actions \eqref{eq:S} and \eqref{eq:S_small_cap} are the basis for our tunneling analysis. We start in the next subsection with the nucleation of a bubble of nothing since having only one bulk side makes the analysis simpler. Building on this, we will then apply the same methods to the dS--to--dS transition.

\subsection{Bubble of nothing}
\label{sec:BON}

In this subsection we compute the nucleation rate of bubbles of nothing (BoNs) at late times. The BoN is a calculationally simpler system, as there is only a non-trivial geometry on one side of the wall. It is then purely the sign of the tension that determines whether the transition is of Type--A 
(for $\sigma<0$) or Type--B (for $\sigma>0$).

\subsubsection{Canonical momentum, Hamiltonian constraint and tunneling rate}

Starting from the action \eqref{eq:S}, we can compute the canonical momentum $p_R$ of $R$
\begin{align}
    \frac{p_R}{4\pi}=2\Mp^2R\left(\arctanh(\dot R -HR)+\frac{\dot R}{1-(\dot R-HR)^2}\right)+ \frac{\sigma R^2(\dot R-HR)}{\sqrt{1-(\dot R-HR)^2}}\,.
\end{align}
Defining $s\equiv\dot R-HR$ and, as before, $\xi\equiv\sigma/(2\Mp^2)$, this simplifies to
\begin{align}
    \frac{p_R}{8\pi\Mp^2}=R\left(\arctanh{s}+\frac{s+HR}{1-s^2}\right)+\frac{\xi R^2s}{\sqrt{1-s^2}}\,.\label{eq:pR_x}
\end{align}
While we cannot solve this to get $\dot R$ as an explicit function of $p_R$ and $R$, we may nevertheless proceed by expressing the Hamiltonian $\mathcal{H}$ as a function of $R$ and $s$:
\begin{align}
    \begin{split}
    \label{eq:Hamiltonian}
        \frac{\mathcal H}{8\pi\Mp^2}=&\frac{R(1+HRs)}{\sqrt{1-s^2}}\left(\frac{(1+HRs)}{\sqrt{1-s^2}}+\xi R\right)\,.
    \end{split}
\end{align}
The constraint $\mathcal H=0$ then reduces to
\begin{align}
    \frac{1+HRs}{\sqrt{1-s^2}}=-\xi R\,.\label{eq:H_constraint}
\end{align}
Substituting \eqref{eq:H_constraint} into \eqref{eq:pR_x}, one finds
\begin{align}
    \begin{split}
        \frac{p_R}{8\pi\Mp^2}=R\left(\arctanh(s)+HR\right)\,.
    \end{split}\label{eq:pR_constraint}
\end{align}
Together with \eqref{eq:H_constraint}, this defines $p_R$ as a function of $R$. We may also solve \eqref{eq:pR_constraint} for $s$ and, plugging this back in 
\eqref{eq:Hamiltonian}, write the Hamiltonian constraint explicitly in terms of $R$ and $p_R$. With $p_R=-i\partial/\partial R$ and choosing an overall factor for convenience, this may be interpreted as the WDW equation
\begin{equation}
\left\{\pm\left[\cosh\left(\frac{p_R}{8\pi\Mp^2 R}-HR\right)+HR\sinh\left(\frac{p_R}{8\pi\Mp^2 R}-HR\right)\right]+\xi R\right\}\,\Psi(R)=0\,.
\label{wdwe}
\end{equation}

We want to use this equation to estimate the tunneling rate, which characterizes the probability of the bubble passing through the off-shell region between $R=0$ and $R=R_{\rm on-shell}$. For Type--A, we expect $R=R_{\rm on-shell}$ to be given by the standard CDL critical radius \eqref{eq:R_crit_BoN}. For Type--B, however, we expect the bubble to stay off-shell up to the dS horizon. The following analysis will confirm these expectations.

Thus, using an appropriate solution 
$\Psi(R)$ of \eqref{wdwe} in the off-shell region, we may estimate the decay rate by
\begin{align}
    \Gamma\sim\frac{|\Psi(R=R_{\rm on-shell})|^2}{|\Psi(R=0)|^2}\,.\label{eq:Gamma_WDW}
\end{align}
Here $|\Psi(0)|^2$ is the probability for creating a very small bubble, which is not exponentially suppressed. 
In standard quantum mechanics, ${\cal H}(p_R,R)$ would be quadratic in $p_R$ and hence the stationary Schr\"odinger equation would have two independent solutions. Of these, we would use the decaying one to estimate the rate according to \eqref{eq:Gamma_WDW}. Here, the Hamiltonian is a complicated function of $p_R$. Of the resulting infinite number of solutions of \eqref{wdwe}, we have to choose the most slowly decaying one.

More concretely, we will solve \eqref{wdwe} using the WKB approximation. Even though our Hamiltonian \eqref{eq:Hamiltonian} does not have canonical form, the standard WKB formula for deriving the leading contribution to the decay rate
\begin{equation}
\label{eq:B_generic}
    \Gamma\sim\exp(-B)\,,\qquad 
    B=\int_0^{R_{\rm on-shell}} 2\Im p_R\, \dd R\,,
\end{equation}
can still be applied, as shown in \cite{CarrilloGonzalez:2017eii}. All that is left is to obtain Im$\,p_R(R)$ explicitly.

\subsubsection{Calculating the tunneling exponent}

To find the tunneling exponent we need to analyze the behavior of the imaginary part of the momentum \eqref{eq:pR_constraint}. A non-trivial imaginary part can only come from the hyperbolic arctangent such that
\begin{equation}
    \frac{\Im p_R}{8\pi\Mp^2}=R\Im \arctanh{s}=\frac{1}{2}R\Im (\log(1+s)-\log(1-s))\,.\label{eq:Im_pR_BoN}
\end{equation}
The square root in the constraint \eqref{eq:H_constraint} as well as the logs in \eqref{eq:Im_pR_BoN} are multi-valued on the complex plane. We may choose the branch cuts to be at $\mathbbm{R} \setminus (-1,1)$ for both.
Crossing a branch cut changes the relevant functions according to
\begin{align}\label{eq:branch_crossings}
    \sqrt{1-s^2}\to -\sqrt{1-s^2}\,,\qquad \arctanh(s)\to \arctanh(s)\pm i\pi\,.
\end{align}
We note that an on-shell geometry is characterized by \eqref{eq:H_constraint} being solved with real values of $s$, with the square root in $\mathcal{H}$ on its fundamental branch, together with $p_R$ being real and the $\arctanh$ on its fundamental branch.

The general procedure is as follows: Moving along the line of real radii $R$, we determine a trajectory $s(R)$ using the constraint. There may be different such solutions related to the different sheets of the square root. One approach is to start at large $R$, with the on-shell solution, and continue analytically to smaller $R$. Then, we use the same trajectory $s(R)$ in the expression \eqref{eq:Im_pR_BoN} to determine the corresponding trajectory $p_R(R)$, the imaginary part of which will eventually have to be integrated.

Before implementing the procedure above explicitly, we collect the relevant equations: The two solutions of \eqref{eq:H_constraint} for $s$ are
\begin{align}\label{eq:s_BoN}
    s_\pm = \frac{-H\pm\xi\sqrt{R^2(H^2+\xi^2)-1}}{(H^2+\xi^2)R}=R_{\rm crit}^2\frac{-H\pm\xi\sqrt{\frac{R^2}{R_{\rm crit}^2}-1}}{R}\,,
\end{align}
with $R_{\rm crit}$ defined in \eqref{eq:R_crit_BoN}. We notice that $s$ is real for $R>R_{\rm crit}$.
For $R<R_{\rm crit}$, the argument of the square root turns negative and we write the solutions as
\begin{align}\label{ima}
    s_\pm =\frac{-H\pm i\xi\sqrt{1-R^2(H^2+\xi^2)}}{(H^2+\xi^2)R}=R_{\rm crit}^2\frac{-H\pm i\xi\sqrt{1-\frac{R^2}{R_{\rm crit}^2}}}{R}\,.
\end{align}

In the region where $s$ is real, $\Im(\arctanh{s})$ is just a constant.
Hence, the only regime where \eqref{eq:Im_pR_BoN} has to be evaluated non-trivially is $R<R_{\rm crit}$.
In order to do so, we note that we can write $\Im(\arctanh{s})$, on its fundamental branch, as
\begin{align}
    \Im\arctanh{(s)}|_{\rm fund}&=\frac{1}{2}\Im\left(\log(\frac{1+s}{1-s})|_{\rm fund}\right)=\frac{1}{2}\arg\left(\frac{1-|s|^2+2i\Im(s)}{|1-s|^2}\right)\\
    &=\arctan(\frac{2\Im(s)}{|1+s||1-s|+1-|s|^2})\,,\label{eq:Im_arctanh_BoN}
\end{align}
with the real arctan function taking values $\arctan(x)\in (-\pi/2,\pi/2)$.\footnote{For the last step one may use that $\tan\left[({\arg z})/2\right]=\Im z/(|z|+\Re z)$ for a complex number $z$.}

\begin{figure}[ht]
        \centering
        \includegraphics[scale=0.18]{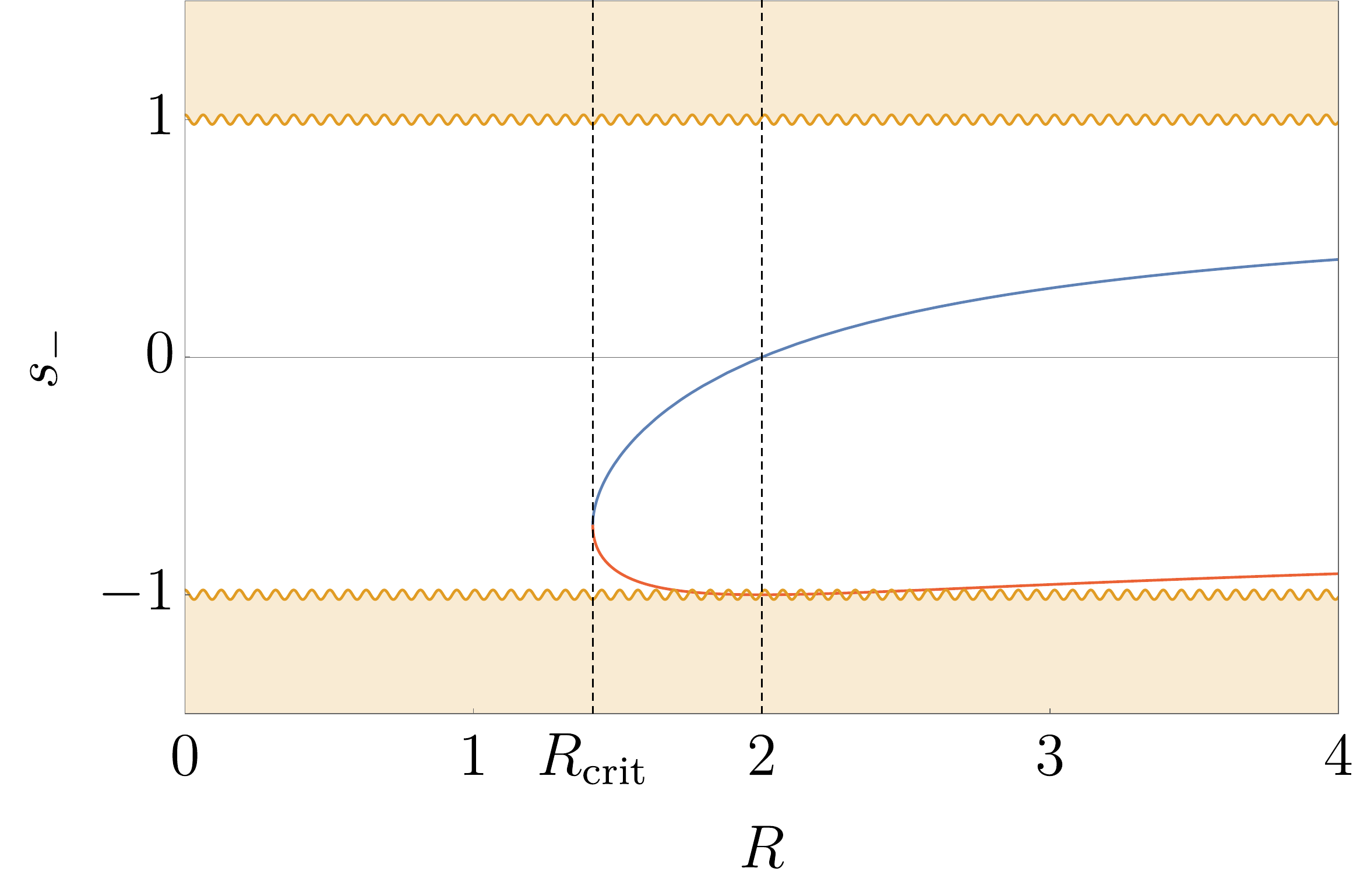}
        \quad
        \includegraphics[scale=0.2]{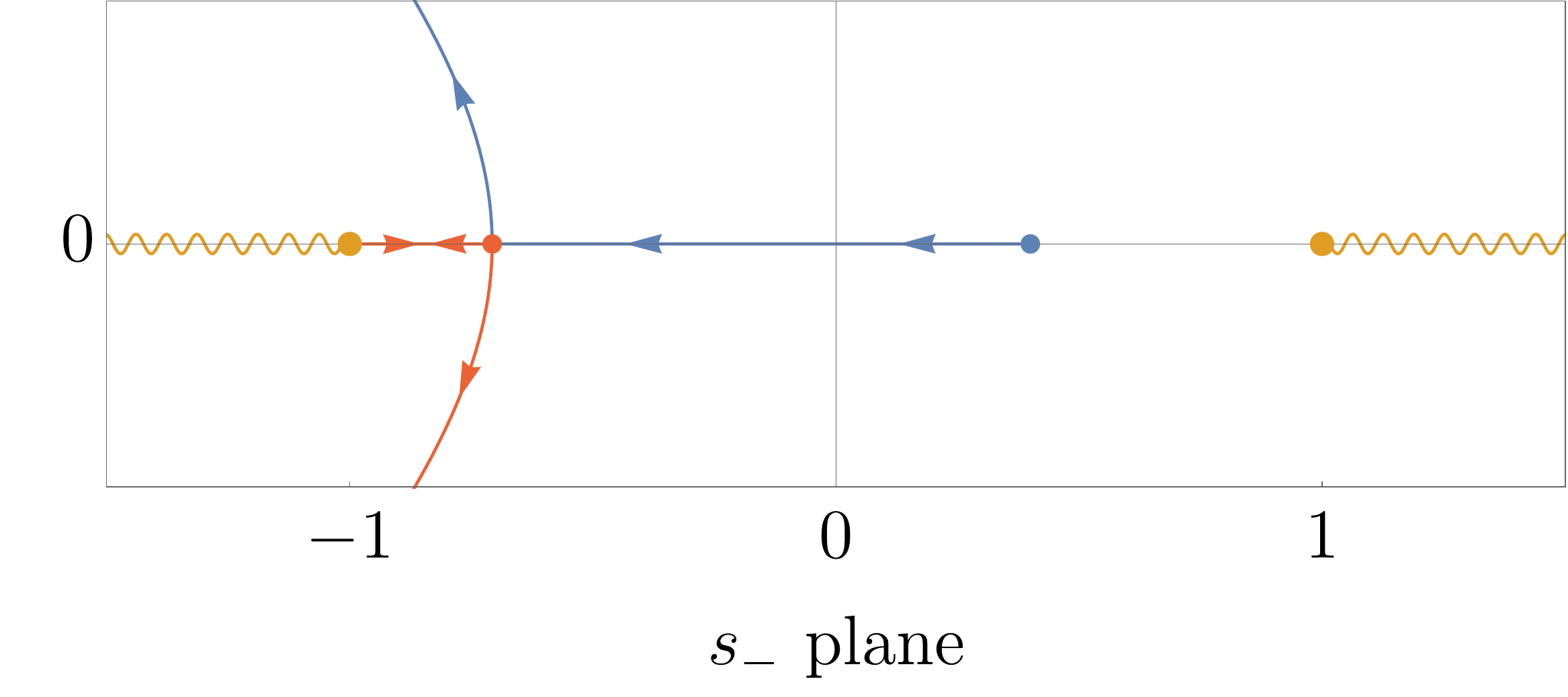}
        \caption{Numerical example of the behavior of $s_-$ ($\Mp=1$, $H=1/2$ and $\sigma=\pm 1$). On the left: $s_-$ for $R\geq R_{\rm crit}$ where it is purely real. The blue curve is Type--A while the red curve is Type--B. On the right: Motion of $s_-$ in the complex plane as $R$ goes from large values to small values. Again the blue curve is Type--A and starts at the blue dot, while the red curve is Type--B and starts at the red dot. The arrows show the direction of decreasing $R$.}
        \label{fig:branches}
\end{figure}

We are now ready to implement the procedure of explicitly determining the trajectories $s(R)$ and $p_R(R)$. We will do so for Type--A and Type--B separately:
\paragraph{Type--A:} For Type--A, we have $\xi<0$. One can then check that, in the on-shell region $R>R_{\rm crit}$, only the solution $s=s_-$ solves the Hamiltonian constraint with the square root on its fundamental branch. This is hence the unique on-shell solution. As $R$ decreases, $s_-$ moves left along the real axis, in the region $(-1,1)$. At $R=R_{\rm crit}$, the value of $s_-$ is still larger than $-1$ such that no transition of sheet according to \eqref{eq:branch_crossings}
can occur along this part of the trajectory (see Fig.~\ref{fig:branches}).
At this point, we switch from the expression \eqref{eq:s_BoN} to \eqref{ima} and now, for $R<R_{\rm crit}$, both expressions $s_\pm$ solve the constraint. We formally have the freedom to keep moving along the $s_-$ trajectory or switch to the $s_+$ trajectory. We select the `tunneling solution', determined by the condition $\Im(p_R)>0$. This turns out to be $s=s_-$. Since we now have a non-zero imaginary part, no transition of sheets according to \eqref{eq:branch_crossings}
can occur in this second part of the trajectory, $0<R<R_{\rm crit}$, either.

\paragraph{Type--B:} Now we have $\xi>0$ and the on-shell region is $R\geq 1/H$. As before, the corresponding on-shell solution requires the choice $s=s_-$, which is again on the real axis in the region $(-1,1)$. As $R$ decreases, the value of $s_-$ also falls and reaches $s=-1$ at $R=1/H$. More precisely, for $R=1/H+\epsilon$ we have $s=-1+{\cal O}(\epsilon^2)$. Thus, $s(R)$ approaches $-1$ from above and then, as $R$ falls below $1/H$, moves up again to larger values (see Fig.~\ref{fig:branches}). As a result of this behavior, the expression $\sqrt{1-s^2}$ in the Hamiltonian constraint can change sheet at $R=1/H$. One may think of this as the freedom to deform the trajectory $s(R)$ such that it encircles the branch point $-1$ a certain number, say $n$, times before leaving it again. One can check that the Hamiltonian constraint is satisfied for $R<1/H$ only if the square root in the constraint changes to the second sheet. This requires $n$ to be odd.

For the square root, all odd $n$ are equivalent since there are only two sheets and encircling the branch point $1$, $3$ or $-1$ times takes us to the second sheet. However, for the $\arctanh(s)$ in the canonical momentum, this choice produces a shift by $in\pi$, with positive $n$ corresponding to counter-clockwise motion around the branch point.
As explained before, the tunneling interpretation requires $\Im(p_R)\geq 0$, with the smallest positive value giving the dominant contribution. Thus, we choose $n=+1$.\footnote{Note that we could have followed a different logic to fix the trajectories $s(R)$ and $p_R(R)$: Starting at very small $R$, we can create a small off-shell bubble and fix the relevant sheets as appropriate for the standard WKB tunneling interpretation. This, in particular, forces us onto the sheet of the $\arctanh$ with the $i\pi$ contribution. Then, by analytic continuation from small to large $R$, the whole trajectory is unambiguously determined, consistently with the above.}

Following the trajectory to even smaller $R$, we reach $R=R_{\rm crit}$, where the square root in 
\eqref{eq:s_BoN} vanishes. We have the freedom to switch from $s_-$ to $s_+$, but it turns out that staying on $s_-$ leads to the smaller value of $\Im(p_R)>0$ and hence to the smaller tunneling suppression.
Thus, the solution $s_-$ again defines  the correct trajectory all along.\\

As we have seen, $s=s_-$ is the relevant solution for both Type--A and Type--B and in all regions of $R$. In the following, we will refer to $R<R_{\rm crit}$ as `Region--1' and to $R_{\rm crit}<R<1/H$ as `Region--2'.
To compute the contribution from Region--1 to the tunneling exponent, we substitute $s_-$ into \eqref{eq:Im_pR_BoN} and use \eqref{eq:Im_arctanh_BoN} to find
\begin{align}
    \frac{\Im(p_R)_{\text{Region--1}}}{8\pi\Mp^2} = R\left[\arctan(-\frac{\sqrt{1-\frac{R^2}{R_{\rm crit}^2}}}{\xi R})+\delta\right]\,.\label{eq:Im_pR_solved_BoN}
\end{align}
Here, the first term should be evaluated on the branch $\arctan(x)\in (-\pi/2,\pi/2)$ and $\delta$ is defined as
\begin{align}\label{eq:delta_BoN}
    \delta=\begin{cases}
        0 & \text{Type--A}\\
        \pi & \text{Type--B}
    \end{cases}\,,
\end{align}
as we have worked out before.

\begin{figure}[ht]
        \centering
        \includegraphics[scale=0.19]{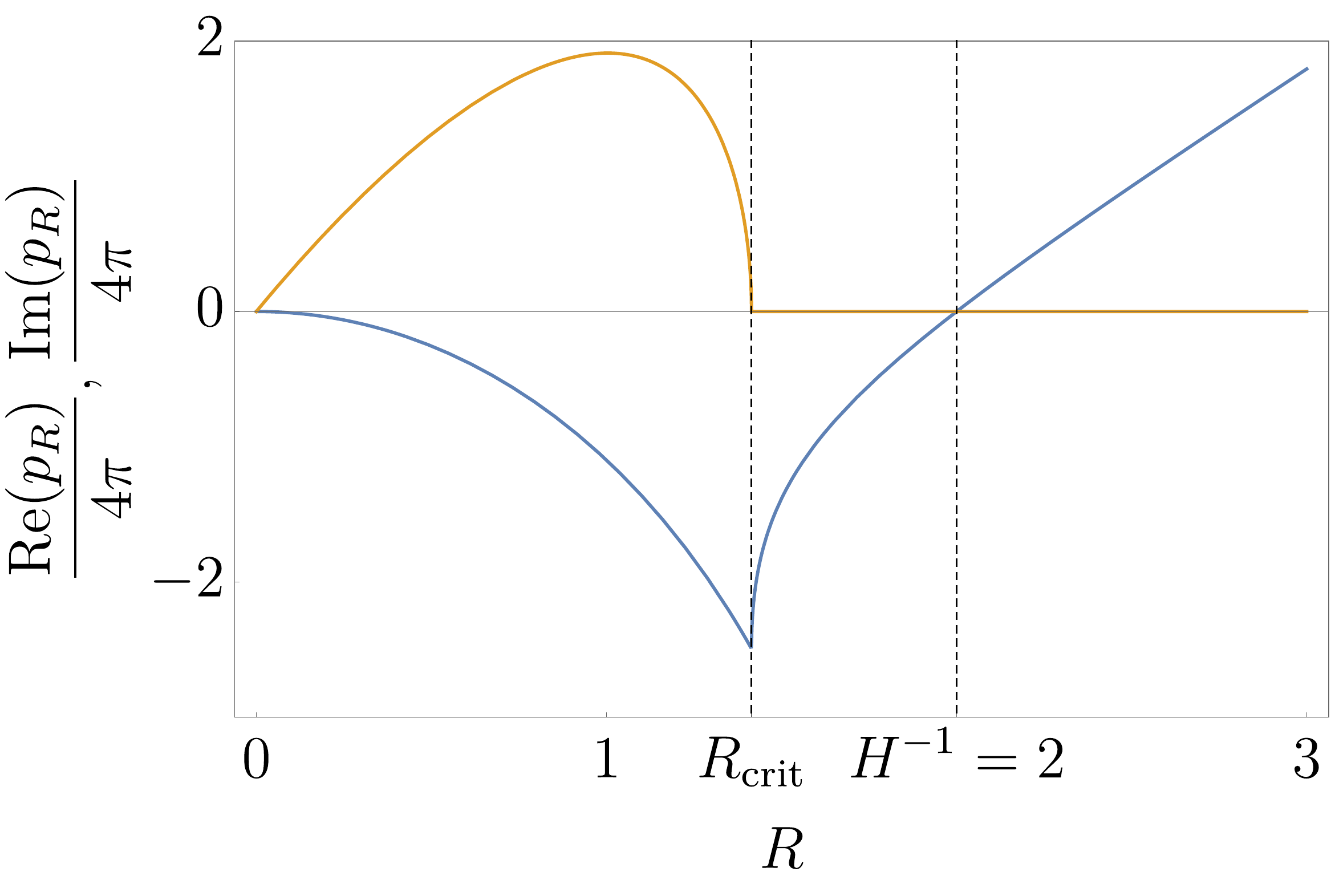}
        \quad
        \includegraphics[scale=0.19]{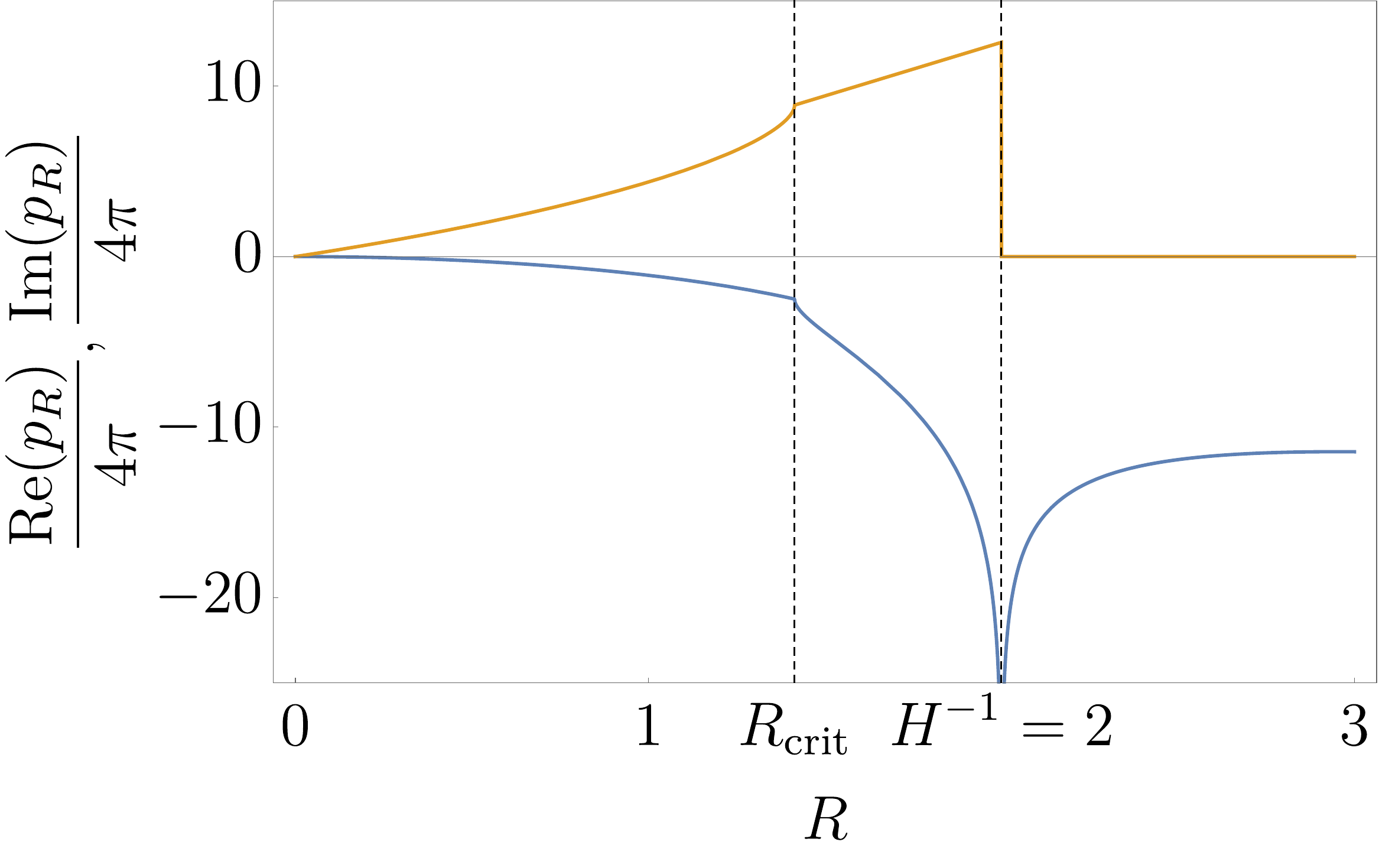}
        \caption{On the left: Real (in blue) and imaginary (in orange) parts of $p_R$ for the Type--A BoN nucleation. To draw these plots we used numerical parameters $\Mp=1$, $H=1/2$ and $\sigma=-1$. On the right: The same color code but for a Type--B BoN, where now the tension has been set to $\sigma=+1$.}
        \label{fig:Re_Im_pR_BoN}
\end{figure}

We now apply \eqref{eq:B_generic} and integrate \eqref{eq:Im_pR_solved_BoN} from $R=0$ to $R=R_{\rm crit}$ to obtain $B_{\text{Region--1}}$.
We obtain
\begin{align}
    &\frac{B_{\text{Region--1}}}{16\pi\Mp^2}=\int_0^{R_{\rm crit}}\dd R R\left[\arctan(-\frac{\sqrt{1-\frac{R^2}{R_{\rm crit}^2}}}{\xi R})+\delta\right]\,\nonumber\\[5pt]
&=\left[\frac{R^2}{2}\arctan(-\frac{\sqrt{1-\frac{R^2}{R_{\rm crit}^2}}}{\xi R})\right]^{R_{\rm crit}}_0+\frac{R_{\rm crit}^2\delta}{2}-\int_0^{R_{\rm crit}}\frac{R^2}{2}\frac{\xi}{\sqrt{1-\frac{R^2}{R_{\rm crit}^2}}(1+\xi^2R^2-\frac{R^2}{R_{\rm crit}^2})}\nonumber\\[5pt]
&=\frac{R_{\rm crit}^2\delta}{2}-\text{sgn}(\xi)\frac{\pi R_{\rm crit}^2}{4(1+|\xi| R_{\rm crit})}\,,
\label{eq:B_Region_1}
\end{align}
where we have used integration by parts.
For Type--A, where $\delta=0$, this already gives the full result
\begin{equation}
B_{\text{Type--A}}=\frac{4\pi^2\Mp^2}{H^2}\left(1+\frac{\xi}{\sqrt{H^2+\xi^2}}\right)\,.\label{eq:B_BoN_A}
\end{equation}
It matches exactly the decay rate \eqref{eq:B_BoN} derived using the instanton method. The real and imaginary parts of $p_R$ for this case are displayed on the left in Fig.~\ref{fig:Re_Im_pR_BoN} for a specific choice of parameters. 
The imaginary part may be thought of as defining a potential under which the ETW brane has to tunnel.
It has a smooth shape and vanishes at $R=R_{\rm crit}$, where the bubble goes on-shell.

For Type--B, we have to add a contribution $B_{\text{Region--2}}$ to the result \eqref{eq:B_Region_1} above.
It reads 
\begin{align}
    \frac{B_{\text{Region--2}}}{16\pi\Mp^2}=\int_{R_{\rm crit}}^{1/H}\dd R\, R\delta = \frac{\pi}{2}\left(\frac{1}{H^2}-R_{\rm crit}^2\right)\,,
\end{align}
giving in total
\begin{equation}
B_{\text{Type--B}}=\frac{4\pi^2\Mp^2}{H^2}\left(1+\frac{\xi}{\sqrt{H^2+\xi^2}}\right)\,.\label{eq:B_BoN_B}
\end{equation}
The above is the same analytical expression as for Type--A (but with $\xi$ now positive) and thus matches again, maybe surprisingly, the decay rate \eqref{eq:B_BoN}. The real and imaginary parts of $p_R$ in this case are displayed on the right in Fig.~\ref{fig:Re_Im_pR_BoN}. The function Im$\,p_R$ has a peculiar shape: As long as $R<R_{\rm crit}$, it grows following some non-trivial functional form. Then its growth becomes linear, as $\Im(\arctanh(s))$ is now constant. Eventually, the imaginary part abruptly goes back to zero at the horizon, where a change of branch occurs. 
It is striking that such a complicated imaginary momentum (or corresponding potential), which characterizes our late-time setup, exactly reproduces the CDL result, which follows from simple sphere-based geometries.

\begin{figure}[t]
    \centering
    \includegraphics[scale=0.25]{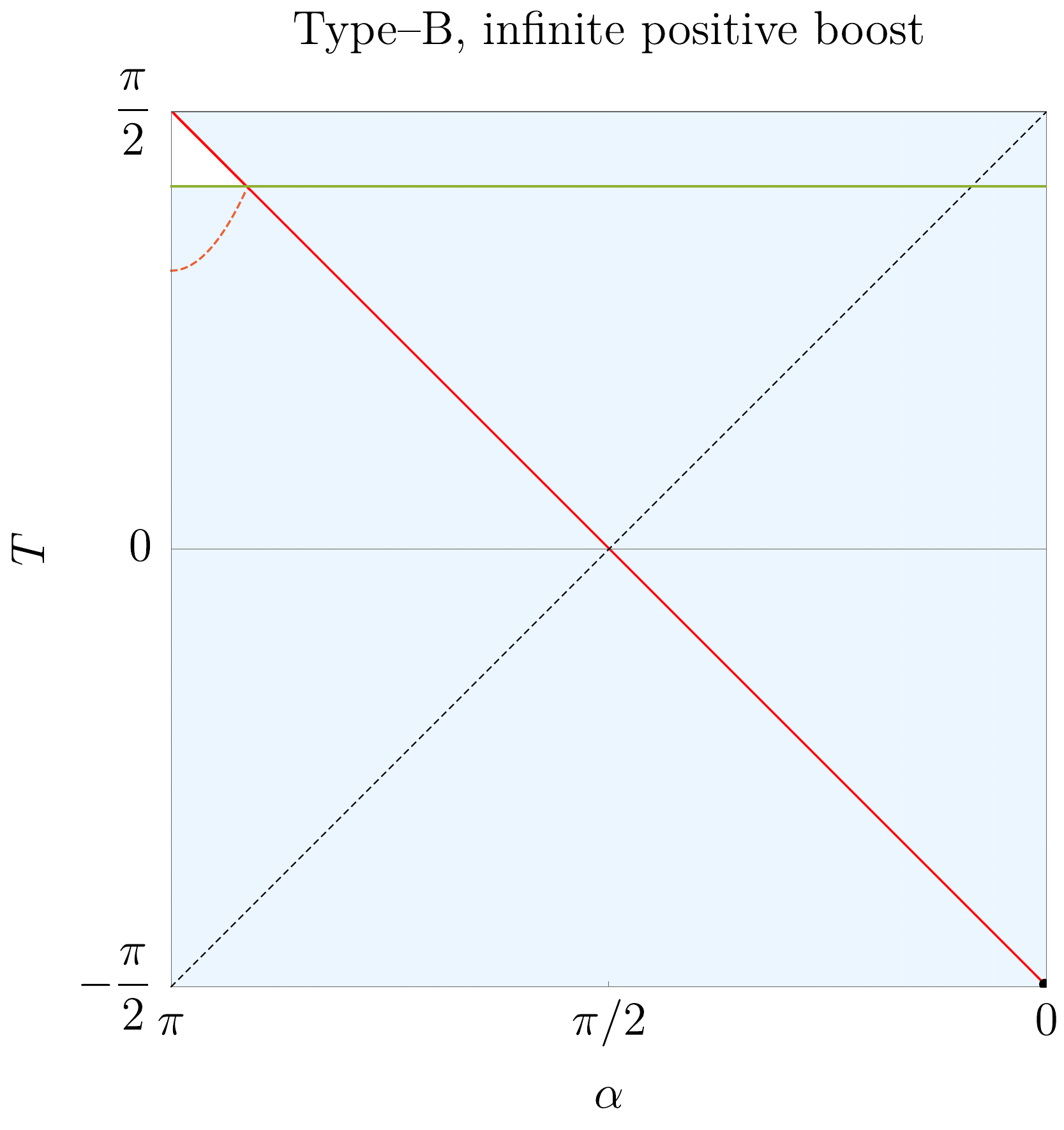}
    \qquad
    \qquad
    \raisebox{0.6cm}{\includegraphics[width=0.4\linewidth]{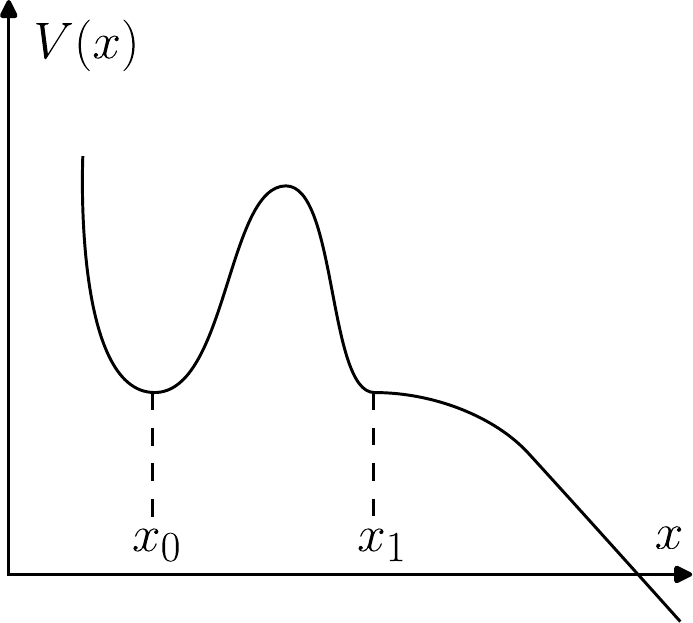}}
    \caption{Left: Maximally boosted CDL geometry where $R$ stays constant at the horizon. Right: Quantum mechanical model of particle tunneling where the final state is marginally unstable.}
    \label{fig:max_boost_and_runaway}
\end{figure}

The calculation above suggests that, for Type--B transitions, the ETW brane goes on-shell at $R=1/H$.
By this we mean that we can connect to a classical solution. Indeed, at $R=1/H$ we have $s=-1$, implying $\dot{R}=0$. Hence the relevant solution is simply $R=1/H$ for all $t$. According to the analysis in Sect.~\ref{sec:trajectories}, this is the ``maximally boosted CDL geometry'', i.e.~the limit $\eta\to\infty$ of \eqref{simp}. 
The trajectory of the corresponding bubble is depicted in Fig.~\ref{fig:max_boost_and_runaway}.

It is important to note, however, that even an arbitrarily small fluctuation of $R$ away from $R=1/H$ leads to a bubble which expands at very late times, as illustrated in the top right panel of Fig.~\ref{fig:typeAB_boosted_bis}.
Moreover, the computation of $B$ was performed using the WKB approximation and hence the value of $R$ where the domain wall goes on shell cannot be determined precisely. 
In the path integral formulation of quantum mechanics, the WKB solution corresponds to the least action path. 
However, there presumably exist further paths with finite action that lead to the domain wall going on-shell at $R>1/H$ and thus to the formation of an expanding bubble.

An analogous situation arises in ordinary quantum mechanical tunneling when the potential barrier takes a form as depicted in Fig.~\ref{fig:max_boost_and_runaway} on the right.
Here, when a particle tunnels from the minimum at $x=x_0$ through the barrier towards $x=x_1$, it can go on shell at $x=x_1$ and stay there forever, due to $V'(x=x_1)=0$.
However, a tiny fluctuation towards larger $x$ lets it roll down the potential. 
Similarly, there exist finite action tunneling trajectories that let the particle go on-shell at $x>x_1$ in the first place. We conclude from this discussion that our calculation estimates the rate for creating not just a bubble at the horizon but an {\it expanding} bubble.

If the potential is locally $-x^2$ there, then $(x-x_1)\sim\dot x$. 
In the instanton solution interpolating between $x_0$ and $x_1$, the particle reaches $x=x_1$ only after an infinite Euclidean time. 
Similarly, if one imagines a particle moving in this potential and starting its motion from a position $x>x_1$ with a negative velocity, it would require an infinite amount of time for the particle to reach the saddle at $x_1$ with zero velocity. This can be easily checked using the explicit solution, which is of the form
$x-x_1=Xe^{-t}$. CDL solutions at asymptotically late time have an analogous equation of motion close to the horizon: $HR-1\sim\dot R$. This relation can also be explicitly derived from our Hamiltonian constraint \eqref{eq:H_constraint} in the limit where both $\dot{R}$ and $HR-1$ are very small. Our off-shell trajectory leading to the marginal, infinitely boosted, bubble solution with $\dot R=0$ and $R=1/H$ is thus ``disconnected'' from an on-shell CDL solution in a way which is completely analogous to the quantum mechanical toy model just discussed.

Finally, let us comment on the reliability of our WKB analysis. As in ordinary quantum mechanics, leading-order WKB is not valid close to the turning points of the potential.
For the quantum mechanical toy model defined by the potential on the r.h.~side of Fig.~\ref{fig:max_boost_and_runaway}, the turning points are $x=x_0$ and $x=x_1$.
Similarly, for our dS decay model, the WKB wavefunction may not be a good approximation very close to the turning points $R=0$ and $R=1/H$. As in quantum mechanics, one may hope that the general expressions \eqref{eq:B_BoN_A}, \eqref{eq:B_BoN_B} nevertheless provide a valid approximation to the decay rate as long as the potential barrier defined by $\Im(p_R)$ is large enough in most of the integration domain.
The dS case does, however, have additional potentially problematic features: Its Hamiltonian does not have canonical form and the imaginary part of $p_R$ is not smooth at $R=1/H$. Both are immaterial for the $R=0$ turning point but become relevant near the horizon. In particular, while the usual breakdown of WKB due to the potential becoming small does not occur, one faces a divergence of Re$\,p_R(R)$ near $R=1/H$. These issues make our analysis questionable in the region where $1-RH\ll 1$. However, most of the integral calculating the tunneling exponent is not dependent on this region, such that we still have a reliable approximation.

\subsection{dS--to--dS transition}
\label{sec:dS-dS}

In this subsection, we study tunneling between dS spaces.
We derive the decay rate for Type--A and Type--B of both down- and up-tunneling transitions.
\subsubsection{Action, momentum and constraint}

For dS--to--dS tunneling, we have to consider the two-sided action \eqref{eq:S_Lorentzian}.
We can use the results \eqref{eq:S} and \eqref{eq:S_small_cap} for the two contributions to the action. 
The bubble radius $R$ is the same for both sides. 
Denoting derivatives with respect to $t_1$ with dots and derivatives with respect to $t_2$ with primes, we find
\begin{align}
        \frac{S_{\rm tot}}{4\pi} &= \int\dd t_1 \left[2\Mp^2\left(H_1^2R^3-R+R\dot R\arctanh(\dot R-H_1R)\right)-R^2\sqrt{1-(\dot R -H_1R)^2}\sigma\right]\nonumber\\
        &-\int\dd t_2\left[2\Mp^2\left(H_2^2R^3-R+R R'\arctanh( R'-H_2R)\right)\right]\,.
\end{align}
We first need to rewrite the action in terms of a unique time variable and we choose the parent global time coordinate $t_1$. Matching the induced metric on both sides leads to the relations
\begin{equation}
    \dd t_2=N\dd t_1\,,\qquad R'=\frac{\dot R}{N}\,,\qquad N=\sqrt{\frac{1-(\dot R-H_1R)^2}{1-(\frac{\dot R}{N}-H_2R)^2}}\,,
\end{equation}
from which we derive
\begin{equation}
    N^2+\frac{2R\dot R H_2 N}{1-H_2^2R^2}+\frac{H_1^2R^2-2R\dot RH_1-1}{1-H_2^2R^2}=0\,.
\end{equation}
We solve this quadratic equation and find
\begin{equation}\label{eq:eta_1}
    N=\frac{-H_2R\dot R+\gamma\sqrt{H_2^2R^2\dot R^2+(1-H_2^2R^2)(1-H_1^2R^2+2H_1R\dot R)}}{1-H_2^2R^2}\,,
\end{equation}
with $\gamma\in\{-1,+1\}$.
In order for both $t_1$ and $t_2$ to move forward together, $N$ should be positive for on-shell configurations. This will determine $\gamma$ in the different cases to be considered below.

Inserting \eqref{eq:eta_1} back into the action, we find
\begin{align}
&\frac{S_{\rm tot}}{4\pi}=\int \dd t_1 \left[2\Mp^2\left(H_1^2R^3-R+R\dot R\arctanh(\dot R-H_1R)\right)-R^2\sqrt{1-(\dot R -H_1R)^2}\sigma\right]\nonumber\\     
&+\int\dd t_12\Mp^2\left[\rule{0em}{10mm}-H_2R^2\dot R+\gamma R\sqrt{(1-H_2^2R^2)\left[1-H_1R(H_1R-2\dot R)\right]+H_2^2R^2\dot R^2}\right.\\
&\left.+R\dot R\arctanh\left(H_2R+\frac{\dot R(1-H_2^2R^2)}{H_2R\dot R-\gamma\sqrt{(1-H_2^2R^2)\left[1-H_1R(H_1R-2\dot R)\right]+H_2^2R^2\dot R^2}}\right)\right]\,.\nonumber
\end{align}
We can then proceed by following the same steps as we did for the BoN case: Compute the momentum $p_R$ and the Hamiltonian $\mathcal{H}$ and use the constraint equation to simplify $p_R$. Introducing $s=\dot R-H_1R$, this leads to
\begin{align}
\begin{split}
\label{eq:Ham_constraint_dS_dS}
\mathcal{H}=0\quad\Longleftrightarrow\quad -\xi R&=\frac{1+H_1Rs-\gamma\sqrt{1+H_1R(2s+H_1R)-H_2^2R^2(1-s^2)}}{\sqrt{1-s^2}}\\
&\equiv\frac{1+H_1Rs-\gamma\sqrt{\mathcal{B}}}{\sqrt{1-s^2}}\,,
\end{split}
\end{align}
where we have defined $\mathcal{B}$ to be the argument of the square root in the numerator.
This is the same expression as under the square root in \eqref{eq:eta_1}.
The canonical momentum, after imposing \eqref{eq:Ham_constraint_dS_dS}, reads
\begin{align}
\label{eq:pR_dS_dS_A}
\frac{p_R}{4\pi}=2\Mp^2R\left[\arctanh(s)+(H_1-H_2)R+\arctanh(\mathcal{A})\right]\,,
\end{align}
where
\begin{equation}
\mathcal A\equiv H_2R+\frac{(1-H_2^2R^2)(s+H_1R)}{H_2R(s+H_1R)-\gamma\sqrt{\mathcal{B}}}\,.
\end{equation}

\subsubsection{Tunneling rate}\label{sec:dS-dS_rate}

We next study solutions to \eqref{eq:Ham_constraint_dS_dS}.
There are now two square roots in $\mathcal{H}$ and two $\arctanh$ functions in $p_R$ that can change sheets.
As before, a sheet transition of $\sqrt{1-s^2}$ goes along with a sheet transition of $\arctanh(s)$, and they take the form \eqref{eq:branch_crossings}.
However, a change of branch of $\sqrt{\mathcal{B}}$ does not necessarily trigger a change of branch of $\arctanh{\cal A}$ in \eqref{eq:pR_dS_dS_A}. In App.~\ref{app:dS_dS_solutions}, the tunneling solutions to \eqref{eq:Ham_constraint_dS_dS} including their branching structure of the complex functions in \eqref{eq:Ham_constraint_dS_dS} and \eqref{eq:pR_dS_dS_A} are carefully studied. As for the bubble of nothing, the tunneling solutions are determined by continuing the on-shell solutions at large $R$ all the way to $R=0$, while demanding that $\Im(p_R)$ is non-negative and as small as possible. For down-tunneling transitions, we distinguish between Region--1 and Region--2, which are defined as $0<R<R_{\rm crit}$ and $R_{\rm crit}<R<1/H_1$ respectively. 
For up-tunneling, there are three regions of interest, defined by
$0<R<R_{\rm crit}$, $R_{\rm crit}<R<1/H_2$ and $1/H_2<R<1/H_1$.
The sheet structure of the complex functions is summarized in Tables~\ref{tab:Down_branches} and~\ref{tab:Up_branches}.
For Type--A up-tunneling transitions, we find that $\gamma=-1$ is required to find a solution, which is consistent with $N>0$ in the on-shell regime. 

\renewcommand{\arraystretch}{1.4}
\begin{table}[t]
\centering
\begin{tabular}{|c|cccc|cccc|}
\cline{2-9}\multicolumn{1}{c|}{} & 
\multicolumn{4}{c|}{\textbf{R--1: $0<R<R_{\rm crit}$}} & 
\multicolumn{4}{c|}{\textbf{R--2: $R_{\rm crit}<R<1/H_1$}} \\
\cline{2-9}\multicolumn{1}{c|}{}
 & $\sqrt{1-s^2}$ & $\sqrt{\mathcal{B}}$ & $\delta_s$ & $\delta_\mathcal{A}$ & $\sqrt{1-s^2}$ & $\sqrt{\mathcal{B}}$ & $\delta_s$ & $\delta_\mathcal{A}$ \\
\hline
\textbf{Type--A} & $+$ & $+$ & 0 & 0 & $+$ & $+$ & 0 & 0 \\
\textbf{Type--B} & $-$ & $-$ & $\pi$ & 0 & $-$ & $-$ & $\pi$ & 0 \\
\hline
\end{tabular}
\caption{The branch structure of down-tunneling transitions in the two relevant regimes for down-tunneling. A $+$ sign for square roots denotes that they are on their principal branch, whereas a $-$ sign denotes the other branch. The parameters $\delta_s$ and $\delta_\mathcal{A}$ indicate the sheets on which both $\arctanh$ functions appearing in $p_R$ are evaluated. The correct sheet is given by the principal sheet shifted by an amount $i\delta_s$ or $i\delta_\mathcal{A}$ respectively.}
\label{tab:Down_branches}
\centering
\vspace{0.4cm}
\begin{tabular}{|c|cccc|cccc|cccc|}
\cline{2-13}\multicolumn{1}{c|}{} & 
\multicolumn{4}{c|}{\textbf{R--1: $0<R<R_{\rm crit}$}} & 
\multicolumn{4}{c|}{\textbf{R--2: $R_{\rm crit}<R<1/H_2$}}&
\multicolumn{4}{c|}{\textbf{R--3: $1/H_2<R<1/H_1$}}\\
\cline{2-13}\multicolumn{1}{c|}{}
 & $\sqrt{1-s^2}$ & $\sqrt{\mathcal{B}}$ & $\delta_s$ & $\delta_\mathcal{A}$ & $\sqrt{1-s^2}$ & $\sqrt{\mathcal{B}}$ & $\delta_s$ & $\delta_\mathcal{A}$ & $\sqrt{1-s^2}$ & $\sqrt{\mathcal{B}}$ & $\delta_s$ & $\delta_\mathcal{A}$ \\
\hline
\textbf{Type--A} & $-$ & $-$ & $\pi$ & $-\pi$ & $-$ & $-$ & $\pi$ & $-\pi$ & $-$ & $-$ & $\pi$ & 0\\
\textbf{Type--B} & $-$ & $-$ & $\pi$ & 0 & $-$ & $-$ & $\pi$ & 0 & $-$ & $-$ & $\pi$ & 0\\
\hline
\end{tabular}
\caption{The branch structure of up-tunneling transitions in the three relevant regimes. As before, we indicate the branches of the functions $\sqrt{1-s^2}$, $\sqrt{\mathcal{B}}$, $\arctanh(s)$ and $\arctanh(\mathcal{A})$ respectively. For Type--A up-tunneling, we require $\gamma=-1$.}
\label{tab:Up_branches}
\end{table}

To compute the tunneling rate, we apply the formula \eqref{eq:B_generic} and make use of App.~\ref{app:dS_dS_solutions}.
Since $H_1,H_2,R$ are real, we see from \eqref{eq:pR_dS_dS_A} that $\Im(p_R)$ can be written as
\begin{align}
\begin{split}
    \frac{\Im(p_R)}{8\pi\Mp^2}=R\big[\Im(\arctanh(s)+\arctanh(\mathcal{A}))\big]\,.
\end{split}
\end{align}
As for the bubble of nothing, the arguments of the $\arctanh$ functions are real for $R>R_{\rm crit}$, cf.~Eq.~\eqref{eq:app:s}.
As a result, their imaginary parts are simply constant in Regions two and three.
As shown in Appendix~\ref{app:dS_dS_solutions},
the contribution to $B$ from Region--1, where the arguments of the $\arctanh$ functions are complex, takes the form
\begin{align}
    \frac{B_{\text{Region--1}}}{16
    \pi\Mp^2}=\int_0^{R_{\rm crit}} \dd R\, R\left[\arctan(\frac{\sqrt{1-\frac{R^2}{R_{\rm crit}^2}}}{R\cdot P})+\delta_s+\arctan(\frac{\sqrt{1-\frac{R^2}{R_{\rm crit}^2}}}{R\cdot Q})+\delta_\mathcal{A}\right]\,,\label{eq:Bfund_dS_dS}
\end{align}
with the arctan functions taking values in $(-\pi/2,\pi/2)$ and
\begin{align}
    P=\frac{H_1^2-H_2^2-\xi^2}{2\xi}\,,\qquad Q=\frac{H_2^2-H_1^2-\xi^2}{2\xi}\,.
\end{align}
Furthermore, $\delta_s$ and $\delta_\mathcal{A}$ are constants that are related to the sheets of $\arctanh(s)$ and $\arctanh(\mathcal{A})$ for the respective transitions. 
Their values can be read from Tables~\ref{tab:Down_branches} and ~\ref{tab:Up_branches}.

We observe that the integrals in \eqref{eq:Bfund_dS_dS} take the same form as \eqref{eq:B_Region_1}.
We hence already know their values and we find
\begin{align}
    \frac{B_{\text{Region--1}}}{16\pi \Mp^2}=\frac{\pi R_{\rm crit}^2}{4}\left(\frac{\sgn (P)}{1+|P|R_{\rm crit}}+\frac{\sgn(Q)}{1+|Q|R_{\rm crit}}\right)+\frac{R_{\rm crit}^2}{2}(\delta_s+\delta_\mathcal{A})\,.
\end{align}
The signs of $P$ and $Q$ are, for the different types of transitions, summarized in Table~\ref{tab:signs}.
We can furthermore express $P,Q,R_{\rm crit}$ in terms of the dimensionless variables \eqref{eq:dim_less_variables} as
\begin{align}
    P=\frac{\xi (1-x)}{2x}\,,\qquad Q=-\frac{\xi (1+x)}{2x}\,,\qquad R_{\rm crit}^2=\frac{4x^2}{\xi^2(x^2+2xy+1)}\,.
\end{align}
We thus find $B_{\text{Region--1}}$, for the four different kinds of transitions, to be
\begin{align}
    \frac{B_{\text{Region--1, A--down}}}{16\pi \Mp^2}&=\frac{\pi x}{\xi^2(y^2-1)}\left[\frac{1+xy-\sqrt{1+2xy+x^2}}{\sqrt{1+2xy+x^2}}\right]\,,\\
    \frac{B_{\text{Region--1, B--down}}}{16\pi \Mp^2}&=\frac{\pi x}{\xi^2(y^2-1)}\left[\frac{1+xy-y\sqrt{1+2xy+x^2}}{\sqrt{1+2xy+x^2}}\right]+\frac{\pi R_{\rm crit}^2}{2}\,,\\
    \frac{B_{\text{Region--1, A--up}}}{16\pi \Mp^2}&=\frac{-\pi x}{\xi^2(y^2-1)}\left[\frac{1+xy-\sqrt{1+2xy+x^2}}{\sqrt{1+2xy+x^2}}\right]+\frac{R_{\rm crit}^2}{2}(\pi-\pi)\,,\label{eq:B1_A_up}\\
    \frac{B_{\text{Region--1, B--up}}}{16\pi \Mp^2}&=\frac{-\pi x}{\xi^2(y^2-1)}\left[\frac{1+xy+y\sqrt{1+2xy+x^2}}{\sqrt{1+2xy+x^2}}\right]+\frac{\pi R_{\rm crit}^2}{2}\,.
\end{align}
Note that for Type--A up-tunneling, we have $\delta_s=-\delta_\mathcal{A}$, such that the final term in \eqref{eq:B1_A_up} indeed vanishes.
\begin{table}[t]
\centering
\begin{tabular}{|c|c|c|}
\hline
 & $\sgn(P)$ & $\sgn(Q)$\\\hline
 \textbf{Down-tunneling Type--A} & $+1$ & $-1$\\\hline
 \textbf{Down-tunneling Type--B} & $-1$ & $-1$\\\hline
 \textbf{Up-tunneling Type--A} & $-1$ & $+1$\\\hline
 \textbf{Up-tunneling Type--B} & $-1$ & $-1$\\\hline 
\end{tabular}
\caption{The signs of $P$ and $Q$ for up- and down-tunneling transitions.}
\label{tab:signs}
\end{table}

For Type--A down-tunneling, the above already provides the full result
\begin{align}
    B_{\text{Type--A, down}}=\frac{16\pi^2\Mp^2 x}{\xi^2(y^2-1)}\left[\frac{1+xy-\sqrt{1+2xy+x^2}}{\sqrt{1+2xy+x^2}}\right]\,,
\end{align}
which is precisely the CDL result \eqref{eq:B_rxy} that we expected to find. 
For Type--B transitions, the contributions from the other regions take the form
\begin{align}
\begin{split}
    \frac{B_{\text{Region--2, B--down}}}{16\pi\Mp^2}&=\frac{B_{\text{Region--2, B--up}}+B_{\text{Region--3, B--up}}}{16\pi\Mp^2}=\int_{R_{\rm crit}}^{1/H_1}\dd R\; \pi R\\
    &= \frac{\pi}{2H_1^2}-\frac{\pi R_{\rm crit}^2}{2}=\frac{\pi x}{\xi^2 (1+y)}-\frac{\pi R_{\rm crit}^2}{2}\,.
\end{split}
\end{align}
In total, we find
\begin{align}
    B_{\text{Type--B, down}}=\frac{16\pi^2\Mp^2 x}{\xi^2(y^2-1)}\left[\frac{1+xy-\sqrt{1+2xy+x^2}}{\sqrt{1+2xy+x^2}}\right]\,,\\
    B_{\text{Type--B, up}}=\frac{-16\pi^2\Mp^2 x}{\xi^2(y^2-1)}\left[\frac{1+xy+\sqrt{1+2xy+x^2}}{\sqrt{1+2xy+x^2}}\right]\,.
\end{align}
For Type--A up-tunneling, we have
\begin{align}
    \begin{split}
        B_{\text{Region--2, A--up}}&=0\,,\\
    \frac{B_{\text{Region--3,A--up}}}{16\pi\Mp^2}&=\int_{1/H_2}^{1/H_1}\dd R\; \pi R = \frac{\pi}{2}\left(\frac{1}{H_1^2}-\frac{1}{H_2^2}\right)=-\frac{2\pi x}{\xi^2(y^2-1)}\,,
    \end{split}\label{eq:A_Regions_2_3}
\end{align}
such that the final result reads
\begin{align}
    B_{\text{Type--A, up}}=\frac{-16\pi^2\Mp^2 x}{\xi^2(y^2-1)}\left[\frac{1+xy+\sqrt{1+2xy+x^2}}{\sqrt{1+2xy+x^2}}\right]\,.
\end{align}
In all cases, we find, maybe surprisingly, the CDL result $B=B^{\rm CDL}$ \eqref{eq:B_rxy}.

\begin{figure}[ht!]
        \centering
        \includegraphics[scale=0.2]{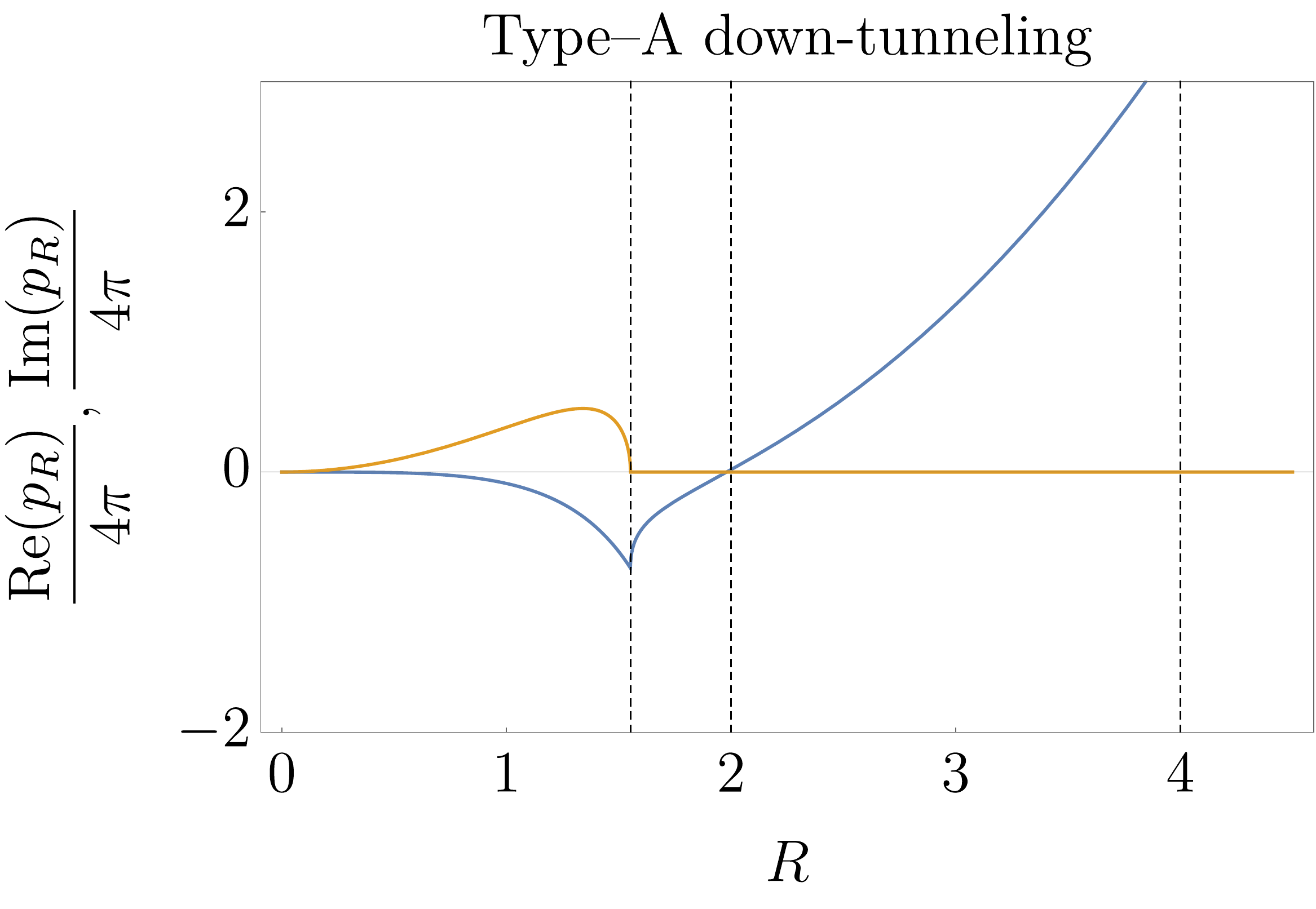}
        \includegraphics[scale=0.2]{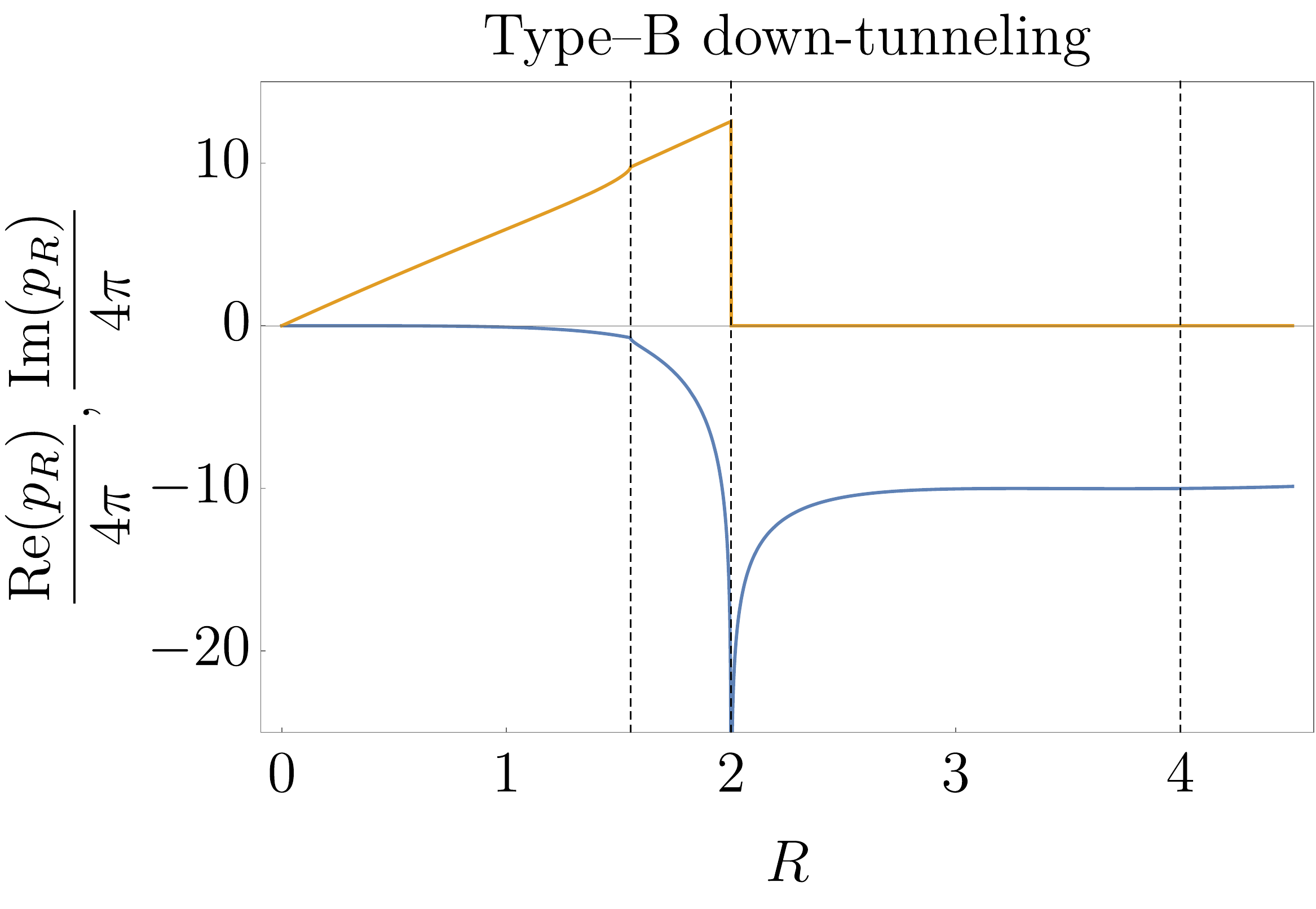}

        \includegraphics[scale=0.2]{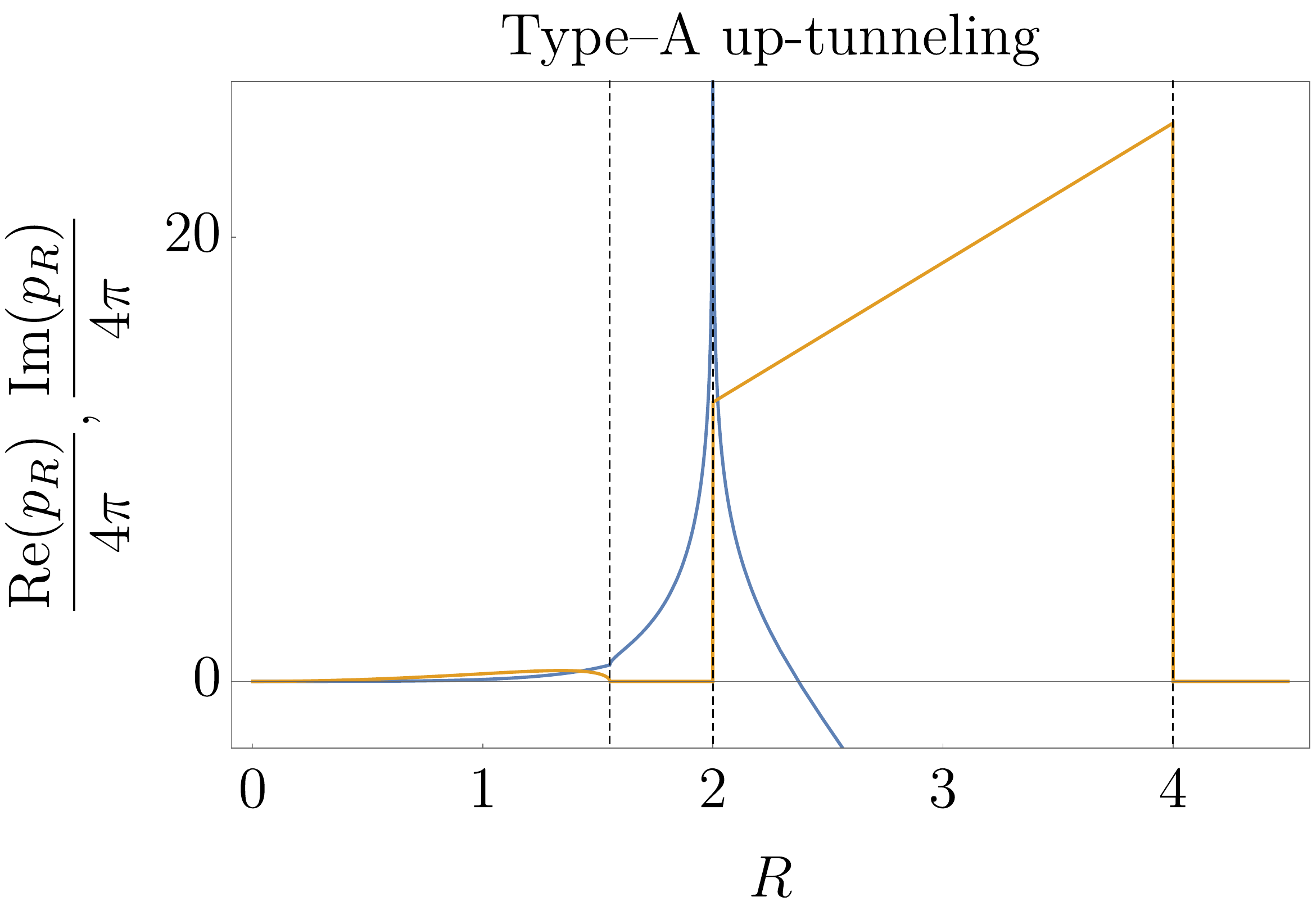}
        \includegraphics[scale=0.2]{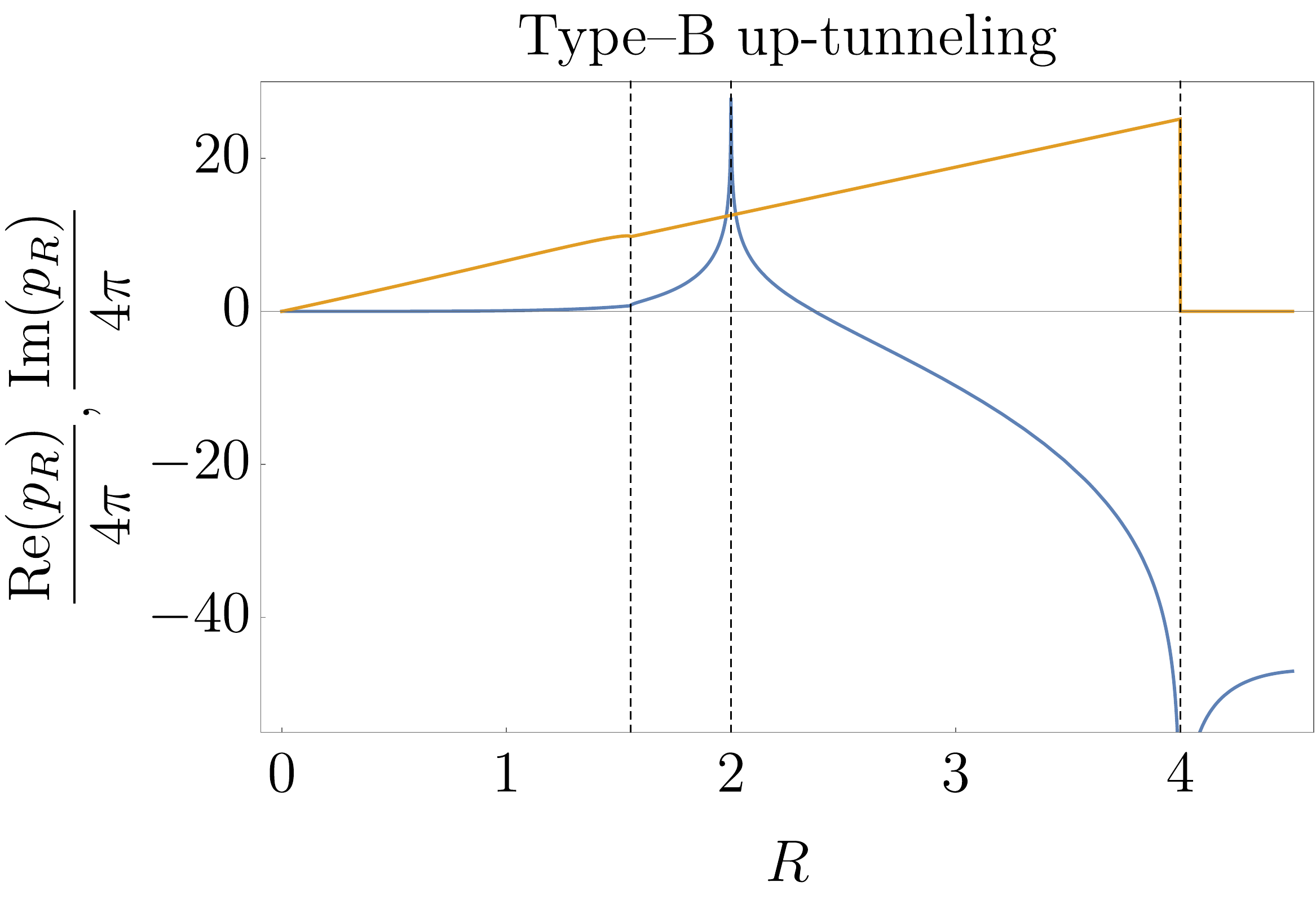}
        \caption{Real part (in blue) and imaginary part (in orange) of $p_R$ for four different transitions. For all examples we set $\Mp=1$ and the two Hubble parameters to $1/2$ and $1/4$. Whether they correspond to $H_1$ or $H_2$ determines if the transition is down-tunneling or up-tunneling. For these values, the critical tension that separates Type--A from Type--B is $\sigma_{\rm crit}=\sqrt{3/4}\approx 0.87$. Top left: Type--A down-tunneling with $H_1=1/2$, $H_2=1/4$ and $\sigma=3/8$. Top right: Type--B down-tunneling with $H_1=1/2$, $H_2=1/4$ and $\sigma=2$. Bottom left: Type--A up-tunneling with $H_1=1/4$, $H_2=1/2$ and $\sigma=1/2$. Bottom right: Type--B up-tunneling with $H_1=1/4$, $H_2=1/2$ and $\sigma=2$.}
        \label{fig:Re_Im_pR}
\end{figure}

To illustrate the situation, we look at various cases of dS--to--dS transitions and plot the real and imaginary parts of $p_R$.
The results are displayed in Fig.~\ref{fig:Re_Im_pR}.
It is quite remarkable that these peculiar functions reproduce exactly the CDL decay rates.

For all up-tunneling transitions and for Type--B down-tunneling transitions, the above computation again suggests that the domain wall goes on-shell at $R=1/H_1$, the parent dS horizon. 
As can be seen in Fig.~\ref{fig:Re_Im_pR}, the imaginary part of $p_R$, which serves as a potential, vanishes when the parent horizon is reached.
For the Type--A up-tunneling transition, the potential vanishes temporarily between $R_{\rm crit}$ and the new dS horizon $1/H_2$. However, we have seen that this is a consequence of leaving the principal sheet of both arctanh functions in $p_R$ in such a way that their contribution to $\Im(p_R)$ cancels.
This means that, despite a vanishing potential, the bubble cannot go on-shell.

Concerning the precise way in which the bubble eventually goes on shell, the situation is analogous to the bubble of nothing: For Type--A down-tunneling, this happens at the critical radius and the bubble immediately starts expanding. In all other cases, the bubble goes on shell at the parent de Sitter horizon and naively remains static. However, a small quantum fluctuation or a minimally non-extremal tunneling path puts it on an expanding trajectory.

\section{Discussion}
\label{sec:discussion}

In this section we discuss how our computation relates to CDL and its different interpretations. We also describe the Larfors--Johnson problem and its potentially disastrous implications for the multiverse.
We then explain how our approach can help determine the true fate of the multiverse. 

\subsection{The physical interpretation of CDL}

The computation in Sect.~\ref{sec:rates} is based on elementary quantum mechanics and has a clear geometric interpretation: The domain wall nucleates at small size and then grows, having to traverse a potential barrier before being able to go on shell.
As we have shown, for up-tunneling and Type--B transitions, it can go on shell at the horizon of the parent dS space. Such decays can occur at arbitrarily late time.
The geometric interpretation of the CDL computation, however, remains elusive. In this section, we explore several potential interpretations and highlight their inherent difficulties.

The CDL computation is based on the gravitational path integral.
The geometry of the CDL instanton is such that the domain-wall radius $R$ never exceeds $R_{\rm crit}<1/H_{1,2}$.
Hence, there is no off-shell configuration of the domain wall with $R>R_{\rm crit}$.
By contrast, in our approach $R$ becomes larger than $R_{\rm crit}$ while remaining off shell for up-tunneling and Type--B down-tunneling transitions.

In the CDL computation, the tunneling exponent $B$ involves two apparently unrelated contributions: one from the bounce or instanton, the other from the parent de Sitter, cf.~\eqref{eq:Decay_rate_def}.
Their interplay in a Lorentzian geometry remains, however, unclear.

One possible interpretation of the CDL result is as follows: the CDL bounce solution is regarded as describing the creation of a small, compact dS--dS universe from nothing. This corresponds to taking the half-Lorentzian geometry in Fig.~\ref{fig:cdl_A_B} at face value and is analogous to the vacuum-creation proposals of \cite{Vilenkin:1982de,Hartle:1983ai,Linde:1983mx,Vilenkin:1984wp}. In this interpretation the domain wall does indeed go on shell at $R=R_{\rm crit}$.
The parent-vacuum contribution to \eqref{eq:Decay_rate_def} then provides the `normalization' of the creation rates.

To be more precise, following \cite{Fischler:1989se,Fischler:1990pk,DeAlwis:2019rxg,Cespedes:2020xpn,Cespedes:2023jdk} the normalization can be interpreted by writing \eqref{eq:Decay_rate_def} as
\begin{align}
\exp(-B)=\frac{\exp(-S_{\rm E,instanton})}{\exp(-S_{\rm E,vacuum})}
\sim \frac{P(\varnothing\to\text{dS$_1$--dS$_2$})}{P(\varnothing\to\text{dS$_1$})}\,,
\label{eq:CDL_interpretation}
\end{align}
i.e.~as the ratio of the probabilities for creating a dS--dS system connected by a domain wall and for creating only the parent dS space from nothing.
How the initially existing  parent dS space fits
into this interpretation at an explicit, geometric level remains mysterious.

Another possibility to give an interpretation to the CDL formula is to consider disconnected geometries in the gravitational path integral.
\begin{figure}[t]
    \centering
    \includegraphics[width=0.6\linewidth]{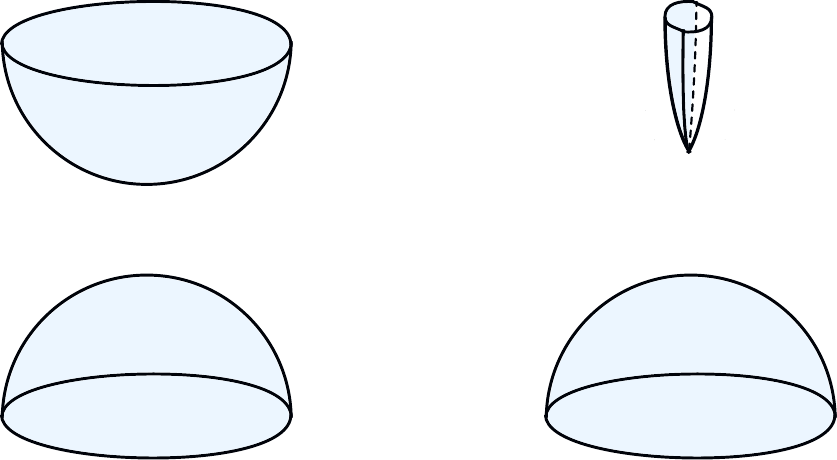}
    \caption{Two disconnected contributions to the path integral with the past and future boundary conditions being spherical. On the left, the destruction of a pure dS space followed by the nucleation of the same dS space is shown. On the right, the destruction of the parent dS space followed by the nucleation of a dS--dS system is displayed.}
    \label{fig:GPI_disconnected}
\end{figure}
In Fig.~\ref{fig:GPI_disconnected}, two such geometries with spherical boundary conditions are shown. 
On the left, a dS space being `destroyed' followed by the nucleation of the same space is shown. On the right, the destruction of the parent dS space followed by the nucleation of the dS--dS system is displayed. 
Assigning a transition probability to each process, one may then try to understand the probability of the CDL transition as the corresponding ratio:
\begin{align}
    \exp(-B)\sim\frac{P(\text{dS$_1$}\to\text{dS$_1$--dS$_2$})}{P(\text{dS$_1$}\to\text{dS$_1$})}\,.\label{eq:P_GPI_disc}
\end{align}
It is not clear how to make a precise argument that the above processes, which all occur at the `dS waist', are relevant at asymptotically late times.

For example, it is well known that the assigned creation rates of universes differ exponentially between the proposals of \cite{Vilenkin:1982de,Hartle:1983ai} and those of \cite{Linde:1983mx,Vilenkin:1984wp}.
The resulting ambiguities in the definition of decay rates have been discussed in \cite{DeAlwis:2019rxg,Cespedes:2020xpn,Cespedes:2023jdk}.

In \cite{Abdalla:2026mxn}, the gravitational path integral was used to revisit the computation of vacuum-creation rates.
The authors claim that all closed universes are equally likely to emerge from nothing, which might mean that late time transition processes are irrelevant. It is, however, unclear to us how the results of \cite{Abdalla:2026mxn} are consistent with standard cosmological predictions.

Decay rates of dS were computed in \cite{Hayashi:2021kro} using the Lorentzian path integral methods of \cite{Halliwell:1988ik,Halliwell:1989vu,Halliwell:1990tu, Lehners:2023yrj}. While agreement with Euclidean results was found\footnote{See \cite{Matsui:2021oio} for an analysis of the consistency of the Lorentzian gravitational path integral with WKB.}, the analysis was restricted to the field-theory regime. Hence Type--B and up-tunneling transitions are not covered. It would clearly be interesting to apply such Lorentzian gravitational path integral techniques to these cases. This could shed light on why our Sect.~\ref{sec:rates} recovered CDL rates. The  approach of \cite{Draper:2023fkz} for treating vacuum decay in time-dependent backgrounds might also prove useful.

A further interpretation of the CDL decay rate can be given in the context of the `Cosmological Central Dogma' (CCD) \cite{Banks:2000fe,Susskind:2003kw}.
It states that dS spaces should fundamentally be viewed from a static-patch perspective and that, in the quantum theory, they are described by a finite-dimensional Hilbert space of dimension $\exp(8\pi^2\Mp^2/H^2)=\exp({\cal S})$, where ${\cal S}$ is the dS entropy.
This perspective is reconciled with the global picture of dS space by regarding different patches of the late-time dS sphere as gauge equivalent. In the CCD approach, the Hamiltonian of a multi-dS system is given by a large Hermitian matrix and tunneling transitions occur due to non-zero matrix elements.
The CDL relation \eqref{eq:detailed_balance} is now understood as a result of Fermi's Golden Rule and the dimension of the Hilbert spaces, cf.~\cite{Hassfeld:2022vzk}. Furthermore, the formula \eqref{eq:Decay_rate_def} then admits a thermal interpretation: The factor $\exp(S_{\rm E, vacuum})=\exp(-{\cal S})$ characterizes the probability of being in one particular microstate of the parent dS. The factor $\exp(-S_{\rm E, instanton})$ counts the number of microstates that describe a dS--dS spacetime and hence enhances the rate correspondingly.

In contrast to the different CDL interpretations above, we feel that our analysis of dS decays more straightforwardly addresses the actual physics problem: Our result is a quantum mechanical wavefunction $\Psi(R)$, determined as the slowest decaying solution to the appropriate WDW equation, from which we estimate a decay rate by applying \eqref{eq:Gamma_WDW}.

\subsection{Relation to  CDL}

In this subsection, we seek to relate the CDL analysis to our late-time tunneling solutions.
The agreement between the decay exponents suggests that the complex wall trajectories of Sect.~\ref{sec:rates} are related to CDL geometries by an analytic deformation of an integration contour.
For simplicity, we here focus on the bubble of nothing scenario, but the following generalizes to dS--dS tunneling.
\subsubsection{Tunneling vs CDL trajectories}
It will be convenient to focus on the bubble wall trajectories as defined by $p_R(R)$, both in the on- and off-shell regimes. We will refer to our analysis of Sect.~\ref{sec:rates} as the `tunneling approach' and correspondingly denote the relevant trajectories $p_R^{\rm Tun}(R)$. They are to be distinguished from the CDL trajectories $p^{\rm CDL}_R(R)$. In the region $R>1/H$, the bubble is on-shell in all cases and the trajectories agree: $p_R^{\rm Tun}(R)=p^{\rm CDL}_R(R)$. They may be taken to be defined explicitly by \eqref{eq:pR_constraint}, which was derived in the standard way, passing from the action to the Hamiltonian formulation.

\begin{figure}[h!]
    \centering
    \includegraphics[scale=0.27]{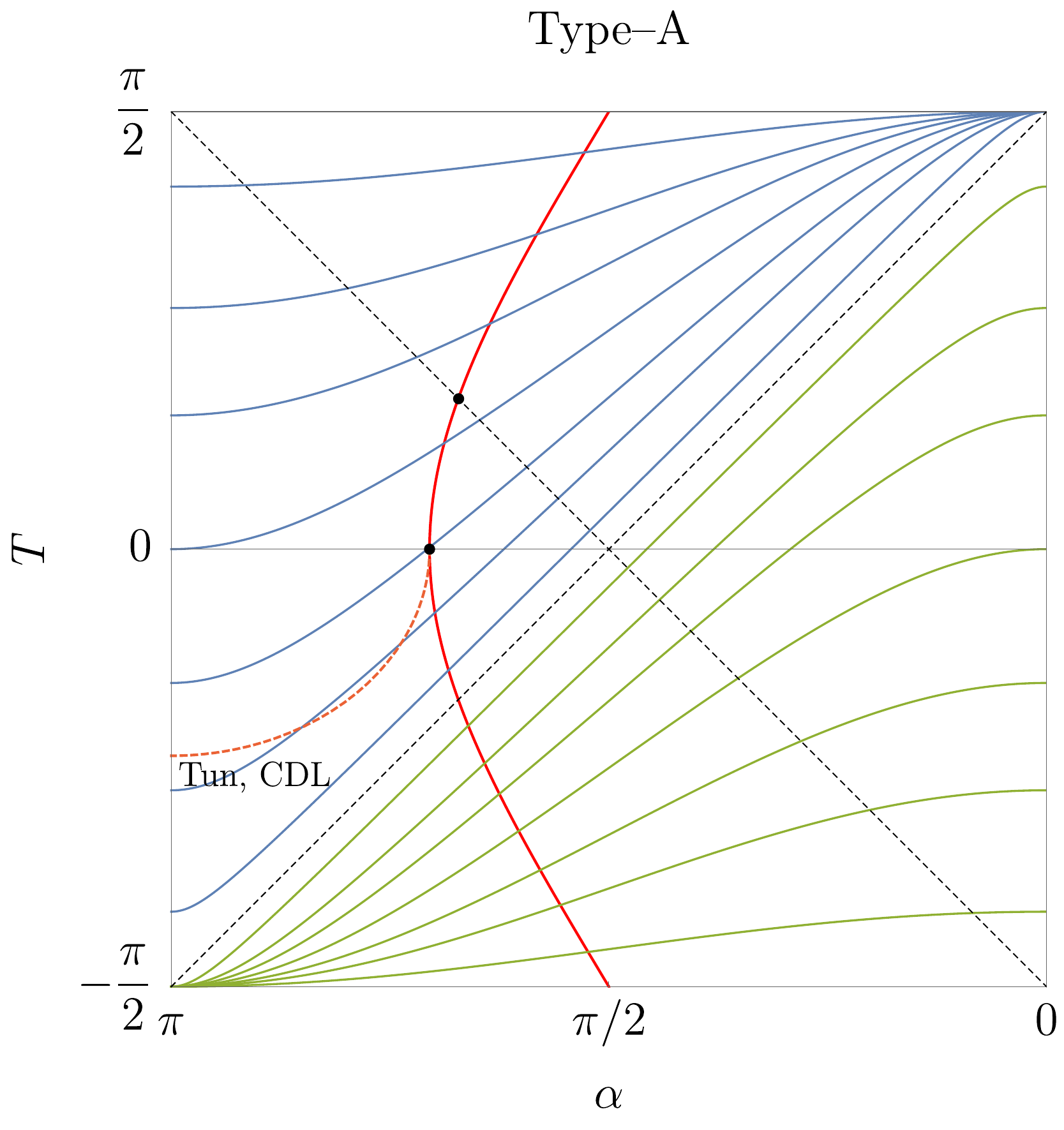}
    \quad
    \includegraphics[scale=0.27]{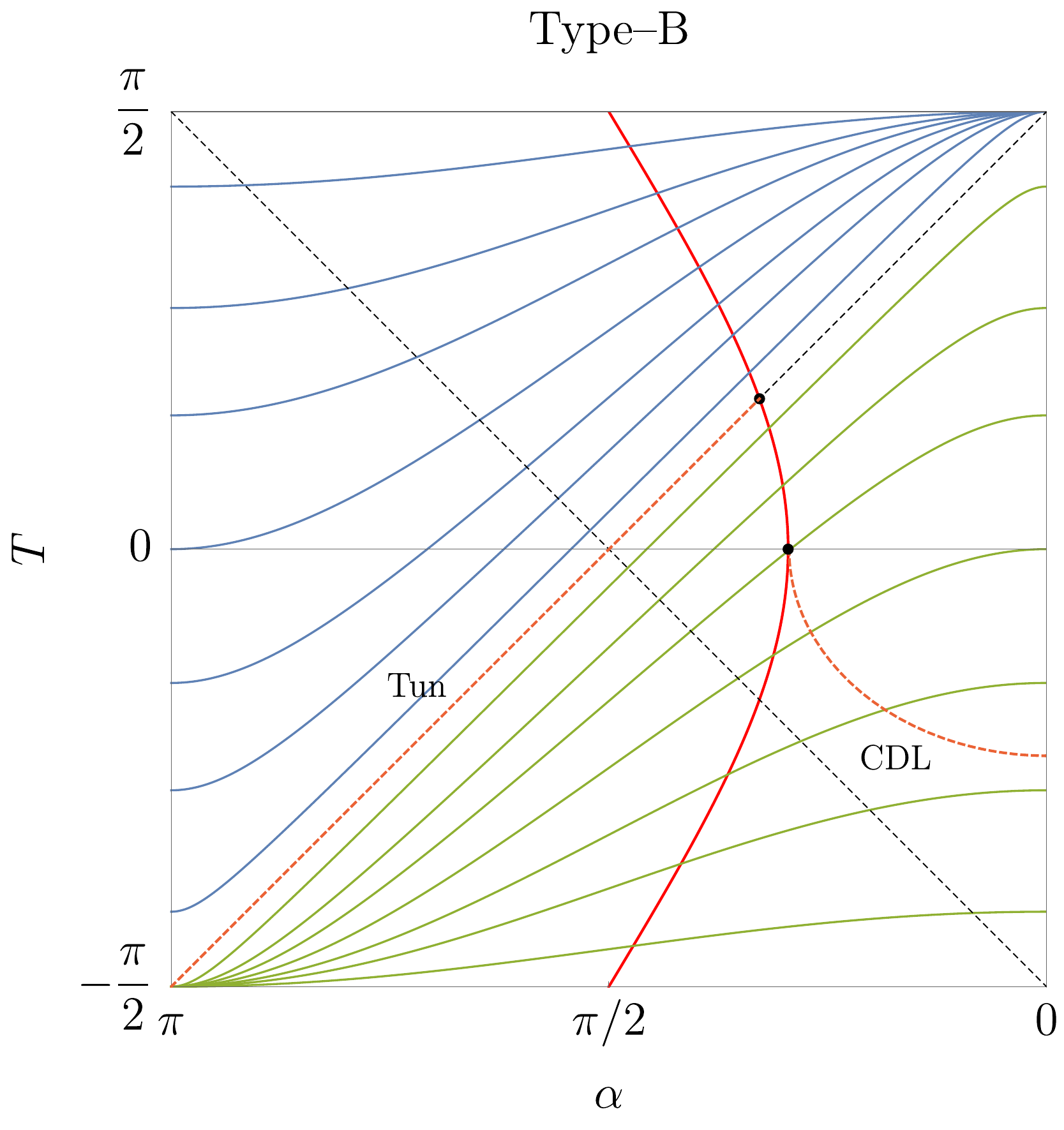}
    \caption{The red curves show unboosted Type--A (left) and Type--B (right) CDL trajectories, while the blue and green curves represent slices of constant $t$ and constant $\tilde t$, respectively. (Note  that, compared to Fig.~\ref{fig:typeAB}, we have drawn the Type--A trajectory on the left of the Penrose diagram for the parent spacetime always to be to the right of the wall trajectories.)
    The portion of the trajectory with $R_{\rm crit}<R_{\rm CDL}<1/H$ lies between the two black dots.
    At $R_{\rm CDL}=R_{\rm crit}$, the CDL trajectory is analytically continued once more toward small radius, as shown by the dashed red curves on the left and on the right of the two trajectories.
    For Type--A this trajectory matches our tunneling evolution described in this paper.
    By contrast, for Type--B, after undoing the large boost, the late-time off-shell trajectory from Fig.~\ref{fig:typeAB_boosted_bis} (top right) becomes approximately the dashed red curve lying along the $45$-degree horizon line.}
    \label{fig:flat_slicing_cdl}
\end{figure}

As a small caveat, we have to recall that uniqueness of the on-shell solution holds only up to boosts, cf.~Sect.~\ref{sec:trajectories}. We find it most convenient to work with the ordinary, `unboosted CDL' solution as depicted in Fig.~\ref{fig:typeAB}. Thus, we have to `unboost' our tunneling solutions accordingly, such that they precisely agree at $R>1/H$. This is illustrated in Fig.~\ref{fig:flat_slicing_cdl},  where we show the trajectories of a Type--A and a Type--B bubble, together with surfaces of constant flat-slicing time, which is equivalent to the late-global-time limit used before. More precisely, the blue lines are surfaces of constant flat-slicing time $t$ centered around $\alpha=\pi$.
This corresponds to the flat slicing \eqref{eq:metric_flat_expanding} used throughout Sect.~\ref{sec:rates}.
The green lines are surfaces of the complementary flat-slicing patch centered around $\alpha=0$ with time coordinate $\tilde t$.
The metric in this patch reads
\begin{align}
    \dd s^2=-\dd\tilde t^2+e^{-2H\tilde t}\left(\dd\tilde r^2+\tilde r^2\dd\Omega_2^2\right)\,.\label{eq:metric_flat_contracting}
\end{align}

For a process as studied in Sect.~\ref{sec:rates} to lead to the trajectories depicted in Fig.~\ref{fig:flat_slicing_cdl}, the decay must correspond to an early-time tunneling transition.
Although we have not emphasized this before, the methods developed there describe such a process as well since the flat-slicing coordinates also cover (part of) the early-time dS space. 
For a Type--B transition, the separation of the domain wall trajectory from the left static patch must be due to a quantum fluctuation needed to obtain an expanding bubble geometry, as discussed below Fig.~\ref{fig:max_boost_and_runaway}.

Consider now the Type--A situation in more detail:
As we have seen in Sect.~\ref{sec:rates}, the solution $s_-(R)$ is the unique solution to the Hamiltonian constraint in the region $R>R_{\rm crit}$, both for $R$ above and below the horizon. The first branch point encountered by the function $p_R$ is at $R=R_{\rm crit}$.
There are two possible analytic continuations of $p_R(R)$ to $R<R_{\rm crit}$, determined by the two solutions $s_+(R)$ and $s_-(R)$, which agree at $R=R_{\rm crit}$. The tunneling solution $p_R^{\rm Tun}$ corresponds to the choice $s_-(R)$.
The exponentially decaying CDL solution also corresponds to this trajectory.\footnote{To determine which of the two solutions $s_\pm$ defines the CDL trajectory for $R<R_{\rm crit}$, one can consider the field theory limit $H\to 0$, which is a perfectly smooth limit for the Type--A scenario. Here, it is clear that the solution $s=s_-$ is the one corresponding to the instanton.}

The situation is different for a Type--B bubble. 
Still, $s=s_-(R)$ is the unique solution to the Hamiltonian constraint at large $R$.
However, $p_R$ reaches a branch point already at $R=1/H$.
Hence, there are multiple possibilities for analytically continuing to $R<1/H$, corresponding to different options of looping around the branch point. 
As discussed in detail in Sect.~\ref{sec:rates}, the tunneling trajectory $p_R^{\rm Tun}(R)$ is obtained by looping around the branch point once, with counter-clockwise orientation. This induces a shift of $+i\pi$ of the function $\arctanh(s)$ appearing in $p_R$, cf.~\eqref{eq:pR_constraint}. In other words, our tunneling trajectory goes off-shell below $R=1/H$.

However, there also exists the option of not going around the branch point. 
We have previously dismissed this because the Hamiltonian constraint is then unsolvable. But there is a caveat: The point $R=1/H$ lies on the boundary of our coordinate chart, see Fig.~\ref{fig:flat_slicing_cdl}. If we allow ourselves to leave this chart, analytically continuing to the patch with coordinates \eqref{eq:metric_flat_contracting}, the constraint can perfectly well be solved and the system remains on-shell.

This solution then corresponds to the CDL trajectory $p_R^{\rm CDL}(R)$.
It remains on-shell in the regime $R_{\rm crit}<R<1/H$ and there now again exist two options for continuing $p_R^{\rm CDL}$ from  $R=R_{\rm crit}$ to $R=0$. They represent the two different solutions of the Hamiltonian constraint, cf.~\eqref{eq:s_BoN}.
Both the tunneling trajectory and the CDL trajectory correspond to the solution with smaller imaginary part of $p_R$.\footnote{In the tunneling case, this is justified by looking for the smallest overall suppression of the wavefunction. In CDL, this corresponds to the `Hartle--Hawking sign choice' if one interprets the instanton as the `creation from nothing' of a ball bounded by the ETW brane.}
As a result, we have that 
\begin{align}
    \label{eq:pR_Tun_CDL_B}
    p_R^{\rm CDL}=p_R^{\rm Tun}-i8\pi^2\Mp^2 R\qquad \text{for}\qquad R<\frac{1}{H}\,,
\end{align}
with the difference between $p_R^{\rm Tun}$ and $p_R^{\rm CDL}$ coming only from the sheet transition of $\arctanh(s)$ that occurs for the tunneling but not for the CDL trajectory.

\subsubsection{Tunneling exponents}

Using the trajectories above, we can now show that the decay exponents following from our tunneling analysis agree with those of CDL without relying on the explicit calculations in 
Sect.~\ref{sec:rates}. In other words, we have a much simpler way to relate the two tunneling exponents computed by \eqref{eq:Decay_rate_def} and by \eqref{eq:B_generic}.

To achieve this, we first observe that
\begin{equation}
    2\Im\left(\int_0^{R_{\rm on-shell}} p_R\dd R\right)=2\Im\left(\int_0^{R_{\rm on-shell}} \frac{\mathcal{L}\dd R}{\dot R}\right)=2\Im\left(\int\mathcal L\dd t\right)\,,\label{eq:Lagrangian_pR}
\end{equation}
where we have used that the constraint $\mathcal{H}=0$ holds along the chosen contour of integration.
Here $\mathcal{L}$ is the Lagrangian of the Lorentzian action \eqref{eq:S_Lorentzian}, expressed in the coordinate system \eqref{eq:metric_flat_expanding}. This implies that the tunneling exponents as introduced in \eqref{eq:B_generic} are directly related to the imaginary part of a complex action (in the simplest case, to a Euclidean action).

Our goal is then to see how the two quantities
\begin{align}
    \tilde{B}^{\rm Tun}\equiv 2\Im\left(\int_0^{R_{\rm on-shell}^{\rm Tun}} p_R^{\rm Tun}\dd R\right)\quad\text{and}\quad \tilde{B}^{\rm CDL} \equiv 2\Im\left(\int_0^{R_{\rm on-shell}^{\rm CDL}} p_R^{\rm CDL}\dd R\right)\label{eq:tunneling_vs_CDL_pR}
\end{align}
relate to each other, thereby connecting the tunneling and the CDL results.

For Type--A, we have 
\begin{align}
    p_R^{\rm Tun}=p_R^{\rm CDL} \qquad \text{and}\qquad  R_{\rm on-shell}^{\rm Tun}=R_{\rm on-shell}^{\rm CDL}=R_{\rm crit}\,,
\end{align}
such that the two integrals in \eqref{eq:tunneling_vs_CDL_pR} are equal.
As we have seen explicitly in Sect.~\ref{sec:rates}, our key object of interest, the late-time tunneling exponent $B$, is calculated by the expression for $\tilde B^{\rm Tun}$ above. We thus have
\begin{align}
    B=\tilde{B}^{\rm Tun}=\tilde{B}^{\rm CDL}\,.
\end{align}
As the last crucial step, we now have to convince ourselves that $\tilde{B}^{\rm CDL}$ equals what is widely known as the `CDL tunneling exponent': $B^{\rm CDL}\equiv S_{\rm E,instanton}-S_{\rm E,vacuum}$. To do so, we first note that \eqref{eq:Lagrangian_pR} relates $\tilde B^{\rm CDL}=\tilde{B}^{\rm Tun}$ to the action integral \eqref{eq:S}. The latter followed from \eqref{eq:S_before_subtraction} after dropping the term $-12\pi^2\Mp^2 H^2 a^3$. Dropping this term corresponds to the subtraction of a background contribution.
This subtraction is necessary to yield a well-defined action, as the slices of constant $t$ extending into the bulk from the domain wall are infinite.
In Fig.~\ref{fig:flat_slicing_cdl} on the left, the infinity corresponds to the blue slices extending from the domain wall all the way to the top-right corner of the diagram. Apart from this background subtraction, our action describes the geometry underlying the CDL analysis.\footnote{
To 
be precise, the `green-sliced' right-lower triangle in the Type--A panel of Fig.~\ref{fig:flat_slicing_cdl} is not part of our integration domain. But since it also cancels between CDL instanton and background, this is immaterial.
} 
It is then clear that $B=B^{\rm CDL}$.

For Type--B, $R_{\rm on-shell}^{\rm Tun}=1/H$ while $R_{\rm on-shell}^{\rm CDL}=R_{\rm crit}$.
From \eqref{eq:pR_Tun_CDL_B}, we then see that 
\begin{align}
     \begin{split}
         B=\tilde{B}^{\rm Tun} &= \tilde{B}^{\rm CDL}+16\pi^2\Mp^2\int_0^{1/H}\dd R\,R= \tilde{B}^{\rm CDL}-S_{\rm E,vacuum}\,.
     \end{split}\label{eq:pR_compare}
\end{align}
Thus, to conclude that $B=B^{\rm CDL}$, we now need to show that $\tilde B^{\rm CDL}= S_{\rm E,instanton}$, without the vacuum subtraction. As in the Type--A case, we first appeal to \eqref{eq:Lagrangian_pR} to relate $\tilde{B}^{\rm CDL}$ to the action of a particular geometry: It is the CDL geometry in the right panel of Fig.~\ref{fig:flat_slicing_cdl} parameterized by the two different flat slicings shown in the figure. Let us start  with the part where $R>1/H$. This region is described by the coordinate system \eqref{eq:metric_flat_expanding} whose constant-time slices are drawn in blue. This is the same slicing used in the Type--A discussion above and it hence includes a background subtraction.
However, this region is on-shell, making the action purely real and giving no contribution to $\tilde B^{\rm CDL}$.

Below the horizon, i.e. for $R<1/H$, but still in the on-shell region $R>R_{\rm crit}$, the geometry is parameterized using the coordinates \eqref{eq:metric_flat_contracting}. The relevant constant-time slices are shown in green in Fig.~\ref{fig:flat_slicing_cdl}.
The parts of these time slices to the right of the domain wall are obviously finite.
Hence the corresponding part of the geometry is {\it not} background subtracted -- it is simply the Lorentzian part of the CDL geometry inside the horizon. The imaginary part of the action arising by further analytic continuation to $R<R_{\rm crit}$ hence gives the familiar CDL instanton result:
\begin{equation}
    \tilde{B}^{\rm CDL}=S_{\rm E, instanton}\,.\label{eq:p_R_CDL_integrated}
\end{equation}
Thus, also in the Type--B case, we have arrived at $B=B^{\rm CDL}$.

We emphasize that, as discussed in detail throughout our paper, the physical interpretation remains different.

\subsubsection{Quick argument for the relation to CDL} 

With the benefit of hindsight, we can give a quicker way to relate our result to CDL. For simplicity, we focus again on the bubble of nothing. As mentioned above, the tunneling exponent can be obtained by calculating the on-shell action along the (complex) tunneling trajectory. Since we will only evaluate the on-shell action along the trajectory fixed by the constraint equation, we can simplify the action by using that the extrinsic curvature term is proportional to the tension term on shell, and the cosmological constant term is similarly proportional to the Ricci scalar term. The on-shell action therefore has only two terms,
\begin{align}
    S_{\rm on-shell} = 2\pi  \sigma \int \dd \tau R^2 +3\Mp^2H^2 \int \dd^4x \sqrt{-g}\,.
\end{align}
This action differs from the action we have analyzed in most of the paper by a total derivative term, related to the integration by parts we used, but must give the same physical answers.

The bulk term in the action will be divergent due to the infinite volume of the spatial slices. However, we really want the difference in action between the bounce and the background. These differ only by the missing piece of de Sitter inside the bubble, which has spatial volume $4 \pi R^3/3$. Therefore, the difference in actions takes the form
\begin{align}
     \Delta S_{\rm on-shell} = 2\pi \sigma \int \dd \tau R^2 - 4\pi\Mp^2 H^2 \int \dd t R^3\,.
\end{align}
This is precisely the formula for the bounce action minus the background action, written in the flat slicing coordinates, as depicted in the right panel of Fig.~\ref{fig:flat_slicing_cdl}.  Note that two different time variables appear in the above formula, the proper time $\tau$ and the flat slicing time $t$.

Now, to relate our calculation to CDL, note that
\begin{itemize}
    \item This form of the action is related to the action analyzed in previous sections by a total derivative term, so it must give the same decay rate.
    \item The constraint equation determines uniquely the complex trajectory $t(R)$, up to physically irrelevant branch choices. To be concrete, the trajectory is
    \begin{equation}
        t(R) = H^{-1} \log \left( H\sqrt{R^2 - R_{\rm crit}^2} - \sqrt{1 - H^2 R_{\rm crit}^2} \right)\,.
    \end{equation}
    The ambiguities related to the choice of branch of the logarithm arise because we chose to use the flat slicing time; these would disappear if we had chosen to use, for example, embedding coordinates to describe the trajectory. Explicitly, the value of the embedding coordinates only depends on $\exp(H t)$, which is unambiguous.
    \item This trajectory \emph{is} the CDL domain wall trajectory, described in flat slicing coordinates. It includes a complex region with $R< R_{\rm crit}$, as sketched in the right panel of Fig.~\ref{fig:flat_slicing_cdl}. 
    \item Since we are evaluating the difference in on-shell actions between the bounce and the background, as established above, along the CDL trajectory, we are guaranteed to get the CDL answer. It is possible but unilluminating to explicitly show this.
\end{itemize}
This argument seems to give a quick way to see that our answer must agree with the CDL result. However, there is a caveat. In previous sections, we analyzed a trajectory where the domain wall tunnels to a size slightly smaller than the horizon, then makes a quantum fluctuation across the horizon, as shown for the tunneling trajectory in the right panel of Fig.~\ref{fig:flat_slicing_cdl}. Due to the presence of the quantum fluctuation, this description seems to have the possibility to deviate from the CDL solution, and therefore have a different action. For this reason, we find the slick argument of this subsection not completely convincing; however, the explicit computations of the previous subsection demonstrate that the action is the same. The `coincidence' between our computation and CDL suggests that there is a more elegant way to relate them.

\subsubsection{A speculative alternative geometry}
\begin{figure}[ht]
    \centering
    \includegraphics[scale=0.34]{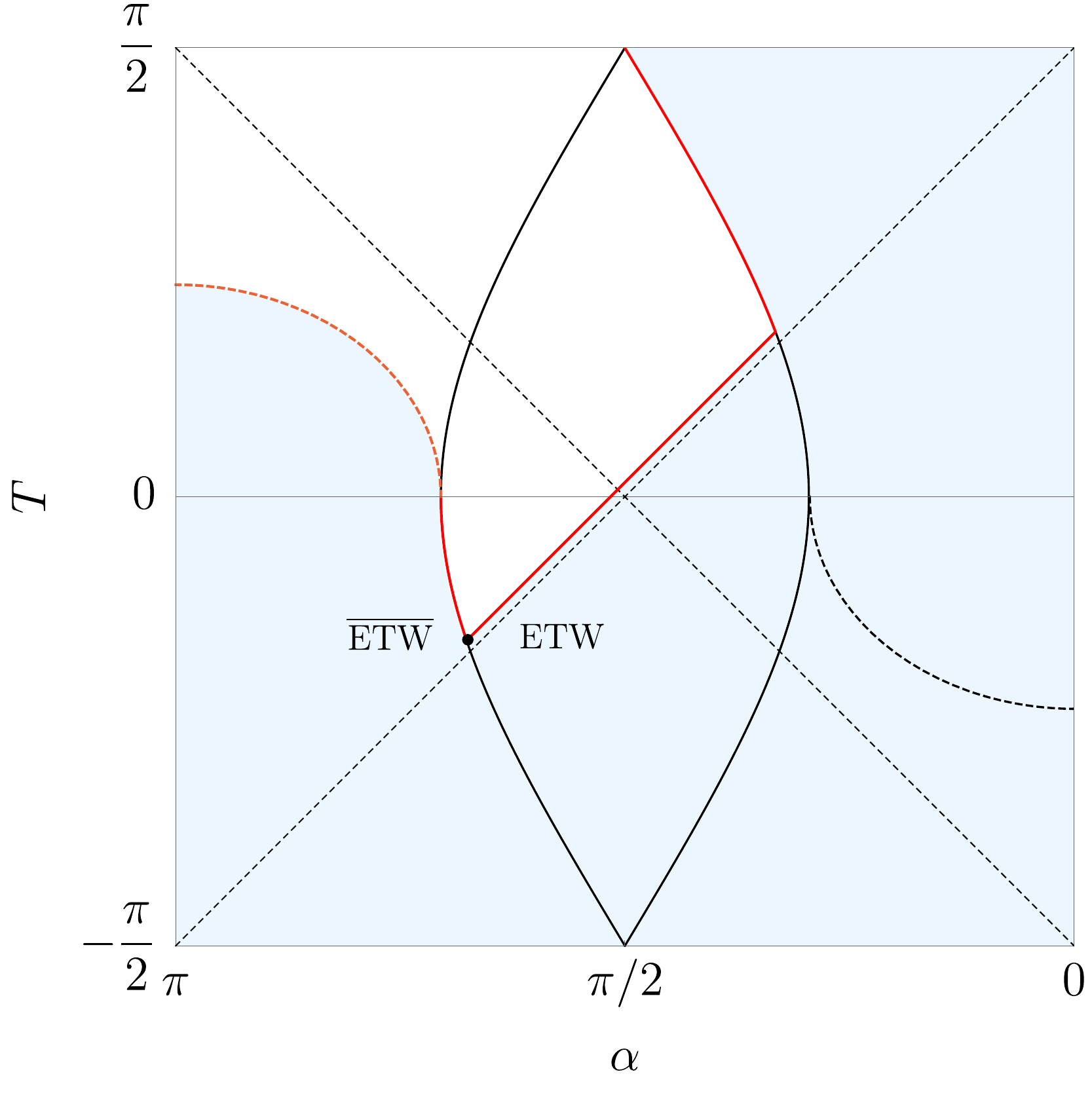}
    \caption{An attempt to illustrate the complex geometry of our tunneling calculation. The black dot stands for the quantum creation of a pair of oppositely oriented ETW branes, one of which quantum-jumps to outside the horizon. See text for more details.}
    \label{fig:tunneling_traj}
\end{figure}

We end this subsection with a 
speculation concerning a possible physical interpretation of the tunneling trajectory $p^{\rm Tun}_R(R)$ and the corresponding complex geometry.
Figure~\ref{fig:tunneling_traj} represents an attempt to illustrate this geometry: The uppermost part of the red line is on-shell CDL outside the horizon. The diagonal straight part of the red line is the quantum jump taking the ETW brane from just inside to just outside the horizon -- cf.~the previous discussion related to Fig.~\ref{fig:max_boost_and_runaway}. The curved red line above the black dot is the off-shell part of the tunneling trajectory between critical radius and horizon size. Here, the momentum is complex but $\dot{R}$ is real and the constraint is solved with the wrong sign of the square root. We can view this as being on shell if we put the parent dS on the other side of the wall, as done in the figure.\footnote{
By 
drawing this part of the line in the Penrose diagram, we identify the time variable in the region $R_{\rm crit}<R<1/H$, as defined implicitly through $\dot{R}=\dot{R}(p_R^{\rm Tun}(R),R)$, with our original flat-slicing time in the on-shell region $R>1/H$. It is not obvious whether this has physical meaning.
} 
Then comes the part which is truly complex, shown as a dashed red line.

Now, this picture invites the following interpretation: A brane--anti-brane pair of ETW branes is created just below the horizon (at the black dot on the figure). 
One brane quantum-jumps to super-horizon size on the right of the diagram and realizes a growing Type--B bubble. The second brane contracts to zero size, first on-shell and then following a tunneling trajectory. The key pair creation process is reminiscent of the naive interpretation of Hawking radiation as pair-creation near the black hole horizon.

It would be interesting to try to derive and interpret the late-time decay of dS starting from this geometry. It is possible that the physical process as described above is the correct way to interpret our analysis, but it could also turn out to be a subleading process. Moreover, it is interesting to consider situations where the tension of the ETW branes involved is small. Then, pair-creation should have small action cost and, dropping also the requirement that the dS region near $\alpha=\pi$ `tunnels to nothing', one may expect a novel and potentially dominant channel for the decay of dS. Clarifying this goes beyond the scope of the present paper. 

\subsection{Larfors--Johnson problem}

In this subsection, we review and expand on a phenomenologically relevant problem arising for decays in the type IIB flux landscape, due to the existence of unstable domain walls \cite{Johnson:2008kc,Johnson:2008vn,Aguirre:2009tp}. 
We will call this issue the `Larfors--Johnson problem'.
We explain how the methods developed here can help address this problem.
\subsubsection{Unstable domain walls}

The series of papers \cite{Johnson:2008kc,Johnson:2008vn,Aguirre:2009tp} investigated domain-wall effects on vacuum transitions within the string-theoretic flux landscape. Their results apply to the KKLT \cite{Kachru:2003aw} and LVS \cite{Balasubramanian:2005zx} de Sitter models, but also more generally to all type IIB vacua where complex structure moduli are stabilized by fluxes at an energy scale above that of Kähler moduli stabilization. In this context, the relevant domain walls are flux-changing. They are realized by D5/NS5-branes wrapping three-cycles inside the compact Calabi--Yau geometry.

To explain the main idea, we recall that the scale at which complex structure moduli are stabilized by fluxes is
\begin{equation}\label{m_flux_V}
m_{\rm flux}\sim g_{\rm s}^{1/2}/\V_0\,.
\end{equation}
Here $g_{\rm s}$ is the string coupling, $\V_0$ is the Calabi--Yau volume measured in units of the 10d Planck scale, and the 4d Planck mass has been set to unity.

Kähler moduli do not obtain a potential from three-form fluxes. They are treated in a low-energy EFT where the complex structure moduli are integrated out. The cutoff of this EFT is set by 
\eqref{m_flux_V}.
Whatever the specific mechanism of Kähler moduli stabilization might be, consistency requires the Kähler moduli mass $m$ to obey
\begin{equation}
m\ll m_{\rm flux} \qquad\text{or, equivalently,} \qquad m\V_0\ll  g_{\rm s}^{1/2}\,.\label{mV0_flux}
\end{equation}
It is easy to see that, again in units with $M_{\rm P}=1$, the flux-changing domain walls mentioned above have tension
\begin{equation}\label{tensionDW}
    T = \frac{A_0}{\mathcal{V}}\,.
\end{equation}
Here $A_0\gtrsim \mathcal{O}(g_{\rm s}^{1/2})$ is independent of the K\"ahler moduli but depends on the value of the complex structure moduli and axio-dilaton in the two adjacent vacua.
Specifically for D5- and NS5-branes, $A_0$ scales as $g_s^{1/2}$ and $g_s^{-1/2}$ respectively.
The quantity ${\cal V}$ appearing in \eqref{tensionDW} is the Calabi--Yau volume at the location of the wall. As we will see, due to domain-wall back-reaction it is in general different from its vacuum value ${\cal V}_0$. From now on, we will for simplicity assume that ${\cal V}$ is the only Kähler modulus and that ${\cal V}_0$ is the same in the two relevant flux vacua.

We note that for certain vacuum pairs the domain wall can accidentally be BPS, such that the numerator $A_0$ is suppressed. However, for two randomly chosen flux vacua the connecting domain wall breaks supersymmetry and the relation $A_0\gtrsim \mathcal{O}(g_{\rm s}^{1/2})$ holds.\footnote{
To 
be more precise on this, consider a basis of special Lagrangian 3-cycles that all preserve the same subset of supercharges. A domain wall made of a single D5- or NS5-brane wrapped on one of these basis cycles would implement a positive unit shift on the associated flux. Decreasing the flux requires wrapping the cycle with opposite orientation, or equivalently, to wrap an anti-brane in the same way, which preserves the complementary subset of supercharges. To be able to connect two different flux vacua with a BPS domain wall all the branes should preserve the same supercharges. We thus conclude that the flux quanta in one vacuum should all at the same time be bigger or smaller than their counterparts in the other vacuum. If one pictures the flux lattice with one vacuum being at the origin, the other vacuum must be in the appropriate quadrant out of the $2^{1+h^{2,1}}$ possibilities.
}\textsuperscript{,}\footnote{
We 
note that flux-changing domain walls have recently been revisited in the Swampland context, see e.g.~\cite{Lust:2022lfc, 
Shiu:2022oti, Basile:2023rvm, Bena:2024are,Montero:2024qtz,Cribiori:2026btb}.}

It will be convenient to replace ${\cal V}$ by the corresponding canonically normalized scalar field $\phi$,
\be
{\cal V}={\cal V}_0e^{B_0\phi}\,,\qquad \mbox{such that}\qquad
T=\frac{A_0}{{\cal V}_0}e^{-B_0\phi}\equiv T(\phi)\,.
\ee
Here $B_0=\sqrt{3/2}$ and, by definition, $\phi=0$ in the vacuum. With this, the 4d effective action with volume modulus and domain wall reads
\begin{equation}
S=\int\dd^4x\left[-\frac{1}{2}(\partial\phi)^2-V(\phi)-\delta(x^3)T(\phi)\right]\,.
\end{equation}
Here we have, for simplicity, taken the domain wall to be localized in the $x^3=0$ hyperplane. Near $\phi=0$, the moduli-stabilizing (e.g.~KKLT or LVS) scalar potential may be approximated as $V(\phi)\approx V_0+m^2\phi^2/2$. 

\begin{figure}[ht]
\centering
\includegraphics[scale=0.75]{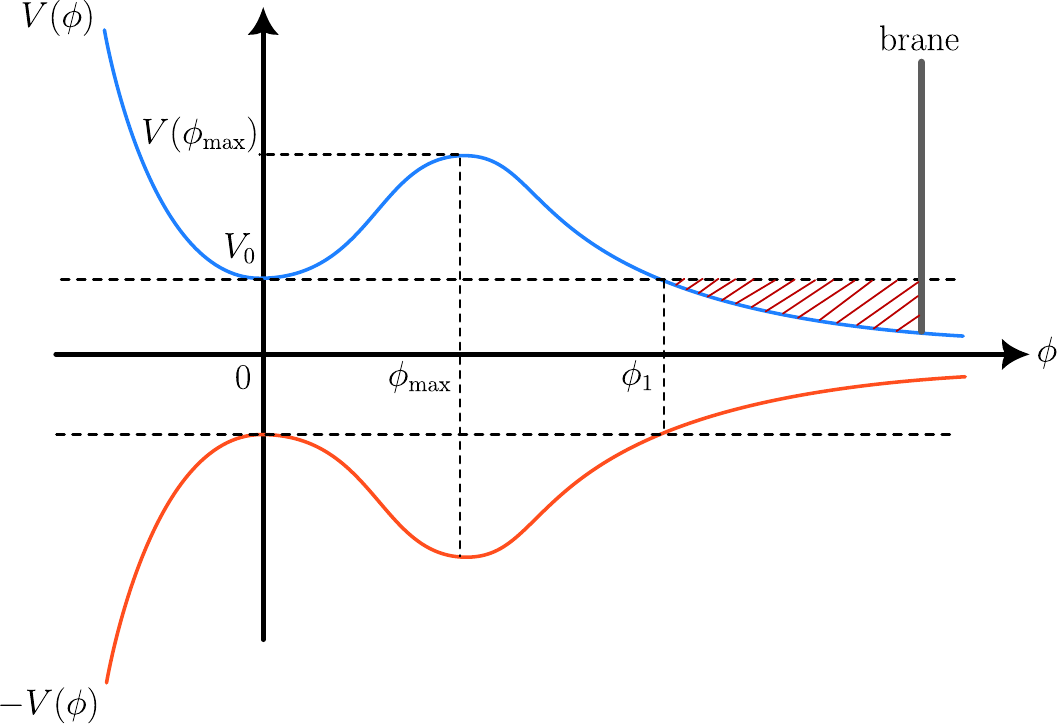}
\caption{A typical potential with a dS metastable minimum and a runaway region, starting at some field value $\phi_{\rm max}$. The field profile in the presence of a domain wall should start at $\phi=0$ and reach the `brane' value at the location of the D5/NS5. However, since this brane value is in the runaway region, such a configuration would be unstable: The vacua on both sides of the wall would decay. Alternatively, a Euclidean domain wall can be described by a particle moving in the inverted potential (with a positive sign for the damping term in the presence of curvature). It is then impossible for the field to go beyond $\phi_1$. As a result, a bounce solution with a D5/NS5 does not exist either.}
\label{fig_instanton}
\end{figure}

Given this action, it is straightforward to calculate the (Euclidean) solution $\phi(x^3)$. The problem is very similar to a Schr\"odinger equation with $\delta$-function potential. One finds $\phi(x^3)=\phi(0)\exp(-m|x^3|)$ and, using the boundary conditions at the domain wall, the domain-wall-induced field displacement
\begin{equation}
    \Delta\phi\equiv\phi(0)-\phi(\pm \infty)=\phi(0)\simeq B_0^{-1}\ln\left(\frac{A_0B_0^2}{m\V_0}\right)\,.
\end{equation}
For $m\V_0\ll 1$, this becomes parametrically large. 
In particular, $\Delta \phi$ should be compared to the field value $\phi_{\rm max}$ that separates the attractor valley of the vacuum from the runaway region (see Fig.~\ref{fig_instanton}). Thus, since $\phi_{\rm max}$ is at best $\mathcal{O}(1/\ln |W_0|)$ in KKLT and $\mathcal{O}(1)$ in LVS, the wall unavoidably drives the vacuum into the runaway regime. Put differently, domain walls interpolating between vacua in the string landscape destabilize those vacua: They simply do not exist as stable solutions. We refer to this issue, which has been analysed in detail in \cite{Johnson:2008kc,Johnson:2008vn,Aguirre:2009tp}, as the `Larfors--Johnson problem'. 

One may also try to take the D5/NS5-based domain wall solution as part of a Euclidean, CDL-type bounce geometry. It is then convenient to consider the inverted potential (see Fig.~\ref{fig_instanton}) and think of a particle that starts at $\phi=0$ and rolls to larger $\phi$. In our setting, this particle can never reach the required large value $\phi(0)$ near the brane since energy conservation forbids it from going beyond the turning point $\phi_1$, which is $\mathcal{O}(1)$. Crucially, in this approach one can go beyond the flat-space approximation and include gravitational or curvature effects. It turns out that they contribute a positive-definite friction term and thus cannot help in that regard. In summary, not only is the flat brane unstable but a bounce solution cannot exist either.

Finally, we note that in KKLT and LVS the dimensionless parameter $x$ defined in \eqref{eq:dim_less_variables} scales as follows:
\begin{equation}
    \text{\underline{KKLT:}}\quad x\sim\frac{1}{|W_0|^2}\gg 1\,,\qquad\qquad\text{\underline{LVS:}}\quad x\sim\V_0\gg 1\,.
\end{equation}
Since $x\gg 1$, such transitions are always of Type--B. We will see below that, as a result, decaying domain walls together with a naive CDL picture of quantum tunneling appear to compromise the whole multiverse.

\subsubsection{Effective theory of unstable domain walls}

\begin{figure}[ht]
        \centering
        \includegraphics[scale=0.6]{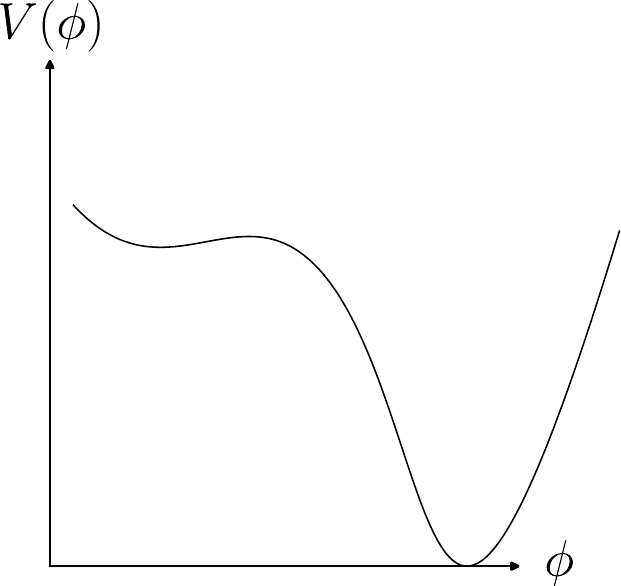}
        \qquad\qquad\qquad\qquad
        \includegraphics[scale=0.7]{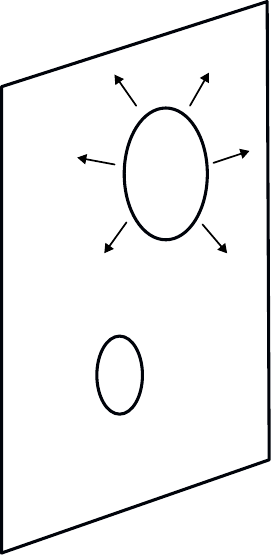}
        \caption{Equivalent descriptions for the decay of a domain wall. On the left: A brane-localized potential allowing for the decay. On the right: Holes bounded by string loops can form in the domain wall and `eat it up'.}
        \label{fig:unstable_DW}
    \end{figure}

Before delving into the multiverse implications of the Larfors--Johnson problem, we find it useful to describe two simple toy models for the main culprit: the unstable domain walls. First, one may consider an effective action
\begin{equation}
    S_{\rm DW}=-\int_{\rm DW}\sqrt{-g}\,\left(\frac{1}{2}\partial_m\phi\partial^m\phi+V_{\rm loc}(\phi)\right)\,,
\end{equation}
where the index $m$ labels coordinates on the domain wall and the potential $V_{\rm loc}(\phi)$ takes the form shown in the left panel of Fig.~\ref{fig:unstable_DW}.
The field $\phi$ is localized at the wall. If $\phi$ sits in the metastable minimum on the left, $S_{\rm DW}$ describes a high-tension domain wall. This wall is unstable to the tunneling of $\phi$ into the zero-tension minimum on the right. After such a tunneling event, the whole domain wall can decay to (e.g.~gravitational) radiation.

A very similar toy model is the following: Consider a 4d effective theory with a four-form field strength $F_4=\dd A_3$ and a three-form field strength $F_3=\dd A_2$. Let `$A_2$ be gauged by $A_3$', such that the gauge part of the 4d action takes the form
\begin{equation}
    S=-\int_{\rm 4d}\sqrt{-g} \left(\frac{1}{4g_4^2}|F_4|^2+\frac{1}{4g_3^2}|F_3-A_3|^2\right)+\int_{\rm DW}A_3+\int_{\rm String}A_2\,.
\end{equation}
This has to be supplemented with standard tension terms with tensions $T$ for the domain wall and $\mu$ for the string.
As a result, any given domain wall can decay through tunneling events in which small holes bounded by string loops form and subsequently grow, cf.~Fig.~\ref{fig:unstable_DW}, right panel. The decay rate is $\sim\exp(-c\mu^3/T^2)$, with $c$ a numerical constant.

Note that in both our models the domain walls are metastable. The classical instability of the type IIB domain walls is recovered only in the limit of tuning the decay time to be small. For example, one may use a small string tension in the last example.

\subsubsection{Problems caused by unstable domain walls in conjunction with CDL}

Now consider a tunneling event in de Sitter which, due to the high tension of the relevant domain wall, falls in the Type--B regime. In addition, assume that the domain wall can decay -- either as in the flux landscape or as described by the effective models above. Moreover, let us take the CDL picture for how the Lorentzian geometry emerges seriously in the most straightforward way. In this case, the newly created spatial geometry, which consists of two 3-balls glued at the boundary, can disappear completely. 
Indeed, after the domain wall has decayed, which may happen fast, there is no longer a constant energy density pulling the total 3-sphere spatial geometry to larger size. Depending on the details of the model, this energy density is replaced by, e.g., radiation. 
The initial `pancake'-shaped universe, which arises for Type--B transitions (cf.~the right-hand side of Fig.~\ref{fig:cdl_A_B}), has thus turned into a simple 3-sphere filled with radiation.
If the energy densities of the cosmological constants $V_1,V_2$ (one may even assume $V_1=V_2$ for simplicity) are at this stage still negligible, the 3-sphere will collapse and crunch. This is the inevitable fate of any closed universe with insufficient vacuum energy.\footnote{The collapse can of course be avoided if the decay of the domain wall happens late enough such that $V_1,V_2$ come to dominate.}

The possibility of such events would have strong implications for the cosmological central dogma and the global de Sitter multiverse.
In this context, with `bubbles within bubbles within bubbles,' observers are counted on a late-time cutoff surface. The semiclassical evolution of this Cauchy surface is supposed to be eternal: Even though local regions may crunch, the surface as a whole never crunches. But this appears to clash with the Type--B transition with unstable domain walls as just discussed. In such transitions, the whole Cauchy surface crunches, implying an apocalypse of the entire multiverse. Even if such transitions are rare, a sufficiently late Cauchy surface in the global landscape picture will contain observers who witnessed such a transition with probability approaching unity. Thus, Cauchy surfaces which entirely crunch would necessarily be part of the landscape and are in fact dominant at late times, in the semiclassical picture. This clashes with the semiclassical global picture of eternal inflation. Alternatively, if one tries to interpret the global picture of de Sitter within the Wheeler--DeWitt framework, the wavefunction cannot be dominated by arbitrarily large spatial surfaces, the hallmark of the measure problem.

One may conclude that a late-time de Sitter multiverse does not exist if Type--B transitions with unstable domain walls can occur. Alternatively, the straightforward interpretation of CDL with analytic continuation or gluing `at the waist' must be abandoned. We prefer the latter, with our analysis providing the right framework for dealing with late-time tunneling events. 

\subsubsection{A Hamiltonian approach to the Larfors--Johnson problem}

In this paper we have shown how to describe a Type--B transition from a late-time perspective. We have seen that the domain wall nucleates, grows and goes on-shell at the horizon of the parent dS. This avoids the multiverse apocalypse which, as described in the previous section, arises if the domain wall instability of Larfors and Johnson is naively combined with the CDL process at the dS waist. However, the instability in itself still persists and likely plays a key role for decay rates in the string landscape. Our late-time analysis and the techniques we have developed should allow for a computation of those rates. Specifically in the Larfors--Johnson context, our analysis should be adjusted by replacing the pure-tension domain wall model with a domain wall that couples to the volume modulus, subject to the characteristic string-derived scalar potential. The corresponding Hamiltonian system should then be studied with the methods developed in this paper. In particular, this should answer conclusively whether such transitions are forbidden or whether a smaller rate always persists, as suggested by \cite{Brown:2011ry,Blanco-Pillado:2019xny}.
This task is left for future work.

\section{Summary and Conclusions}
\label{sec:conclusions}

The Coleman--De Luccia (CDL) result \cite{Coleman:1980aw} for vacuum transition rates in de Sitter (dS) space relies on the Euclidean gravitational path integral. The analytic continuation required for physical interpretation connects half of the corresponding CDL instanton to a Lorentzian de Sitter at its `waist' (where the spatial dS sphere reaches its minimal radius, cf.~Fig.~\ref{fig:cdl_A_B}). 
As explained in more detail in the Introduction and the bulk of this paper, this leads to interpretational problems in the `gravity-dominated case', when most of the spatial dS sphere is replaced by the new vacuum: One finds that the entire late universe is in the causal future of each single bubble describing such a transition.
This is particularly hard to accept if one recalls that we expect an infinity of such transitions to occur within the enormously large late-time dS sphere. To improve this situation, we have tried to quantify whether and how small bubbles of new vacuum form locally, at very late global dS time.

To achieve this, we studied dS decays from a Lorentzian perspective, focusing explicitly on the late-time limit.
We based our analysis on the Lorentzian Einstein--Hilbert action and worked in the flat slicing of dS.
We derived a quantum-mechanical Hamiltonian depending on a single variable, the domain-wall radius $R$.
This approach is justified because, at late times, the spatial dS sphere has become very large and its fluctuations can be neglected.
We then considered a process in which a domain wall nucleates at vanishing size, traverses a potential barrier off shell, and subsequently goes on shell.
We derived the WKB tunneling exponent for all possible types of decay and found, perhaps surprisingly, agreement with the CDL results.

Our method thus provides a clear geometric interpretation of dS vacuum decay, including situations in which the decay is dominated by gravitational effects and without the pathologies apparently implied by the Euclidean path integral approach. While we succeeded in understanding, at a technical level, why the CDL calculation and ours give the same rate, an a priori physical reason for trusting the CDL instanton in predicting late-time transition rates is still not available.

Our methods can now be used to address further open questions in the dS landscape. We explained this using the Larfors--Johnson problem \cite{Johnson:2008kc,Johnson:2008vn,Aguirre:2009tp}
as an example. The problem is that unstable domain walls inhibit typical flux transitions in the string landscape. We demonstrated, relying on the causal structure noted above, that if one trusts the CDL approach this problem threatens the existence of the whole multiverse.
In our treatment, this is avoided. Moreover, the question whether some highly suppressed residual rate for flux transitions exists can hopefully be addressed at the quantitative level in the future. 

One may furthermore hope that the theory of late-time de Sitter decays studied here is helpful for obtaining a better understanding of the quantum nature of de Sitter space itself, as studied e.g.~in \cite{Witten:2001kn,Chandrasekaran:2022cip,Anninos:2022ujl,Chakraborty:2023los,Ivo:2024ill,Kaimakkamis:2024lkb,Collier:2025lux}.

Moreover, there are important open questions concerning our approach itself: First, of the many solutions of our WDW equation we picked the one suggested by quantum mechanical tunneling intuition. 
But to what extent is this justified if we really want to ask an initial-condition-dependent question in global de Sitter? Second, even when the parent dS sphere is infinitely large, local bulk fluctuations exist.
Are we certain that their effect on our result is subdominant?
Third, in the region where our bubble goes on shell, the WKB approximation clearly breaks down. Can one quantify the error introduced? 
Fourth, it would certainly be desirable to avoid our `complete gauge fixing before quantization', the main goal being to achieve a form of the WDW equation which is quadratic in the derivative operator(s).

Further tasks include the generalization of our analysis beyond the thin-wall approximation, the estimation or even calculation of the non-exponential prefactor, and the study of deviations from spherical symmetry.\footnote{See e.g.~\cite{Masoumi:2012yy,Avraham:2026mgk} for analyses of less symmetric bounces.} In particular, the effect of such deviations on the tunneling rate and the subsequent Lorentzian evolution of the domain wall after tunneling is of interest.

\subsection*{Acknowledgements}

We are very grateful for useful discussions with Max Wiesner. We are thankful for being allowed to listen to a conversation between Sagredo and Salviati about the uses of instantons. This work was supported by Deutsche Forschungsgemeinschaft (DFG, German Research Foundation) under Germany’s Excellence Strategy EXC 2181/1 - 390900948 (the Heidelberg STRUCTURES Excellence Cluster).
B.H.~acknowledges the hospitality of the Aspen Center for Physics, where this work was performed in part.
The Aspen Center for Physics is supported by National Science Foundation grant PHY-2210452 and by a grant from the Simons Foundation (1161654, Troyer).

\appendix
\numberwithin{equation}{section}

\section{Solutions for dS--dS tunneling}
\label{app:dS_dS_solutions}
In this appendix, we study solutions to the Hamiltonian constraint for dS--dS tunneling.

We begin by recalling the constraint \eqref{eq:Ham_constraint_dS_dS}:
\begin{align}
\begin{split}
\label{eq:app:Ham_constraint_dS_dS}
\mathcal{H}=0\quad\Longleftrightarrow\quad -\xi R&=\frac{1+H_1Rs-\gamma\sqrt{1+H_1R(2s+H_1R)-H_2^2R^2(1-s^2)}}{\sqrt{1-s^2}}\\
&\equiv\frac{1+H_1Rs-\gamma\sqrt{\mathcal{B}}}{\sqrt{1-s^2}}\,,
\end{split}
\end{align}
and solve it for $s$. The solutions are given by
\begin{align}\label{eq:app:s}
    s_\pm &=R_{\rm crit}^2\frac{-2H_1\xi\pm (H_1^2-H_2^2-\xi^2)\sqrt{\frac{R^2}{R_{\rm crit}^2}-1}}{2R\xi}\,,
\end{align}
with $R_{\rm crit}$ defined in \eqref{eq:R_crit_dS_dS}.
In the region $R<R_{\rm crit}$, we can rewrite this as 
\begin{align}\label{eq:app:s_im}
    s_\pm &=R_{\rm crit}^2\frac{-2H_1\xi\pm i(H_1^2-H_2^2-\xi^2)\sqrt{1-\frac{R^2}{R_{\rm crit}^2}}}{2R\xi}\,.
\end{align}
We see that for $R>R_{\rm crit}$, the solutions $s_\pm$ are real.
In this regime, they obey $|s_\pm(R)|\leq 1$ with one of the two functions, depending on the sign of the second term, touching $s=-1$ at $R=1/H_1$.\footnote{This can be seen by first noting that $|H_1^2-H_2^2-\xi^2|$ can be expressed as $2\xi\sqrt{1-H_1^2R_{\rm crit}^2}/R_{\rm crit}$ and then using the Cauchy-Schwarz inequality in \eqref{eq:app:s}.}
For Type--A down-tunneling, the solution touching $s=-1$ at $R=1/H_1$ is given by $s_-$, whereas for up-tunneling and Type--B down-tunneling, the corresponding solution is given by $s_+$.
Reality of the $s_\pm$ for $R>R_{\rm crit}$ does not mean that the system can go on-shell at $R=R_{\rm crit}$, as the above expressions are only solutions to \eqref{eq:Ham_constraint_dS_dS} with a certain choice of branches of the square roots. 
In order to go on-shell, we require real values of $s$ that solve \eqref{eq:Ham_constraint_dS_dS} with both square roots on their fundamental branches.

We recall from \eqref{eq:pR_dS_dS_A} that the imaginary part of $p_R$ takes the form
\begin{align}\label{eq:app_Im_PR}
    \frac{\Im(p_R)}{8\pi\Mp^2 R}=\Im(\arctanh(s)+\arctanh(\mathcal{A}))\,,
\end{align}
with
\begin{align}
    \begin{split}
        \mathcal{A}&\equiv H_2R+\frac{(1-H_2^2R^2)(s+H_1R)}{H_2R(s+H_1R)-\gamma\sqrt{1+H_1R(2s+H_1R)-H_2^2R^2(1-s^2)}}\\
    &=H_2R+\frac{(1-H_2^2R^2)(s+H_1R)}{H_2R(s+H_1R)-\gamma\sqrt{\mathcal{B}}}\,.
    \end{split}\label{eq:app:A}
\end{align}
As already discussed in Sect.~\ref{sec:BON}, $\arctanh(z)$ has a branch cut given by $|z|\geq 1$ on the real line.
As for the bubble of nothing, we note that $\sqrt{1-s^2}$, which appears in $\mathcal{H}$, and $\arctanh(s)$, which appears in $p_R$, have the same branch cut structure.
Hence, a sheet transition of one function goes together with a sheet transition of the other.

Let us now study, as we did for the bubble of nothing, the branch structure for our four cases of interest individually.
\paragraph{Down-tunneling:}
For all down-tunneling transitions, the bubble can go on-shell at $R<1/H_2$.
Then, it follows immediately from \eqref{eq:eta_1} that we need to consider $\gamma = +1$.
\begin{itemize}
    \item \textbf{Type--A}:
    For $R>R_{\rm crit}$, $s_+$ is a solution to the Hamiltonian constraint with the square roots on their fundamental sheets.
    Hence, the system can be on-shell.
    None of the complex functions cross a branch cut at $R=R_{\rm crit}$, as $|s_+|<1$, $|\mathcal{A}|<1$ and $\mathcal{B}>0$ hold there.
    At $R=R_{\rm crit}$, the two solutions $s_+$ and $s_-$ coincide.
    We determine the tunneling solution in the region $R<R_{\rm crit}$ by requiring  $\Im(p_R)\geq 0$ to hold.
    With both $\arctanh$ function on their principal branches, as we have just worked out, this determines the solution to again be $s=s_+$.
    In the entire region $R<R_{\rm crit}$, the arguments of both square root functions and both arctanh functions are not real, such that no branch crossings can occur.
    \item \textbf{Type--B}:
    For Type--B, the constraint \eqref{eq:Ham_constraint_dS_dS} can only be solved on the principal sheets of the square roots for $R>1/H_1$, and only the solution $s=s_+$ then does so.
    Since $s_+\neq s_-$ for $R>R_{\rm crit}$, we have to consider the solution $s=s_+$ in the entire regime $R>R_{\rm crit}$ by continuity.
    For $R_{\rm crit}<R<1/H_1$, the Hamiltonian constraint is then only satisfied when both square roots are on their second sheet.
    The transitions of sheets can occur since at $R=1/H_1$, $s_+=-1$, which defines a branch point for both square roots.
    Encircling $-1$ in the complex $s$-plane then leads to a transition of sheets for both roots.
    Since $\sqrt{1-s^2}$ crosses a branch cut, we know that the $\arctanh(s)$ function changes sheets as well.
    Like for the bubble of nothing, a contour that encircles $s=-1$ $n$ times, for $n$ an odd integer, leads to a transition of both square roots to their second branch.
    For such a contour, $\arctanh(s)$ picks up a term $in\pi$.
    The tunneling solution is characterized as the slowest decaying wavefunction with $\Im(p_R)>0$.
    This condition selects the contour associated to $n=+1$.
    The function $\arctanh(\mathcal{A})$ does not cross a branch cut at $R=1/H_1$, as $|\mathcal{A}|<1$ there. 
    No branch transitions occur at $R=R_{\rm crit}$ either.
    In the regime $R<R_{\rm crit}$, we again determine the relevant solution by demanding $\Im(p_R)\geq 0$, which fixes the solution to be $s=s_+$ in this region as well.
    As before, no sheet transitions can occur for $R<R_{\rm crit}$.
\end{itemize}

\paragraph{Up-tunneling:}
For up-tunneling, a further complication arises since $\gamma = -1$ can occur, as the bubble must go on-shell at $R>1/H_2$.
Let us begin with Type--B this time.
\begin{itemize}
    \item \textbf{Type--B}:
    It can be checked that the branch structure is the same as for down-tunneling Type--B transitions.
    For $R>1/H_1$, $s_+$ solves the Hamiltonian constraint with all functions on their fundamental sheets and $\gamma=+1$.
    It can be verified from \eqref{eq:eta_1} that $N>0$ then still holds.
    At $R=1/H_1$, both roots must transition to their second sheet in order to find a solution to the Hamiltonian constraint.
    The sheet transition can occur since again $s_+(R=1/H_1)=-1$, which is a branch point for both roots.
    The function $\arctanh(\mathcal{A})$, however, does not touch a branch cut at $R=1/H_1$. 
    As $\sqrt{1-s^2}$ crosses a branch cut at $R=1/H_1$, $\arctanh(s)$ does so as well. As before, the tunneling solution corresponds to a transition where $\arctanh(s)$ picks up a term $+i\pi$.
    In the regime $R<R_{\rm crit}$, we again have to choose $s=s_+$, as a result of demanding the tunneling condition $\Im(p_R)>0$ to hold.
    No branch transitions can occur in this region.
    \item \textbf{Type--A}:
    For Type--A up-tunneling, the Hamiltonian constraint can only be solved for $R>1/H_1$ on the fundamental sheets of the square roots if $\gamma=-1$. 
    Then, $s=s_+$ solves the constraint for $R>1/H_1$ and by continuity, it must be the solution of choice in the entire regime $R>R_{\rm crit}$.
    For $R_{\rm crit}<R<1/H_1$, \eqref{eq:Ham_constraint_dS_dS} is solved by $s=s_+$ with both square roots on their second sheet.
    As before, the transition of sheets can occur at $R=1/H_1$ as $s_+(R=1/H_1)=-1$, which is a branch point for both roots.
    As before, this means that $\arctanh(s)$ must transition sheet as well, and the tunneling solution corresponds to the transition leading to a shift of $+i\pi$.
    At $R=1/H_1$, $|\mathcal{A}|<1$, such that no sheet transition of $\arctanh(\mathcal{A})$ occurs.
    However, $\arctanh(\mathcal{A})$ touches its branch point at $R=1/H_2$, cf.~\eqref{eq:A_sp_real} and below.
    At this point, it can transition to another sheet.
    A transition such that $\arctanh(\mathcal{A})\to \arctanh(\mathcal{A})-i\pi$ is consistent with the tunneling condition $\Im(p_R)\geq 0$ and leads to the smallest tunneling suppression.
    Then, in the regime $R<R_{\rm crit}$, the tunneling condition $\Im(p_R)\geq 0$ again determines the relevant solution to be $s=s_+$.
    As before, no branch transitions can occur in this region.
\end{itemize}
\vspace{.5cm}

We thus find that for all transitions and in all regions, we have to consider the solution $s=s_+$. 
The branching structure for the various transitions is summarized in Tables~\ref{tab:Down_branches} and~\ref{tab:Up_branches}.
\paragraph{Decay exponent:}
To compute the decay exponent, we have to evaluate \eqref{eq:app_Im_PR} with $s=s_+$ and with the branches of the arctanh functions as worked out above.
For $R>R_{\rm crit}$, $s_+(R)$ is real and hence the imaginary parts of the arctanh functions in $p_R$ are simply constants.
Their value can be read off from Tables~\ref{tab:Down_branches} and~\ref{tab:Up_branches}.
Let us hence focus on $R<R_{\rm crit}$.
We recall from \eqref{eq:Im_arctanh_BoN} that $\Im(\arctanh(z))$, on its fundamental branch, can be computed as
\begin{align}
    \Im(\arctanh{z})=\arctan(\frac{2\Im(z)}{|1+z||1-z|+1-|z|^2})\,,\label{eq:app:Im_arctanh}
\end{align}
with $\arctan(x)\in (-\pi/2,\pi/2)$.
We thus now want to compute \eqref{eq:app:Im_arctanh} for both $z=s_+$ and $z=\mathcal{A}(s_+)$.

We start with
\begin{align}
    2\Im(s_+)=\frac{R_{\rm crit}^2(H_1^2-H_2^2-\xi^2)}{R\xi}\sqrt{1-\frac{R^2}{R_{\rm crit}^2}}\,,
\end{align}
and
\begin{align}
    |1\pm s_+|^2 &= \left(1\mp \frac{H_1R_{\rm crit}^2}{R}\right)^2+\frac{R_{\rm crit}^4(H_1^2-H_2^2-\xi^2)^2}{4R^2\xi^2}\left(1-\frac{R^2}{R_{\rm crit}^2}\right)\\
    &=R_{\rm crit}^2\left(\frac1R\mp H_1\right)^2\,.
\end{align}
Here, we have used (cf.~\eqref{eq:R_crit_dS_dS})
\begin{align}
	\frac1{R_{\rm crit}^2}=\frac{(H_1^2-H_2^2-\xi^2)^2+4H_1^2\xi^2}{4\xi^2}=\frac{(H_1^2-H_2^2+\xi^2)^2+4H_2^2\xi^2}{4\xi^2}\,.
\end{align}
Hence, we find
\begin{align}
    |1+s_+||1-s_+|=R_{\rm crit}^2\left(\frac{1}{R^2}-H_1^2\right)\,.
\end{align}
Using 
\begin{align}
    |s_+|^2=\frac{R_{\rm crit}^2}{R^2}-\frac{R_{\rm crit}^2(H_1^2-H_2^2-\xi^2)^2}{4\xi^2}\,,
\end{align}
we evaluate \eqref{eq:app:Im_arctanh} and arrive at
\begin{align}
    \Im(\arctanh(s_+))=\arctan(\frac{2\xi\sqrt{1-\frac{R^2}{R_{\rm crit}^2}}}{R(H_1^2-H_2^2-\xi^2)})+\delta_s\,,\label{eq:app:Im_arctanh_sp}
\end{align}
where $\delta_s=0$ or $\delta_s=\pi$ to account for the sheet of $\arctanh(s_+)$ we are on.

Next, we compute $\mathcal{A}(s_+)$ for $R<R_{\rm crit}$. 
We start with 
\begin{align}
s_++H_1R&=\frac{R_{\rm crit}^2}{R}\sqrt{1-\frac{R^2}{R_{\rm crit}^2}}\left(i\frac{H_1^2-H_2^2-\xi^2}{2\xi}-H_1\sqrt{1-\frac{R^2}{R_{\rm crit}^2}}\right)\,,\label{eq:app:Rdp}\\
\sqrt{1-s_+^2}&=\frac{R_{\rm crit}^2}{R}\left(\frac{H_1^2-H_2^2-\xi^2}{2\xi}+iH_1\sqrt{1-\frac{R^2}{R_{\rm crit}^2}}\right).\label{eq:app:sqrt_s}
\end{align}
Using these results, the Hamiltonian constraint \eqref{eq:app:Ham_constraint_dS_dS} then gives
\begin{align}
\gamma \sqrt{\mathcal{B}}=-R_{\rm crit}^2\frac{H_2^2-H_1^2-\xi^2}{2\xi}\left(\frac{H_1^2-H_2^2-\xi^2}{2\xi}+iH_1\sqrt{1-\frac{R^2}{R_{\rm crit}^2}}\right)\,.\label{eq:app:B_os}
\end{align}
Inserting \eqref{eq:app:Rdp} and \eqref{eq:app:B_os} into \eqref{eq:app:A} yields, after simplification, 
\begin{align}
\mathcal A(s_+)&=\frac{H_2R_{\rm crit}^2}{R}+i\frac{R_{\rm crit}^2(H_2^2-H_1^2-\xi^2)}{2R\xi}\sqrt{1-\frac{R^2}{R_{\rm crit}^2}}.
\end{align}
Similarly, for $R>R_{\rm crit}$, we find
\begin{align}
    \mathcal{A}(s_+)=\frac{H_2R_{\rm crit}^2}{R}+\frac{R_{\rm crit}^2(H_2^2-H_1^2-\xi^2)}{2R\xi}\sqrt{\frac{R^2}{R_{\rm crit}^2}-1}\,.\label{eq:A_sp_real}
\end{align}
We see that $\mathcal{A}(s_+)$ takes a very similar form to $s_+$ itself.
It has the analogous property that for $R>R_{\rm crit}$, $|\mathcal{A}(s_+)|\leq 1$ and, for Type--A situations, $|\mathcal{A}(s_+)|=1$ at $R=1/H_2$.
A computation similar to the previous one then shows that for $R<R_{\rm crit}$
\begin{align}
    \Im(\arctanh{\mathcal{A}(s_+)})= \arctan(\frac{2\xi\sqrt{1-\frac{R^2}{R_{\rm crit}^2}}}{R(H_2^2-H_1^2-\xi^2)})+\delta_\mathcal{A}\,,\label{eq:app:Im_arctanh_A}
\end{align}
with $\delta_\mathcal{A}=0$ or $-\pi$ determined by the sheet of the arctanh function.
The decay exponent itself is computed in the main text in Sect.~\ref{sec:dS-dS_rate}.

\bibliographystyle{utphys}
\bibliography{references}

\end{document}